\documentclass[11pt]{article}
\usepackage{fullpage}
\usepackage{graphicx} %
\usepackage{amsfonts,amsmath,amssymb,amsthm}
\usepackage{thmtools}
\usepackage{mathrsfs}
\usepackage{complexity}
\usepackage{enumitem}
\usepackage[normalem]{ulem}
\usepackage{color}
\usepackage[olditem,oldenum]{paralist} %
\usepackage{bbm} 
\newcommand{\sP}{\mathscr{P}}
\newcommand{\Z}{\mathbb{Z}}
\renewcommand{\C}{\mathbb{C}}
\newcommand{\sfS}{\mathsf{S}}
\newcommand{\AS}{\mathsf{AS}}
\newcommand{\Id}{\mathbb{I}}

\newcommand{\sL}{\mathscr{L}}
\newcommand{\sA}{\mathscr{A}}

\newcommand{\AugSigmoid}{\Sigmoid\circ(\Sigmoid\|\Poly)}

\usepackage{hyperref}
\usepackage{cleveref}
\hypersetup{hypertexnames=false}

\newcommand{\defermark}[2]{\hypertarget{deferred:#2}{\hyperlink{resolved:#2}{#1}}}
\newcommand{\resolvemark}[2]{\hypertarget{resolved:#2}{\hyperlink{deferred:#2}{#1}}}

\newcommand{\tO}{\widetilde{O}}

\newcommand{\Cor}{\mathsf{Cor}}

\newtheorem{theorem}{Theorem}[section]
\newtheorem{lemma}[theorem]{Lemma}
\newtheorem{proposition}[theorem]{Proposition}
\newtheorem{corollary}[theorem]{Corollary}
\theoremstyle{definition}
\newtheorem{definition}[theorem]{Definition}

\usepackage[most]{tcolorbox}
\usepackage{marginnote}
\newtheoremstyle{remarkbox}{0pt}{0pt}{\normalfont}{}{\bfseries}{.}{5pt plus 1pt minus 1pt}{}
\theoremstyle{remarkbox}
\newtheorem{remark}{Remark}
\tcolorboxenvironment{remark}{enhanced, breakable, colback=black!1, colframe=black!35,
  coltext=., boxrule=0.4pt, arc=2pt, left=6pt, right=6pt, top=5pt, bottom=5pt,
  before skip=10pt, after skip=10pt, before upper={\small}}

\newtheoremstyle{examplebox}{0pt}{0pt}{\normalfont}{}{\bfseries}{.}{5pt plus 1pt minus 1pt}{}
\theoremstyle{examplebox}
\newtheorem{example}{Example}
\tcolorboxenvironment{example}{enhanced, breakable, colback=black!1, colframe=black!35,
  coltext=., boxrule=0.4pt, arc=2pt, left=6pt, right=6pt, top=5pt, bottom=5pt,
  before skip=10pt, after skip=10pt, before upper={\small}}

\DeclareMathOperator*{\Ex}{\mathbb{E}}

\newfunc{\sgn}{sgn}
\newfunc{\MAJ}{MAJ}
\newfunc{\THR}{THR}
\newfunc{\GapTHR}{GapTHR}
\newfunc{\AND}{AND}
\newfunc{\OR}{OR}
\newfunc{\XOR}{XOR}
\newfunc{\SYM}{SYM}
\newfunc{\PSigmoid}{PSigmoid}
\newfunc{\LTF}{LTF}
\newfunc{\PTF}{PTF}
\newfunc{\Exp}{Exp}
\newfunc{\Var}{Var}
\newfunc{\Poly}{Poly}
\newfunc{\ETHR}{ETHR}
\newfunc{\WETHR}{WETHR}
\newfunc{\ELIN}{ELIN}
\newfunc{\IPtwo}{IP2}

\newcommand{\cR}{\mathcal{R}}

\newclass{\CC}{CC}

\hypersetup{
colorlinks=true,
linkcolor=red,
citecolor=blue,
urlcolor=red
}

\newcommand{\diff}[1]{{\color{blue}#1}}

\newcommand{\zo}{\{0,1\}}
\newcommand{\zoR}{[0,1]}
\renewcommand{\R}{\mathbb{R}}
\newcommand{\eqdef}{:=}
\newcommand{\eps}{\varepsilon}
\newfunc{\LTH}{LTH}
\newfunc{\Sigmoid}{Sigmoid}

\newfunc{\ReLU}{ReLU}
\newfunc{\SiLU}{SiLU}
\newcommand{\N}{\mathbb{N}}

\title{New Gate, Weight and Wire Complexity Bounds\\ for Sigmoid Neural Networks
}
\author{
  Jiatu Li\thanks{Supported by MIT Schwarzman College of Computing PhD Fellowship. Part of work is undertaken during the program Logical Foundations of Complexity Theory (LFC) organized by the Isaac Newton Institute for Mathematical Sciences, Cambridge; the LFC program was supported by EPSRC
grant EP/Z000580/1.} \\
  MIT \\
  \texttt{jli@mit.edu}
  \and
  Daniel Reichman \\
  WPI \\
  \texttt{dreichman@wpi.edu}
  \and
  Ryan Williams\thanks{Work supported in part by NSF CCF-2420092 and the AI Security Institute (AISI).} \\
  MIT \\
  \texttt{rrw@mit.edu}
  \and
  Tony Wu \\
  MIT \\
  \texttt{tonyyzwu@mit.edu}
}
\date{}

\begin{document}

\pagenumbering{roman}

\maketitle


\begin{abstract}
    Feedforward neural networks with sigmoid activations are of fundamental importance in neural computation. However, our understanding of the circuit complexity of such networks for implementing Boolean functions remains limited. The analog output of the sigmoid, along with its transcendental nature, leads to unique mathematical challenges for proving lower bounds.

    Introducing new techniques, we establish new upper and lower bounds on the complexity of sigmoidal networks computing Boolean functions: 
    \begin{itemize}
    \item We prove nearly tight bounds for the wire complexity of computing PARITY with such networks of constant depth and polynomial-bounded weights. 
    \item We extend the size lower bound of Kane and Williams [STOC'16] against depth-$3$ threshold circuits, to depth-3 sigmoidal networks with polynomial weights, proving a $\Omega(n^{1.5}/ (\log n)^{O(1)})$ size lower bound for computing an explicit function in $\sf P$.
    \item We prove that every symmetric function on $n$ bits can be computed by a depth-$3$ sigmoidal network with $O(n)$ wires and polynomial weights. This stands in stark contrast with threshold circuits of $d=O(1)$ depth, where $n^{1+\alpha_d}$-wire lower bounds are known for $\alpha_d > 0$.
    \end{itemize}
    We also answer a few fundamental problems related to sigmoidal networks, including proving new simulations and separations between threshold circuits and sigmoidal networks. 
    
    We establish these results using a variety of techniques, including random restrictions for sigmoid gates, analyzing the average sensitivity of networks with sigmoids, simulation arguments, and using tools from approximation theory, complex analysis, and algebraic geometry. 
\end{abstract}


\newpage 

\setcounter{tocdepth}{2}
\enlargethispage{2\baselineskip}%
\tableofcontents

\newpage 

\pagenumbering{arabic}

\section{Introduction}

For decades, researchers have studied the expressive power of neural networks~\cite{minsky1988perceptrons,mcculloch1943logical} with respect to natural parameters, such as depth, size (number of neurons or number of connections), and the activation functions used. 
Our focus here is on neural networks computing Boolean functions (i.e.,~binary classification problems on discrete inputs).

The basic building block of a neural network is a \emph{gate} (or computation unit, or neuron). A fan-in $m$ gate has $m$ incoming wires, where the $i$-th wire has a real \emph{weight} $w_i\in\R$. Fix a real function $\phi:\mathbb{R}\rightarrow \mathbb{R}$, commonly referred to as the \emph{activation function} or simply activation. The output of the gate is $\phi(\sum_{i=1}^mw_ix_i+\theta)$ where $\theta\in\R$ is called \emph{bias}. From now on, the number of input variables in the network is denoted as $n$.

There are many possible functional forms the activation function can have, such as polynomial, piecewise linear and sigmoidal. Two of the most studied activations are the \emph{threshold} (or \emph{Heaviside}) activation $\tau(x)$ and the \emph{sigmoid} activation
\[
\tau(x) = \Id[x\ge 0], \qquad \Sigmoid(x):=\frac{1}{1+e^{-x}}.
\]
Given an activation function $\sigma$, a $\sigma$-\emph{circuit} is defined as a circuit with gates computing functions of form $\vec x\mapsto \sigma(\langle \vec w,\vec x\rangle + \theta)$. In particular, the ordinary linear threshold circuit (see \cite[Section~11]{Jukna12}) is a $\sigma$-circuit with $\sigma(x)$ being a threshold function. %
For a $\sigma$-circuit $C:\zo^n\to\R$ and a Boolean function $f:\zo^n\to\zo$, we say $C$ computes $f$ with $\eps$ separation, if there exists $\tau\in[0,1]$ such that 
\begin{equation}
f(x) = 0 ~\Longrightarrow~ C(x)\le \tau \qquad \text{and} \qquad f(x) = 1 ~\Longrightarrow~ C(x) > \tau+\eps. \label{equ: compute def}
\end{equation}
In particular, $C$ computes $f$ \emph{with no separation} if $\eps = 0$, and it is said to compute $f$ \emph{with noticeable separation} if $\eps=\Omega(n^{-k})$ for some $k\ge 1$.
Our focus in this work is proving upper and lower bounds for various complexity measures (number of gates, wires and magnitude of weights) for neural networks with sigmoid activations.

Beyond their foundational role in machine learning, sigmoidal networks serve as the canonical testbed for a central question in neural computation:
\begin{quote}
\emph{When do continuous, smooth analog neurons confer a provable advantage over discrete neurons?} \cite{MSS94, DS96, Koiran96, KS97} 
\end{quote}
Unlike threshold gates (which are discrete) or ReLU and polynomial activations (which are piecewise-linear or algebraic), the sigmoid combines local curvature with the exponential function $e^{-x}$. %
On the one hand, this makes sigmoidal gates a core primitive across modern machine learning architectures and verification problems---underpinning Boltzmann machines \cite{hinton1986learning, martens2013representational}, gating mechanisms, softmax and attention layers \cite{aggarwal2022optimal, merrill2022saturated, dooley2026vc}, graph neural networks \cite{grohe2024descriptive, d2025vc}, and formal DNN verification \cite{isac2023dnn}. %
On the other hand, from a circuit complexity perspective, passing continuous real values across wires and introducing exponential non-linearities fundamentally alters the landscape of bounded-depth computation, rendering classical tools for threshold circuits ($\TC^0$) challenging to apply and raising basic questions about whether smooth analog networks can bypass long-standing Boolean circuit barriers \cite{SRK94, MSS94, Koiran96}. %

\paragraph{Circuit Lower Bounds for Neural Networks.} While we have built up a decent (yet still very incomplete) understanding of circuits with threshold gates (see, e.g., \cite{haastad1991power,GHR92,HMPSMT93,ROS94,Hastad94,PS94,IPS97,HP10,KW16,KKL17,CSS18,HHTT23,FLY22}), sigmoidal networks have been a far more challenging representation to understand \cite{Koiran96,SRK94}.
This is perhaps not surprising, as sigmoidal networks are at least as powerful as threshold circuits %
(see \Cref{lmm: threshold by sigmoidal}). 
Indeed, it is also known that for certain Boolean functions, sigmoidal networks can compute these functions with \emph{asymptotically few} gates compared to threshold circuits implementing the same functions \cite{MSS94,DS96,Koiran96}.

Additionally, lower bound techniques for threshold circuits %
simply do not apply in the sigmoidal setting. At a high level, the Boolean output of threshold gates makes them amenable to a variety of methods from computational complexity such as communication complexity \cite{Nisan93,ROS94}, the discriminator lemma \cite{HMPSMT93}, and random restrictions \cite{IPS97,CSS18,HHTT23,FLY22,KKL17}. There seems to be no portable analog of these in the sigmoidal case. 

\begin{example}
An important component in establishing lower bounds for threshold circuits is \emph{upper bounding} the magnitude of weights of threshold circuits: it is known that if a Boolean function is computed by a threshold circuit $C$ then without loss of generality the size of every weight in $C$ is at most $2^{O(n \log n)}$ (and weights of order $2^{\Omega(n \log n)}$ are sometimes necessary); see, e.g., \cite{Hastad94}. Devising weight upper bounds for sigmoidal networks in terms of the input dimension is not known, limiting the applicability of techniques to prove lower bounds that rely on the boundedness of the weights, e.g., Williams' algorithmic approach \cite{Tamaki16,ACW16,CTW26}. 
\end{example}

In the zero separation case, the transcendental nature of the sigmoid function is an additional complicating factor. While simulation results of neural networks with piecewise linear (such as ReLUs) or piecewise polynomial activation of constant depth using threshold circuits with a polynomial blowup are known~\cite{Maass97,MSS94}, such results are unknown for sigmoidal networks in the zero separation case. %

\paragraph{Motivating Questions.} These challenges translate into noticeable gaps in terms of known lower bounds for sigmoidal networks versus threshold circuits. While the study of the latter has seen progress in recent years, the former has not seen progress since the 90's. %
\begin{compactitem}
\item While superlinear gate lower bounds have been established for \emph{threshold circuits} of depth $2$ (with arbitrary weights) \cite{Tamaki16,ACW16,CTW26} and depth $3$ (with polynomial weights) \cite{KW16}, comparable lower bounds are not known even for sigmoidal networks with polynomial weights computing Boolean functions with a noticeable separation. %
\item Several works have successfully applied the random restriction method to prove lower bounds on the wire complexity of well studied functions such as parity for threshold circuits both in the worst and average case \cite{PS94,IPS97,CSS18} as well as the more general case of circuits with polynomial threshold functions \cite{KKL17}. However, we are not aware of any lower bounds on the wire complexity of sigmoidal networks for computing parity or other Boolean function.
\item Finally, a linear lower bound is known on the gate complexity of threshold circuits (with arbitrary depth and weights) computing the IP Mod 2 function using communication complexity \cite{ROS94}. For sigmoidal networks computing with zero separation the best size lower bound is $\Omega(n^{1/4})$ for the Boolean selection function relying on upper bounds of the VC dimension of such networks~\cite{karpinski1997polynomial} (which relies on results from algebraic geometry~\cite{khovanskiui1991fewnomials}) \cite{Koiran96}. 
\end{compactitem}

We establish new lower and upper bounds for sigmoidal networks computing Boolean functions related to some (though not all) of the open questions above.

\subsection{Our Results}

We are interested in proving upper and lower bounds for the number of gates (i.e.~neurons), number of wires (i.e.~non-zero parameters), and magnitude of weights in the following parameter regimes:
\begin{description}[itemsep=0pt,topsep=6pt]
\item[Robust sigmoidal circuits:] Sigmoidal circuits with \emph{polynomially bounded weights and biases} that compute Boolean functions with \emph{noticeable separation} (see \cite{MSS94,PS94,SRK94,DS96}). 
\item[Unrestricted-weight sigmoidal circuits:] Sigmoidal circuits with arbitrary weights computing Boolean functions with \emph{no separation\footnote{While computing without separation allows arbitrarily small margins that may be infeasible under finite-precision hardware, we view this as a natural theoretical regime worthy of study. Indeed, the same convention is standard in unrestricted threshold circuits, where gate margins can be exponentially small in the input dimension.}} (see \cite{Koiran96}), which is a generalization of $\THR$ circuits.
\item[Small-weight sigmoidal circuits:] Sigmoidal circuits with \emph{polynomially bounded weights} that compute Boolean functions with \emph{no separation}, which is a generalization of $\MAJ$ circuits.
\end{description}

Robust sigmoidal circuits are indeed equivalent to $\MAJ$ circuits of the same depth up to a polynomial overhead (see \Cref{lmm: sigmoidal by threshold,lmm: threshold by sigmoidal}). Therefore, existing lower bounds against depth-$2$ $\MAJ$ circuits \cite{HMPSMT93,Forster02,FKLMSS01} applies directly. The more interesting question is whether we can prove lower bounds for depth three and beyond given, e.g., the $\tilde{\Omega}(n^{1.5})$ lower bound\footnote{The notation $\tilde{\Omega}(\cdot)$ means that we ignore poly log factors} against depth-$3$ $\MAJ$ circuits from \cite{KW16} and the super-linear wire lower bound \cite{IPS97,CSS18} against $\TC^0$. These lower bounds do not translate directly due to the polynomial overhead. 

The unrestricted-weight model might be surprisingly strong: it is not immediately clear whether we can simulate them by even $\P/\poly$. On the other hand, however, it is also unclear whether it is much stronger than threshold circuits.

\paragraph{Wire Complexity of Parity.} A classical question in circuit complexity and also machine learning is the complexity of computing the parity function $\oplus_n(x_1,\dots,x_n)\eqdef x_1\oplus \dots\oplus x_n$ by neural networks. We fully resolved (up to poly-log factors) the wire complexity of polynomial weight sigmoidal circuits computing the parity function with noticeable separation.

\begin{theorem}[see \Cref{thm: symmetric upper bound}, \Cref{prop: depth two upper parity}, and \Cref{cor: sigmoid lower bound}]\label{thm: wire intro}
Every symmetric Boolean function (including parity) is computable by a depth-$3$ robust sigmoidal circuit with $O(n)$ wires. 

Any symmetric function (including parity) is computable by a depth-2 robust sigmoidal circuit with $O(n^2)$ wires, and least $\Omega(n^2/\log^2n)$ wires are required for parity in the same model. 
\end{theorem}

The linear wire upper bound for $d\ge 3$ stands in sharp contrast with the known super-linear wire complexity lower bound for parity against threshold circuits for \emph{any} $d\ge 1$ (see, e.g., \cite{IPS97,KKL17,CSS18}). It shows that there is a provable advantage for real-valued activation functions (compared to Boolean-valued activation) in wire complexity measure.

\paragraph{Lower Bounds for Constant-Depth Polynomial Weight Sigmoidal Circuits.} 
Given the separation between sigmoidal and threshold circuits in wire complexity, it is natural to consider whether known threshold circuit lower bounds can be adapted to sigmoidal circuits.

\smallskip\noindent\emph{Average-case Lower Bound for Parity.} The number of gates and wires required to compute the parity function is a classical question in circuit complexity. For worst-case computation, almost optimal wire lower bounds are known for both threshold circuits \cite{IPS97} (following \cite{PS94}), and almost optimal gate lower bounds are known for polynomial-weight sigmoidal circuits \cite{SRK94}. Chen, Santhanam, and Srinivasan \cite{CSS18} extends \cite{IPS97} to show a near-optimal average-case lower bound on the number of wires of threshold circuits computing parity. 

We prove an average-case lower bound against polynomial weight sigmoidal circuits.\footnote{Average case lower bounds for computing Boolean functions by neural networks are desirable as they can be used to achieve lower bounds for approximating functions in the $\ell^2$-norm that is common in machine learning applications~(see, e.g., \cite{vardi2020neural}).}

\begin{theorem}[see \Cref{thm: correlation parity gate}]\label{thm: correlation intro}
For every $d\ge 2$, and any function $f:\zo^n\to\zo$ that is computable by a depth-$d$ robust sigmoidal circuit using $O(n^{1/(2(d-1))-\Omega(1)})$ gates, the correlation between $f$ and $\oplus_n$ is at most $O(n^{-\Omega(1)})$. 
\end{theorem}

The lower bound matches that in \cite{CSS18} and is thus also near optimal. Indeed, it is a corollary of an upper bound on the \emph{average sensitivity} of functions computable by sigmoidal circuits, which is of independent interests. It is worth noting that the results of \cite{CSS18} are obtained to analyze the \emph{noise sensitivity} of threshold circuits; we suspect both techniques (with appropriate adaption) would work for both settings.

\begin{theorem}[see \Cref{thm: AS sigmoid}]\label{thm: AS sigmoid intro}
    For any $d\ge 1$, and any function $f:\zo^n\to\zo$ that is computable by a depth-$d$ robust sigmoidal circuit using $s$ gates, the average sensitivity of $f$ is at most $O(s^{1-1/d}\cdot n^{1-1/(2d)+o(1)})$. 
\end{theorem}

To derive this lower bound, we need to analyze an extension of sigmoidal circuits, where the bias of each gate is replaced by a low-degree polynomial. This is possible due to a strong random restriction lemma for \emph{polynomial threshold circuits} \cite{KKL17}; see \Cref{sec: intuition} and the technical sections.

\smallskip\noindent\emph{Explicit Lower Bounds for Depth-3 Circuits.} Recall that Kane and Williams \cite{KW16} proved that there is a polynomial-time computable function that requires at least $\Omega(n^{1.5}/(\log n)^{O(1)})$ gates and $\Omega(n^{2.5}/(\log n)^{O(1)})$ wires to compute by $\THR\circ\THR$. In this work, we show that this lower bound can be adapted to sigmoidal circuits with \emph{polynomial weights} and \emph{noticeable separation}. 

\begin{theorem}[see \Cref{thm: depth-3 lower bound,thm: depth-3 wire lower bound}]\label{thm: depth-three intro}
    There is a polynomial-time computable function $f:\zo^n\to\zo$ that requires $\Omega(n^{1.5}/(\log n)^{O(1)})$ gates and $\Omega(n^{2.5}/(\log n)^{O(1)})$ wires to compute by depth-three robust sigmoidal circuits. 
\end{theorem}

We note that the lower bound is nontrivial in that it cannot be obtained by simply simulating sigmoidal circuits with threshold circuits: the simulation involves a gate overhead corresponding to weights (see \Cref{lmm: sigmoidal by threshold}). The wire lower bound is particularly interesting because of the surprising upper bound in \Cref{thm: wire intro}. 

We prove the lower bound by generalizing the technique in \cite{KW16}. Instead of directly using their results, we have to re-engineer the entire proof \cite{KW16} in white box as most parts of the proof are not applicable to sigmoidal circuits. The main issue is that the key random restriction lemma does not work for sigmoidal circuits: unlike threshold gates, sigmoidal gates have continuous output and cannot be well approximated by constants after random restrictions. To circumvent this problem, we prove a random restriction lemma for sigmoidal gates that further uses \emph{Taylor approximation}, and shows that it suffices for the framework in \cite{KW16}.

\paragraph{Sigmoidal Circuits with Unbounded Parameters.} The results above work only for robust sigmoidal circuits. A natural question then is to study the nature of unrestricted-weight sigmoidal circuits. This is indeed a surprisingly tough model to analyze: very few upper and lower bounds are known. In this work, we answer a few fundamental questions related to this model.

\smallskip\noindent\emph{A Weight Upper Bounds for Sigmoidal Circuits.} For threshold circuits, it is known that $\THR$ gates with unbounded parameters can be simulated by $\THR$ gates with integer weights at most $2^{O(n\log n)}$ (see \cite{Hastad94} and references therein). As $\THR$ is Boolean valued, it further implies that any threshold circuit can be simulated by one with the same topology but weights bounded by $2^{O(n\log n)}$. This serves an fundamental tool to analyze threshold circuits (see, e.g., \cite{GHR92,HP10}). 

However, this result does not easily generalize to sigmoidal circuits as it is real-valued and we thus cannot deal with each gates individually. We prove a weight upper bound for depth-two sigmoidal circuits using tools from computational algebraic geometry \cite{BPR96,BSP06}. 

\begin{theorem}[see \Cref{thm: Sigmoid weight bound depth 2}]
    Suppose that $f:\zo^n\to\zo$ is computable by a depth-two unrestricted-weight sigmoidal circuit with $s$ gates. It is computable by a circuit that has the same architecture with weights and biases having absolute values at most $2^{O(sn\log(sn))}$.
\end{theorem}

\noindent\emph{Explicit Lower Bounds for Depth-2 Sigmoidal Circuits.} Proving explicit lower bound against unbounded weight sigmoidal circuits appears to be a surprisingly hard problem. Prior to this work, the best known lower bound against constant-depth sigmoidal circuits with unbounded weights is only $\Omega(n^{1/4})$ \cite{Koiran96}. Following Koiran's approach, we observe that there is a near-linear lower bound\footnote{This is nearly optimal as there is a depth-two sigmoidal circuit of linear size computing selection~\cite{Koiran96}.} for depth-two sigmoidal circuits.

\begin{theorem}[see \Cref{thm:depth2-lb}]
The selection function $f(x,y)\eqdef y_x$ requires $\Omega(n/\log n)$ gates to compute by any depth-two unrestricted-weight sigmoidal circuits. 
\end{theorem}

\noindent\emph{Advantage of Sigmoids.} Sigmoidal circuits are known to be asymptotically stronger than threshold circuits in various settings \cite{MSS94,DS96,Koiran96}. For circuits with unbounded weight, it is known that the Selection function $f(x,y)=y_x$ requires $\Omega(n/\log n)$ threshold gates, while there is an $O(\sqrt n)$ upper bound by Sigmoidal circuits \cite{Koiran96}. We show that there is a similar upper bound for another well-studied function $\IPtwo(\vec x,\vec y)\eqdef \langle \vec x,\vec y\rangle\bmod 2$. 

\begin{restatable}[see \Cref{sec: IP mod 2 lower bound}]{theorem}{IPtwoUpper}\label{thm: IP2 sublinear Sigmoid}
    $\IPtwo$ can be computed by an unrestricted-weight sigmoidal circuit with $O(\sqrt{n\log n})$ gates.
\end{restatable}

This upper bound is particularly interesting as an $\Omega(n)$ lower bound is known for threshold circuits \cite{ROS94}, and a $\Omega(n/\log n)$ lower bound is known for ReLU networks \cite{VRPS21}; in contrast, the Selection function can be computed by ReLU networks of size $O(\sqrt n)$ \cite{Koiran96}. It therefore demonstrates an asymptotic separation between sigmoidal and ReLU networks. Finally we note that the improved upper bound above for the Inner Product function is the only result in this paper concerning a sigmoidal circuit whose depth grows with $n$.

\paragraph{An Equivalence between Sigmoidal and Threshold Circuits.} We are also interested in the intermediate small-weight model that generalizes $\MAJ$ circuits: sigmoidal circuits with small weights that computes Boolean functions with no separation. It is tempting to ask whether this model is strictly stronger than its discrete counterpart, i.e., $\MAJ\circ\MAJ$. 

We show that depth-two circuits in this model is at least as strong as $\THR\circ\MAJ$, and indeed, a further restriction to integer weights make it exactly equivalent to $\THR\circ\MAJ$. 

\begin{restatable}{theorem}{SigmoidEquiv}\label{thm: equiv sigmoid threshold}
    Let $n\in\N$ be sufficiently large and $s=s(n),W=W(n)\ge 1$ be functions, and $f:\zo^n\to\zo$ be a Boolean function. The following two statements are equivalent:
    \begin{compactenum}
    \item[(1)] $f$ can be computed by a depth-two sigmoidal circuit of size $s\cdot \poly(n,W)$ with no separation, where all weights are integers in $[-W,W]$ and all biases are real numbers in $(-1,1)$.
    \item[(2)] $f$ can be computed by a $\THR\circ\MAJ$ circuit of size $s\cdot \poly(n,W)$, where the bottom $\MAJ$ gates have weight taken from $\{-W,\dots,W\}$. 
    \end{compactenum}
    Moreover, the output gate of the circuit in (1) can be assumed to have weights in $\{\pm1\}$.
\end{restatable}

The implication (2) $\Rightarrow$ (1) gives an interesting structural characterization of $\THR\circ\MAJ$. It shows that sigmoidal circuits can be surprisingly strong even with small integer weights and small real biases. In particular, as exponential separations between $\THR\circ\MAJ$ and $\MAJ\circ\MAJ$ is known \cite{GHR92}, this also indicates that real-valued activation functions are exponentially stronger than threshold activation if we restrict to the polynomial weight regime.

Following a similar approach, we also prove a structural upper bound for simulating $\THR\circ\THR$ with sigmoidal circuits with top weights in $\{\pm 1\}$. Since $\MAJ\circ\THR\subsetneq\THR\circ\MAJ\subsetneq\THR\circ\THR$ \cite{GHR92,CM22}, sigmoidal circuit with small top weights is stronger than its counterpart for threshold sigmoidal circuits ($\MAJ\circ\THR$) and also the restricted sigmoidal circuits in \Cref{thm: equiv sigmoid threshold}.

\begin{restatable}{theorem}{SigSimTHR}\label{thm: sigmoid simulates THR THR}
Suppose that $f:\zo^n\to\zo$ is computed by a $\THR\circ\THR$ circuit of size $s$. Then $f$ is computed with no separation by a depth-two sigmoidal circuit of size $s\cdot\poly(n)$ whose output gate has weights in $\{\pm 1\}$.
\end{restatable}

We also note that the simulation of threshold circuits using sigmoidal circuits in \Cref{thm: equiv sigmoid threshold,thm: sigmoid simulates THR THR} also generalize to larger depth; see \Cref{sec: further simulation} for more details.

\section{Intuition}
\label{sec: intuition}

\subsection{Wire Complexity for Symmetric Functions}

Recall that Parity requires $\widetilde\Theta(n^2)$ wires for $\MAJ\circ\MAJ$ \cite{PS94}, and super-linear wires for constant depth threshold circuits \cite{IPS97,CSS18}. Assume that $n$ is even for simplicity. The depth-two construction is as follows: for each $k\in\{1,3,5,\dots,n-1\}$, create a gate $G_k(\vec x)$ that outputs $1$ iff $|\vec x|=k$, and take an OR of all such gates. Each gate needs $O(n)$ wires, and therefore there are $O(n^2)$ wires in total. %

The depth-three sigmoidal circuit for Parity (and indeed any symmetric function) is built on the same construction. The key observation is, instead of letting $G_1,G_3,\dots,G_{n-1}$ read all inputs by themselves, we can create a gate $G_{\sf sum}$ that outputs $\Sigmoid(\frac{1}{n}\sum_{i=1}^nx_i)$, so $G_1,G_3,\dots,G_{n-1}$ only need to read the output of $G_{\sf sum}$ and collect $|\vec x|$ by appropriate scaling. The total number of wires is $O(n)$ as $G_1,G_3,\dots,G_{n-1}$ have fan-in $1$, and the weights and biases of all gates are $\poly(n)$.

The wire complexity upper bound for depth-two sigmoidal circuits follows by simulating the $\MAJ\circ\MAJ$ circuit gate by gate. The matching lower bound follows by generalizing the \emph{rational approximation} technique in \cite{PS94}. 

In more detail, the proof of \cite{PS94} shows that if the activation function can be well approximated by a low-degree rational function, then every function computed by a depth-two network with $o(n^2/\log^2 n)$ wires and polynomial weights and biases with noticeable separation can be sign-represented by a rational function of degree less than $n/2$ (see \Cref{thm: PS94 restated}). The lower bound then follows from the fact that a rational function $P/Q$ sign-representing Parity has degree at least $n/2$ \cite{PS94}. Subsequently, it suffices to show that $\Sigmoid(x)$ can be well approximated by a low-degree rational function. Fortunately, this has been studied in \cite{SRK94}, and there is an approximation by combining Taylor expansion and an approximation of the sign function in \cite{PS94}.  

\subsection{Lower Bounds from Random Restriction} 

We now explain the intuition behind \Cref{thm: depth-three intro,thm: correlation intro,thm: AS sigmoid intro}, which all use the random restriction method (but in slightly different manners).

\paragraph{Random Restriction Method, and Why it is Nontrivial for Sigmoid.} One of our main technical contributions is to show that the random restriction method, which has been widely used to prove lower bounds against threshold circuits (see, e.g., \cite{IPS97,KW16,CSS18,KKL17,HHTT23,FLY22}), can be adapted to analyze sigmoidal circuits. 

A random restriction lemma (see, e.g., the one in \cite{KW16}) for $\THR$ states that for any threshold gate $\THR(\vec x)\eqdef\sgn(\langle\vec w,\vec x\rangle + \beta)$, after applying a random assignment to all but a few random bits, the function likely becomes a constant function on all (or almost all) assignments on free variables. It is often used to simplify a circuit by removing the bottom layer, as many gates become constants and can be absorbed into the gates above it. 

In our setting, gates involve real-valued activation functions (e.g.~$\Sigmoid$), so we cannot hope for them to become a constant under random restriction. A natural attempt is to say, instead, that a gate will likely become close to a constant. This is indeed possible by approximating $\Sigmoid(x)$ by a piece-wise constant function; specifically, letting $t$ and $\delta$ be two parameters, we write:
\begin{equation}
\Sigmoid(x)\approx \Id[x \ge (t+1)\delta] + \sum_{-t\le j\le t} \Sigmoid(j\delta)\cdot \Id[j\delta \le x< (j+1)\delta],\label{equ: piecewise constant}
\end{equation}
and subsequently approximate a sigmoid gate by a linear combination of $O(t)$ $\THR$ gates. Then we can apply the random restriction lemma for $\THR$ gates, and use a \emph{union bound} to say that it also works for sigmoid gates: if every threshold gate (used to test $\Id[j\delta\le x< (j+1)\delta]$) is trivialized, the sigmoidal gate will be close to a constant up to an error $\delta$.  

However, this appears to be insufficient to prove lower bounds: The random restriction lemmas for threshold circuits have an error probability $n^{-\Omega(1)}$ if we want to leave $n^{\Omega(1)}$ variables free (which is usually required for lower bound techniques), so we could afford at most $t=n^{\eps}$ for a constant $\eps < 1$, and subsequently $\delta\ge n^{-\eps}$ (as otherwise the error for $x<-t\delta$ or $x\ge(t+1)\delta$  will be too large). An approximation of a gate in the bottom layer with error $n^{-\eps}$ can \emph{not} be ignored, as the error might be amplified by higher layers of the circuit and affect the final output. 

\paragraph{Random Restriction Lemma.} To prevent the aforementioned issue, we indeed need a random restriction lemma for $\Sigmoid$ with an approximation error at most $n^{-c}$ for \emph{any} $c$. This appears to be impossible if we approximate it by a constant after random restriction: $\Sigmoid(\sum_{i=1}^n x_i / n^2)$ is not close to any constant after any partial restriction.  

We therefore use approximation by \emph{polynomials}. Instead of using the constant $\Sigmoid(j\delta)$ in \Cref{equ: piecewise constant}, we use a degree-$k$ Taylor approximation of $\Sigmoid(x)$ at $x_0\eqdef j\delta$. For any $x\in[j\delta,(j+1)\delta)$, this estimation has an error about $\delta^{k}$; if we choose $t=\polylog(n)$, $\delta \eqdef 1/\polylog(n)$, and $k=O(\log n/\log\log n)$, the error can be boosted to $n^{-c}$ for any fixed $c\ge 1$. Subsequently, we can show that for any sigmoidal gate, after a random restriction, it becomes close to a low-degree polynomial with error small enough to avoid affecting the final output. 

\paragraph{Lower Bounds for Depth-$3$.} Recall that \Cref{thm: depth-three intro} shows lower bounds comparable to results of Kane and Williams \cite{KW16} for $\MAJ\circ\THR\circ\THR$ circuits. Indeed, our main contribution is an adaption of their proof to work for sigmoidal circuits. 

\medskip\noindent\emph{Kane-Williams Lower Bounds.} We first review the result in \cite{KW16}, focusing on the $n^{1.5}/(\log n)^{O(1)}$ gate lower bound. The hard function $F$ is called the Generalized Andreev's function. It takes $(\vec x,\vec y)\in\zo^n\times\zo^m$ as input and    satisfies the following property: For every set $\sL$ of Boolean functions on $m$ variables, if $|\sL|\le 2^{o(n)}$, there exists an $\vec x\in\zo^n$ such that for every $f\in\sL$, the correlation between $f$ and $F(\vec x,\cdot)$ is extremely small. 

Suppose for contradiction that a small $\MAJ\circ\THR\circ\THR$ circuit $C$ computes $F(\vec x,\vec y)$. Then by a random restriction $\rho$ to some variables in $\vec y$, many gates in the first layer will be trivialized. In that case, $C|_\rho$ can be computed by a $\MAJ\circ\THR\circ\THR$ circuit $C'$ where there are only \emph{a few bottom layer gates}. We then apply a standard tool called the discriminator lemma \cite{HMPSMT93} to show that there exists a second layer gate $G$ of $C'$, which is a $\THR\circ\THR$ circuit with a few bottom gates that has nontrivial correlation with $F(\vec x,\cdot)$ for every $\vec x$. We will show that this is impossible.  

Let $\sL$ be the set of all depth-two circuits with a few bottom gates. By counting the total number of parameters, we can argue that $|\sL|\le 2^{o(n)}$, and subsequently there exists $\vec x$ such that $F(\vec x,\cdot)$ has very low correlation with every function in $\sL$. Fix this $\vec x$. Then in particular, the gate $G$, which is a $\THR\circ\THR$ circuit with a few bottom gates, has very low correlation with $F(\vec x,\cdot)$. This leads to a contradiction. 

\medskip\noindent\emph{Adaption to Sigmoidal Circuits.} There are two main challenges to adapt the Kane-Williams lower bound to sigmoidal circuits. First, the discriminator lemma \cite{HMPSMT93} is used to remove the top $\MAJ$ gate, while we need to remove a $\Sigmoid$ gate. To address this, we observe that as we assume polynomial weights for all $\Sigmoid$ gates, it is more similar to $\MAJ$ compared to $\THR$, so we can discretize the top $\Sigmoid$ gate and apply the same proof idea. 

The second challenge is that the random restriction lemma for $\Sigmoid$ gates simplifies them only to low-degree polynomials rather than constants, and in particular, each of the low-degree polynomials may have large weights. We call the subcircuits corresponding to the second-layer gates (in the simplified circuit) $\AugSigmoid$ circuits, as the bottom layer consists of both $\Sigmoid$ gates and low-degree polynomials. 

We may not use the same counting argument in \cite{KW16} as there are too many polynomials (one for each simplified gate). Nevertheless, as the top $\Sigmoid$ gates compute a linear combination of bottom gates, and polynomials are \emph{closed} under linear combination, we can merge all polynomials into one. Therefore, the total number of such polynomials can be bounded in terms of the number of free inputs and the degree, both of which are independent of the number of gates in the original circuit. We use this to conclude that the total number of possible $\AugSigmoid$ circuits with certain parameters is small enough to for the argument in \cite{KW16}. 

\begin{remark}
    We note that another technique often used in analyzing neural networks is rational approximation (see, e.g., \cite{PS94,SRK94}). Within the same degree, rational approximation could have exponentially better precision (see, e.g., \cite{Newman64} and \Cref{sec: sigmoid rational approx}). However, the proof above crucially relies on the fact that polynomials are closed under linear combination \emph{without degree blow up}, which is not satisfied by rational approximation; therefore it is unclear whether we can use rational approximation to improve this lower bound.
\end{remark}

\begin{remark}
It can be observed that the random restriction lemma works for any sigmoid gate (i.e.~with unbounded weights). The reason that the lower bounds work only for bounded weights is that otherwise (1) we cannot control the propagation of error (caused by the approximation of the bottom layer by polynomials) to the final output, and (2) we cannot apply the discriminator lemma. Indeed, the lower bound in \Cref{thm: depth-three intro} works even if the bottom layer has unbounded weights.
\end{remark}

\paragraph{Correlation Bound with Parity.} For threshold circuits, Chen, Santhanam, and Srinivasan \cite{CSS18} proved an almost-tight average-case lower bound on both the number of gates and wires. We show that the gate lower bound can be adapted to sigmoidal circuits with polynomial weights.

Our proof follows from a somewhat different approach compared to that in \cite{CSS18}. The lower bound in \cite{CSS18} is proved by analyzing the \emph{noise sensitivity} of threshold circuits with random restriction; we also use the random restriction method, while we instead prove an \emph{average sensitivity} upper bound for functions computable by sigmoidal circuits. (Nevertheless, we suspect both ours and the approach in \cite{CSS18} work in both cases.)

We first recall some basic concepts from Boolean function analysis. Let $f:\zo^n\to\zo$ be a function and $x\in\zo^n$. The sensitivity of $f$ at $x$ is defined as the number of $i$ such that flipping the $i$-th bit of $x$ will change $f(x)$. The average sensitivity of a function $f:\zo^n\to\zo$, denoted $\AS(f)$, is defined as the expected sensitivity at a random input $x\gets\zo^n$. The relative average sensitivity of $f$ is defined as $\AS(f)/n$. We need an observation of Boppana \cite{Boppana97} that over a random restriction $\rho$, the expected relative average sensitivity of $f|_\rho$ is exactly the relative average sensitivity of $f$. 

We prove an average sensitivity upper bound for functions computable by depth-$d$, size-$s$ sigmoidal circuits by induction on $d$. The base case $d=1$ can be derived from the average sensitivity upper bound for linear threshold functions. Consider the induction case. Suppose we are given a circuit $C$ of depth $d+1$. By the observation of Boppana \cite{Boppana97}, we only need to upper bound the expected average sensitivity of $C|_\rho$ over a random restriction $\rho$. We hope to simplify $C$ by the random restriction lemma and apply the induction hypothesis for depth $d$. The only problem is that the sigmoidal gates simplify only to low-degree polynomials, and we cannot convert the resulting circuit into a depth $d$ sigmoidal circuit.

To circumvent this issue, we strengthen the induction hypothesis to work for extended sigmoidal circuits, in which the bias of each gate is replaced by a low-degree polynomial in input variables (rather than a constant). It can then absorb any polynomials attached to the bottom layer. As a result, the base case should be changed to work with polynomial threshold functions ($\PTF$), and the random restriction lemma must work for $\PSigmoid$ gates. Fortunately, the average sensitivity of $\PTF$ has been resolved by Kane \cite{Kane14}, and the random restriction lemma for $\PSigmoid$ can be proved by a random restriction lemma for $\PTF$ in \cite{KKL17} (in place of \cite{KW16}).

\subsection{Unrestricted Weight Sigmoidal Circuits}
\label{sec: overview unrestricted}

The lower bounds above all require small weights (and possibly also biases) for various technical reasons. Now we explain our results on depth-two sigmoidal circuits with \emph{unbounded parameters}. 

\paragraph{Parameter Upper Bound for Depth-$2$.} The main difficulty to generalize the weight bounds (see \cite{Hastad94}) for threshold functions to sigmoidal circuits is that
\begin{compactenum}
\item the gates are real-valued, so we cannot hope to replace them by bounded weight gates with bounded parameters without incurring an approximation error; 
\item even for depth-two circuits, approximation error in the bottom layer will be amplified by the next layer, so it may change the function computed by the circuit. 
\end{compactenum}
Therefore, we cannot hope to prove an upper bound for parameters by adjusting the parameters gate-by-gate, or even layer-by-layer. 

The key idea is to consider the depth-two circuits as a whole. Suppose that a function $f:\zo^n\to\zo$ is computed by a depth-two sigmoidal circuit with no separation. As the $\Sigmoid$ function is monotone, we can remove the output gate so that 
\begin{equation}
f(\vec x) = \Id \left[c + \sum_{i=1}^{h}v_{i}\Sigmoid \left(b_i + \sum_{j=1}^{n}w_{ij}x_j\right) > 0\right].\label{equ: sigmoid depth two}
\end{equation}
As $\Sigmoid(x)=1/(1+e^{-x})$, we can observe that if we write $W_{ij} = e^{-w_{ij}}$ and $B_{i}=e^{-b_i}$, for every \emph{fixed} input $\vec x\in\zo^n$, the RHS of \eqref{equ: sigmoid depth two} is the indicator of a \emph{polynomial} inequality in $c,\{v_i\},\{B_i\},\{W_{ij}\}$. We can then write a system of $2^n$ polynomial inequalities with small integer weights (by enumerating all $2^n$ inputs) that defines the feasible set of parameters such that the circuit computes $f$.\footnote{Indeed, we also need $B_i,W_{ij}>0$ due to the change of variables.} 

Our theorem then follows from a generic result due to Basu, Pollack, and Roy \cite{BPR96} that upper bounds the bit complexity of solutions for systems of polynomial inequalities (see also \cite{BSP06} for a detailed explanation). Remarkably, their upper bound depends only on the number of variables, the weights and degree of polynomials, and is independent of the number of inequalities (which is $2^n$); for this reason, we can upper bound all weights and biases by a single exponential in $n$ and $s$ (rather than exponential in $2^n$).  

\paragraph{Lower Bound for Selection.} The only technique for proving lower bounds against unrestricted sigmoidal circuits is due to Koiran \cite{Koiran96}. Fix a topology of a sigmoidal circuit. Let $N_{\vec w}(x;\vec v)$ be the circuit where some of the parameters are fixed as $\vec w$, while other parameters (called \emph{programmable parameters}) are left unspecified. We can define the family of real functions computed by the $N_{\vec w}$ as the set of functions computed by $N_{\vec w}(x;\vec v)$ for any possible $\vec v$. 

Koiran \cite{Koiran96} observed that if a sigmoidal circuit $C$ of size $s$ computes the Selection function $f(x,y)\eqdef y_x$, where $x\in\zo^{\log n}$ and $y\in\zo^{n}$, by making $\le s$ parameters (corresponding to $y$) programmable, the family of functions could have VC dimension at least $n$. Combining this with an $O(s^2m^2)$ VC dimension upper bound for sigmoidal circuits with $s$ gates and $m$ programmable parameters, he proved an $\Omega(n^{1/4})$ lower bound for computing Selection. 

Our lower bound is proved by showing a stronger VC dimension upper bound for depth-two Sigmoidal circuits. Specifically, for any depth-two sigmoidal circuit of size $s$, the family of functions computed by making the biases programmable is at most $O(s \log s)$. This is proved using a VC dimension bound for polynomial threshold functions in \cite{AB02,BW96}. Plugging the upper bound to Koiran's approach, we can obtain the $\Omega(n/\log n)$ lower bound. 

\paragraph{Upper Bound for $\IPtwo$.} Recall that $\IPtwo(\vec x,\vec y)\eqdef \langle\vec x,\vec y\rangle\bmod 2$. We explain the upper bound with $O(\sqrt{n\log n})$ gates; the other one follows from a similar idea. Let the input length be $2n$. The upper bound relies on two tricks: 
\begin{compactitem}
\item Multiplication of two real numbers $x,y$ can be approximated by a linear combination of four functions of form $\Sigmoid(ax+by+c)$, where $a,b,c\in\R$ \cite{DS96}. 
\item Let $C(\vec x)$ be any sigmoidal circuits and $\eps\in(0,1)$, then there is a sigmoidal circuit $D$ with one input and has $k$ gates such that $|(D\circ C)(\vec x)-b_k|\le \eps$, where $b_k$ denotes the $k$-th bit in the binary expansion of $C(\vec x)\in(0,1)$ (see, e.g., \cite{Koiran96}). 
\end{compactitem}

Let $\vec x=(\vec x_1,\dots,\vec x_k)$ and $\vec y=(\vec y_1,\dots,\vec y_k)$ be a partition of the input into $k$ blocks of size $r=n/k$. The key observation is that we can compute $\langle \vec x_i,\vec y_i\rangle$ by looking into the binary expansion of 
\[
p_i \eqdef \left(\sum_{j=0}^{r-1}2^{jl}\vec x_{i,j}\right)
    \left(\sum_{j=0}^{r-1}2^{(r-1-j)l}\vec y_{i,j}\right)
\]
where $\vec x_i=(\vec x_{i,0},\dots,\vec x_{i,r-1})$ and $\vec y_i$ likewise. In more detail, the coefficient for the term $2^{(r-1)l}$ is exactly $\langle\vec x_i,\vec y_i\rangle$. When $l=O(\log n)$ is large enough, $\langle\vec x_i,\vec y_i\rangle\bmod 2$ is exactly the $(r-1)l$-th bit of $p_i$; indeed, we can even read out $\sum_{i=1}^k \langle \vec x_i,\vec y_i\rangle\bmod 2=\IPtwo(\vec x,\vec y)$ from the $(r-1)l$-th bit of $\sum_{i=1}^k p_i$. 

Therefore, we can compute $\IPtwo(\vec x, \vec y)$ by having a layer of $O(k)$ sigmoidal gates computing $p_i$, taking the sum of them, and then read out the $(r-1)l$-th bit using $O(rl)$ gates. By taking $k=\sqrt{n\log n}$, the circuit is of size $O(\sqrt{n\log n})$.

\subsection{Equivalence between Sigmoidal and Threshold Circuits}

We first explain the intuition behind \Cref{thm: equiv sigmoid threshold}. For simplicity, we assume $s(n),W(n)=n^{O(1)}$; the theorem then states that depth-two sigmoidal circuits with polynomial integer weights and real biases in $(-1,1)$ compute, with no separation, the same functions as $\THR\circ\MAJ$ of polynomial size.

\smallskip\noindent\emph{Simulating Sigmoid with $\THR\circ\MAJ$.} The simulation of sigmoidal circuits follows from the observation that each gate in the bottom layer has $\poly(n)$ possible pre-activation values. Therefore, we can replace each sigmoidal gate in the bottom layer by a linear combination of $\MAJ$ gates (with no error). The top sigmoidal gate can be replaced by a threshold gate as $\Sigmoid$ is monotone, so the circuit computes the same Boolean function with no separation even before activation.

\smallskip\noindent\emph{Simulating $\THR\circ\MAJ$ with Sigmoid.} The main nontrivial direction is to simulate $\THR\circ\MAJ$ with small integer weight sigmoidal circuits. Without the constraint to have small integer weights, we can simulate $\THR\circ\MAJ$ as follows. Let $T\subseteq\R$ be a finite set of size $t$. For any distinct $\beta_1,\dots,\beta_t\in(-1,1)$, it can be proved that the matrix $(\Sigmoid(u+\beta_j))_{u\in T,j\in[t]}$ is invertible. Subsequently, every function $\phi:T\to\R$ can be represented as a linear combination of $f_j(u)\eqdef \Sigmoid(u+\beta_j)$. For each $\MAJ$ gate $G_i(\vec x)$ in the bottom layer, letting $w_i\in\R$ be its weight in the top threshold gate, we can replace $w_i\cdot G_i(\vec x)$ by a linear combination of at most $t$ sigmoidal gates (by taking $T$ as a $\poly(n)$-sized set). The top threshold gate can thus be replaced by a sigmoidal gate applied to the sum of all sigmoidal gates.

This simulation does not work directly as the real coefficients to represent $w_i\cdot G_i(\vec x)$ as linear combinations of sigmoidal gates are not guaranteed to be small integers; we can make the weights small by normalization, but they may no longer be integers. The key idea is to apply the above idea with the \emph{derivative} of $\Sigmoid$. With essentially the same argument, we can represent $w_i\cdot G_i(\vec x)$ as a linear combination of $g_j(u)\eqdef \Sigmoid'(u+\beta_j)$. We can further represent
\begin{equation}
\delta\cdot \Sigmoid'(u+\beta_j)\approx \Sigmoid(u+\beta_j+\delta) - \Sigmoid(u+\beta_j) \label{equ: sigmoid derivative}
\end{equation}
when $\delta$ is small. We then can encode the coefficients into the biases of sigmoidal gates, so the weights of the top gate are only $\pm 1$. (With a more careful argument via the inverse function theorem, we can obtain an exact version of \eqref{equ: sigmoid derivative}, see \Cref{lmm: exact difference}.)

\smallskip\noindent\emph{Extension to $\THR\circ\THR$.} The main obstacle to extend this argument to simulate $\THR\circ\THR$ is that the bottom gates may have large weights and thus the pre-activation values no longer come from a small set. To resolve this issue, we use a structural result that $\THR\subseteq\mathsf{DOR}\circ\ETHR$ \cite{HP10}, where $\mathsf{DOR}$ (disjoint OR) denotes an OR with a promise that at most one of the inputs is $1$. Using a slightly different construction, we can simulate each of the $\ETHR$ gates approximately as a linear combination of $\Sigmoid'$ functions. By applying the construction as in \eqref{equ: sigmoid derivative}, we can use differences of sigmoidal gates to simulate $\Sigmoid'$, which leads to the final simulation. 

\subsection{Related Work}

\paragraph{Complexity of Threshold Circuits.} The rising popularity of neural networks in the 80's and 90's have led researchers to study the complexity of threshold circuits. This study has proven challenging, and several basic questions remain wide open despite much attention. For depth-2 threshold circuits with polynomial integer weights either for gates in the first or second layer, exponential size lower bounds are known for the IP Mod $2$ function \cite{HMPSMT93,Nisan93,Forster02,FKLMSS01}. However, the best lower bound against depth-$2$ threshold circuits without weight restriction is only $\Omega(n^{2.5}/\log^{O(1)}(n))$ \cite{CTW26}, which improves upon the near-quadratic lower bound \cite{Tamaki16,ACW16}; these are built on Williams' algorithmic method (see \cite{Williams14}). 

For depth-3, no super polynomial size lower bounds are known, even if \emph{all} gates have polynomial integer weights. In fact, establishing super linear lower bounds for depth-3 circuits with polynomial weights was proved relatively recently \cite{KW16}; their $\Omega(n^{3/2}/\log^{O(1)} n)$ gate lower bound remains the best known to date. %

\paragraph{Threshold vs Sigmoid networks.}
Both~\cite{maass1997networks,DS96} pioneered the study of how threshold and signoid circuits compare in computing Boolean functions. With respect to sigmoid circuits, their focus is on \emph{upper bounds} and they do not consider lower bounds for such networks. 
For their upper bounds they rely on the following: While threshold gates only depend on the sign of their pre-activation, a smooth activation $\gamma(x)$ possesses a curved ``middle zone'' around points $x_0$ where $\gamma''(x_0) \neq 0$, allowing a constant number of gates to locally approximate degree-$2$ polynomials. This, as well as the analog output of sigmoid gate has led researchers to believe that the use of sigmoids in neural architectures should lead in some cases to more efficient representations~\cite{KS97}. 

Several works have supported this belief. 
Maass, Schnitger, and Sontag \cite{MSS94} used this property to show that $f(x,y) = \mathsf{Majority}(x) \oplus \mathsf{Majority}(y)$ can be computed with noticeable separation by a depth-$2$ sigmoidal circuit with $5$ gates, whereas depth-$2$ threshold circuits with noticeable separation require $\omega(1)$ gates, by exploiting the middle zone of $\Sigmoid$ function.
Their lower bound holds for $\THR \circ \THR$ circuits and is proved using random restrictions for threshold gates. 
DasGupta and Schnitger \cite{DS96} consider stronger separation against arbitrary threshold circuits (no depth or weight restriction): they showed that unary squaring, $\mathrm{SQ}(x,y) = \mathbb{I}[|x|^2 \le |y|]$ for $x \in \{0,1\}^n$ and $y \in \{0,1\}^{n^2}$ (where $|z|$ is the Hamming weight of $z$), is computable by a depth-$2$ sigmoidal circuit of constant size, whereas $\mathsf{TC}^0$ requires $\Omega(\log n)$ gates (which is nearly tight).

The ``middle zone'' idea is used (along with additional arguments) to establish the improved upper bound on the Inner Product function (see \Cref{thm: IP2 sublinear Sigmoid}). 
On the other hand, our proofs or equivalence between depth-2 sigmoidal circuits computing with zero separation and $\THR \circ \MAJ$ circuits (e.g., \Cref{thm: equiv sigmoid threshold}) rely on different ideas than those in~\cite{MSS94,DS96}. 

\paragraph{Wire and gate complexity of Parity and Andreev's function.}
We are not aware of previous work studying the wire complexity of sigmoidal networks implementing Boolean functions (neither for upper or lower bounds). We are also not aware of previous study of the study of random restrictions in the context of sigmoidal networks with Boolean inputs and outputs.

For depth-$2$ majority circuits computing $\mathrm{Parity}$, Paturi and Saks \cite{PS94} used Newman's rational approximation of $|x|$ \cite{Newman64, PP87} to prove an $\Omega(n^2/\log^2 n)$ wire lower bound and $\Omega(n)$ gate lower bound. %
Building on \cite{PS94}, Siu, Roychowdhury, and Kailath \cite{SRK92, SRK94} proved nearly matching $\Theta(d n^{1/(d-1)})$ worst-case and $\Theta(n^{1/(2(d-1))})$ average-case \emph{gate} lower bounds for depth-$d$ circuits computing $\mathrm{Parity}$ (see also \cite{Bruck90, ABFR94, BBGL08, SO15, AT19} for related polynomial sign-representation bounds). %
Because the lower bounds of \cite{SRK92, SRK94} rely on rational approximation, they apply only when weights and biases are polynomially bounded. 

For general (unbounded-weight) threshold circuits of depth $d$, worst-case gate ($\Omega(n^{1/(2(d-1))})$) and super-linear wire ($n^{1+\exp(-O(d))}$) lower bounds for $\mathrm{Parity}$ were later proved by Impagliazzo, Paturi, and Saks \cite{IPS97} via 
random restrictions. %
Chen, Santhanam, and Srinivasan \cite{CSS18} strengthened \cite{IPS97} to the average case under the uniform distribution: they proved $n^{-\Omega(1)}$ correlation bounds for $\mathrm{Parity}$ against depth-$d$ threshold circuits with $o(n^{1/(2(d-1))})$ gates or $n^{1+\epsilon_d}$ wires, and $\exp(-n^{\Omega(1)})$ correlation bounds for the Generalized Andreev function \cite{Andreev87, IN93, PZ93, Hastad98, IMZ12, KR13}.
Concurrently, Kane and Williams \cite{KW16} proved $n^{1.5-o(1)}$ gate and $n^{2.5-o(1)}$ wire lower bounds against $\mathsf{MAJ}\circ\mathsf{THR}\circ\mathsf{THR}$ circuits. Kabanets, Kane, and Lu \cite{KKL17} extended super-linear wire lower bounds to depth-$d$ PTF circuits. %

Our work shows that while the gate and depth-$3$ wire lower bounds of \cite{KW16, CSS18} extend to polynomial-weight sigmoidal circuits with noticeable separation, 
the super-linear wire lower bound of \cite{IPS97, CSS18} for $\mathrm{Parity}$ fails for sigmoidal circuits of depth $d \ge 3$ (Theorem~\ref{thm: wire intro}). 

\paragraph{VC dimension.}
When weights are unrestricted and separation is zero, VC dimension and related shattering bounds are the primary known tools for proving gate lower bounds \cite{Koiran96, KS97, maass1997networks, AB02}. Using Khovanskii's fewnomial bounds \cite{khovanskiui1991fewnomials}, Karpinski and Macintyre \cite{karpinski1997polynomial} upper bounded the VC dimension of sigmoidal and Pfaffian networks with $s$ gates and $w$ programmable parameters by $O(s^2 w^2)$ (recent works have extended such VC dimension and expressivity analyses to Graph Neural Networks \cite{grohe2024descriptive, d2025vc, dooley2026vc}). 

Koiran \cite{Koiran96} applied \cite{karpinski1997polynomial} to prove an $\Omega(n^{1/4})$ gate lower bound for the Boolean Selection (indexing) function on $\log_2 n + n$ bits against unrestricted sigmoidal circuits of arbitrary depth, while showing that Selection has depth-$3$ sigmoidal circuits of size $O(\sqrt{n})$ but requires $\Omega(n/\log n)$ gates for unrestricted threshold circuits (Maass \cite{maass1997networks} proved an analogous $\Omega(n^{1/4})$ sigmoidal and $\Omega(n/\log n)$ threshold gate lower bound for the Boolean Coincidence Detection function $\mathrm{CD}_n(x,y) = \bigvee_{i=1}^n (x_i \wedge y_i)$).

\subsection{Discussion}

Our results reveal an interesting landscape for expressing Boolean functions using neural networks varying activations: for certain complexity measures and Boolean functions, sigmoidal and threshold circuits have similar circuit complexity, while for others, sigmoidal networks lead to significant savings. %
Some of our results apply to additional activation functions (e.g., ReLU activations), demonstrating noticeable differences of lower and upper bounds in terms of implementation of Boolean functions depending on whether the activation function used is threshold, sigmoid or ReLU. For example, our upper bound for representing the inner product function using sigmoid networks is provably better than what can be achieved by neural networks with ReLUs~\cite{VRPS21}.

Our investigation of the complexity of sigmoidal circuits reveals several intriguing open problems:

\begin{compactenum}
\item Prove super-linear wire complexity lower bound for sigmoidal circuits with polynomial weights, large constant depth that computes a function with noticeable separation. The function is necessarily not a symmetric function by \Cref{thm: wire intro}. 
\item Prove a weight upper bound for depth-three unrestricted sigmoidal circuits and beyond; our weight upper bound uses change-of-variable trick that do not seem to work for depth three. Such weight upper bounds could be a first step in extending our lower bounds for depth $3$ networks with polynomial weights (\Cref{thm: depth-three intro}) to depth $3$ networks with arbitrary weights. Alternatively, it might be that sigmoidal networks with unbounded weights and bounded depth (say, depth $3$) in the zero separation model are significantly stronger than threshold circuits and can compute functions that are believed to be outside $\TC_0$. Obtaining new results in this regard is an interesting future direction. 
\end{compactenum}


\addtocontents{toc}{\protect\setcounter{tocdepth}{1}}
\section{Preliminaries}
\label{sec: preliminaries}

\textbf{Notation.} Let $\Sigmoid(x)\eqdef (1+e^{-x})^{-1}$ and $\Sigmoid_{\vec w,\theta}(\vec x)\eqdef \Sigmoid(\langle \vec x,\vec w\rangle+\theta)$. Let $\sigma:\R\to\zoR$ be a function, a $\sigma$-circuit is a circuit with unbounded fan-in gates of form $\sigma_{\vec w,\theta}(\vec x)\eqdef \sigma(\langle \vec w,\vec x\rangle+\theta)$, where $\vec w=(w_1,\dots,w_k)\in\R^k$ is said to be the \emph{weight} of the gate, and $\theta\in\R$ is said to be the \emph{bias} of the gate. In particular, a Sigmoidal circuit is a $\sigma$-circuit with $\sigma\eqdef \Sigmoid$. 

We also use $\Sigmoid$ to denotes a layer of Sigmoid gates, and thus $\Sigmoid\circ\Sigmoid$ denotes depth-two sigmoidal circuits.

For $x\in\zo^n$, we use $|x|$ to denote the Hamming weight of $x$. Similarly, for $x,y\in\zo^n$, we use $|x-y|$ to denote the Hamming distance between $x$ and $y$. Let $\sgn:\R\to\zo$ where $\sgn(x)=1$ iff $x> 0$.

\begin{definition}[separation]
Let $\eps=\eps(n)$. We say that a $\sigma$-circuit $C$ computes $f:\zo^n\to\zo$ with $\eps$-separation, if there exists $\theta\in[0,1]$ such that for every $x\in\zo^n$, 
\begin{align*}
& f(x) = 0 \quad \Longrightarrow \quad C(x)\le \theta  \\ 
& f(x) = 1 \quad \Longrightarrow \quad C(x)> \theta + \eps
\end{align*}

It is said to compute $f:\zo^n\to\zo$ with \emph{noticeable separation} if it has separation $\Omega(n^{-k})$ for some $k\ge 1$. It is said to compute $f$ with \emph{no separation} if it has separation $0$. It is said to compute $f$ \emph{exactly} if $C(x)=f(x)$ for every $x\in\zo^n$. 
\end{definition}

\subsection{Lipschitz Property of \texorpdfstring{$\sigma$}{sigma}-circuits}

\begin{definition}
    Let $f_1,f_2:\zo^n\to\R$ be real valued functions on Boolean inputs. They are $\delta$-almost $\eps$-close to each other, if 
    \[
    \Pr_{\vec x\in \zo^n}[|f_1(x)-f_2(x)|\le \eps ] \ge 1-\delta, 
    \] 
    and are said to be $\eps$-close to each other if they are $0$-almost $\eps$-close to each other. 
\end{definition}

\begin{definition}[Lipschitz functions]
    A function $\sigma:\R^n\to\zoR$ is said to be $\kappa$-Lipschitz if for every $\vec x,\vec y\in\R^n$, 
    \[
    |\sigma(\vec x) - \sigma(\vec y)|\le \kappa\cdot \|\vec x-\vec y\|_1.
    \] 
\end{definition}

\begin{lemma}[Lipschitz property of $\sigma$-circuits]\label{lmm: sensitivity}
    Let $\sigma:\R\to\zoR$ be a $\kappa$-Lipschitz function. For every depth-$d$ $\sigma$-circuit $C$ with $n$ inputs, $s$ wires, and the maximum absolute value of weights is $W\in\R$ such that $\kappa sW > 1$, it computes a $(\kappa s W)^d$-Lipschitz function. 
\end{lemma}

\begin{proof}
    Let $\kappa_d\eqdef (\kappa s W)^d$, we prove this lemma by induction on $d$. The base case $d=0$ is trivial, so it suffices to consider the induction case. 
    
    Let $G$ be the output gate and $C_1,\dots,C_k$ be the sub-circuits that feed $G$, $k\le s$, each of size at most $s$. Let $\vec x_1,\vec x_2\in\R^n$ be inputs and $\delta=\|\vec x_1-\vec x_2\|_1$, we need to prove that 
    \[
    \left|G(\vec c_1) - G(\vec c_2)\right| \le \kappa_d\cdot \delta, 
    \]
    where $\vec c_i\eqdef (C_1(\vec x_i),\dots,C_k(\vec x_i))$ for $i\in\{1,2\}$. By the induction hypothesis, each sub-circuit $C_j$ computes a $\kappa_{d-1}$-Lipschitz function, and thus $\|\vec c_1-\vec c_2\|_1\le \kappa_{d-1}\cdot \delta k\le \kappa_{d-1}\cdot \delta s$. 

    It then suffices to prove that $G$ is a $(\kappa W)$-Lipschitz function. Recall that each gate of a $\sigma$-circuit is of form $\sigma(\langle \vec w,\vec x\rangle+\theta)$, where $\vec w=(w_1,\dots,w_k)\in[-W,W]^k$. Then for $\vec c_1,\vec c_2\in\R^k$, 
    \begin{align*}
    \left|G(\vec c_1)-G(\vec c_2)\right| &\le \left|\sigma(\langle \vec w,\vec c_1\rangle+\theta) -\sigma(\langle \vec w,\vec c_2\rangle+\theta)\right| \\
    &\le \kappa \cdot \left|\langle \vec w,\vec c_1\rangle-\langle \vec w,\vec c_2\rangle\right| \tag{$\sigma$ is $\kappa$-Lipschitz}\\ 
    &= \kappa \cdot |\langle w,\vec c_1-\vec c_2\rangle|\\ 
    &\le \kappa W\cdot \|\vec c_1-\vec c_2\|_1. \tag{$\vec w\in[-W,W]^k$}
    \end{align*}
    This completes the proof. 
\end{proof}

\subsection{Analysis of Boolean Functions}

\begin{definition}[correlation]
    The \emph{correlation} between Boolean functions $f,g:\zo^n\to\zo$ is the bias of their agreement probability over a uniformly random input:
    \[
    \Cor(f,g)\eqdef \left|\Pr_{\vec x\gets\zo^n}[f(\vec x)=g(\vec x)] - \tfrac{1}{2}\right|.
    \]
\end{definition}

\begin{definition}[$\eps$-balanced]
    A Boolean function $f:\zo^n\to\zo$ is said to be \emph{$\eps$-balanced} if $\bigl|\Pr_{\vec x\gets\zo^n}[f(\vec x)=1]-\tfrac{1}{2}\bigr|\le \eps$.
\end{definition}

We will need the following lemma that converts a discriminator-type lower bound into a lower bound on $\Cor(f,g)$.

\begin{lemma}\label{lmm: discriminator to correlation}
    Let $f:\zo^n\to\zo$ be $\eps$-balanced, and suppose $g:\zo^n\to\zo$ satisfies
    \[
    \left|\Pr_{\vec x\gets\zo^n}[g(\vec x)=1\mid f(\vec x)=1]-\Pr_{\vec x\gets\zo^n}[g(\vec x)=1\mid f(\vec x)=0]\right| \ge \delta.
    \]
    Then $\Cor(f,g)\ge \delta/2-\eps$.
\end{lemma}

\begin{proof}
    Define $p_b\eqdef \Pr_{\vec x\gets\zo^n}[g(\vec x)=1\mid f(\vec x)=b]$ for $b\in\zo$, and $\eta\eqdef \Pr_{\vec x\gets\zo^n}[f(\vec x)=1]-\tfrac{1}{2}$; by the assumption, $|p_1-p_0|\ge\delta$ and $|\eta|\le\eps$. We can see that 
    \begin{align*}
    \Pr_{\vec x\gets\zo^n}[f(\vec x)=g(\vec x)]
    &= \Pr_{\vec x\gets\zo^n}[f(\vec x)=1]\cdot p_1 + \Pr_{\vec x\gets\zo^n}[f(\vec x)=0]\cdot (1-p_0) \\
    &= \bigl(\tfrac{1}{2}+\eta\bigr)p_1 + \bigl(\tfrac{1}{2}-\eta\bigr)(1-p_0) \\
    &= \tfrac{1}{2} + \tfrac{1}{2}(p_1-p_0) + \eta\,(p_1+p_0-1).
    \end{align*}
    Subsequently, by $|p_1+p_0-1|\le 1$, we have
    \[
    \Cor(f,g)\ge \tfrac{1}{2}|p_1-p_0| - |\eta|\cdot |p_1+p_0-1| \ge \tfrac{\delta}{2}-\eps. \qedhere
    \]
\end{proof}

We need the definition of average sensitivity of Boolean functions, and a theorem from \cite{Kane14} on the average sensitivity of polynomial threshold functions.  

\begin{definition}[average sensitivity]
    Let $f:\zo^n\to\zo$. The \emph{sensitivity} of $f$ at the input $x$ is defined as 
    \[
    \sfS(f,x)\eqdef \sum_{i=1}^n|f(\vec x)-f(\vec x^{(i)})|,
    \]
    where $\vec x^{(i)}$ denotes the input with the $i$-th bit flipped. The \emph{average sensitivity} of $f$ is defined as $\AS(f)\eqdef\mathbb{E}_{\vec x}[\sfS(f,x)]$, where $\vec x$ is sampled uniformly from $\zo^n$.  
\end{definition}

\begin{theorem}[\cite{Kane14}]\label{thm: PTF AS}
    Let $p(\vec x)$ be a degree-$d$ multilinear polynomial and $f(\vec x)\eqdef \sgn(p(\vec x))$, then $\AS(f)\le \sqrt n (\log n)^{O(d\log d)} 2^{O(d^2\log d)}$.  
\end{theorem}

\subsection{Relationship with \texorpdfstring{$\TC^0$}{TC0} Circuits}
\label{sec: relation to TC}

We now consider the relationship between Sigmoidal circuits and $\TC^0$ circuits, i.e., linear threshold circuits. It is known that these two circuit model (assuming polynomial weights) simulates each other within polynomial overhead. However, from a more fine-grained view, simulation of Sigmoidal circuits with $\TC^0$ incurs an substantial size blowup, while the other direction is almost for free. 

\paragraph{Simulating Sigmoidal circuits with $\TC^0$.} The result of Maass, Schnitger, and Sontag \cite{MSS94} shows that $\poly(n)$-weight depth-$d$ Sigmoidal circuits can be simulated by depth-$d$ threshold circuits with $\poly(n)$ size overhead.

This simulation is obtained by replacing Sigmoid gates in depth $1, 2,\dots,d$, where depth $1$ is the layer fed by input variables, by a step function (using a linear combination of threshold gates) that approximates Sigmoid. The correctness of the simulation is derived from the sensitivity of Sigmoidal circuits, see, \Cref{lmm: sensitivity}. By a close inspection of the proof, we can obtain: 

\begin{lemma}[implicit in \cite{MSS94}]\label{lmm: sigmoidal by threshold}
    Let $C(x)$ be a Sigmoidal circuit with depth $d$, size $s$, and weight within $[-W, W]$ that computes the Boolean function $f:\zo^n\to\zo$ with separation $\eps$. Then there exists a $\TC^0$ circuit $C'$ that computes the same Boolean function $f$ of depth $d$, size $\tO((W s)^d \cdot \eps^{-1})$, and with integer weights of $O_d\!\left(\log\!\left((n+s)W/\varepsilon\right)\right)$ bits. 
\end{lemma}
As a corollary, sigmoidal circuits with polynomial weights (i.e.~$W\le \poly(n)$) can be simulated by Majority circuits with polynomial size overhead. In particular: 

\begin{corollary}\label{corr: simulation}
    Any function $f:\zo^n\to\zo$ computable by depth-two, polynomial weights, and polynomial sized Sigmoidal circuits with noticeable separation can also be computed by polynomial-sized $\MAJ\circ\MAJ$ circuits. 
\end{corollary}

\begin{corollary}[\cite{HMPSMT93}, see also \cite{Razborov92-THR}]
Inner Product function $\IP(x,y)\eqdef \langle x,y\rangle \bmod 2$ is not computable by polynomial sized $\Sigmoid\circ\Sigmoid$ circuits with polynomial weights and noticeable separation. 
\end{corollary}

\begin{remark}\label{remark: no lb via simulation}
    Note that the simulation incurs a size overhead that is polynomial in the weight $W$ (i.e.~exponential in the bit length of weights). As a result, more fine-grained lower bounds, such as the gate and wire complexity lower bounds against threshold circuits due to Impagliazzo, Paturi, and Saks \cite{IPS97}, do not immediately yield nontrivial lower bounds against Sigmoidal circuits. 
\end{remark}

\paragraph{Simulating $\TC^0$ with Sigmoidal circuits.} We can also simulate $\TC^0$ with Sigmoidal circuits at almost no cost. Formally: 

\begin{lemma}\label{lmm: threshold by sigmoidal}
    Let $C(x)$ be a $\TC^0$ circuit with depth $d$, size $s$, and integer weight at most $W$ that computes Boolean function $f:\zo^n\to\zo$. Then there is a Sigmoidal circuit with depth $d$, size $s$, and weight at most $\tO(sW)$ that computes $f$ with noticeable separation. 
\end{lemma}

\begin{proof}[Proof Sketch]
    We replace each threshold gate $G(x) := \sgn(\langle \vec w,\vec x\rangle + \theta)$ by the following sigmoid gate $G'(x) := \Sigmoid(\langle\vec w', \vec x\rangle + \theta')$, where 
    \[
    \vec w' = (\log^2(sW))\cdot \vec w,\quad \theta' = \log^2(sW)\cdot (\theta-1/2), 
    \]
    shifting the bias by $1/2$ to account for $\langle\vec w, \vec x\rangle + \theta = 0.$
    
    Fix any input $\vec x$ to the circuit. We can prove by induction on the topological order of gates that for each gate $G'$ in the new Sigmoidal circuit, the output of $G'$ differs from the output of $G$ by at most $1/(10sW)$, and in particular, the output is within $b\pm 1/(10sW)$ for some $b\in\zo$. Therefore, the output of $C'$ is within $C(x)\pm 1/(10sW)$, which implies that it computes $f$ with noticeable separation. 
\end{proof}

Note that the proof above utilizes the property that $\Sigmoid(x)$ converges exponentially to $0$ (resp.~$1$) for $x\to-\infty$ (resp.~$x\to+\infty$).

\begin{remark}\label{remark: lb with separation is hard}
Subsequently, super-polynomial lower bounds for depth-two Sigmoidal circuits with unbounded weights would imply super-polynomial lower bounds for depth-two $\TC^0$ circuits, a major open problem in circuit complexity. 
\end{remark}

Another corollary is that the size blowup in \Cref{lmm: sigmoidal by threshold} is likely unavoidable if we require the threshold circuits to have small weights. Recall that there is an explicit function in $\THR\circ\THR$ that requires $2^{\Omega(n^{1/4})}$-size $\THR\circ\MAJ$ circuits \cite{CM22}. As $\THR\circ\THR$ can be simulated by $\Sigmoid\circ\Sigmoid$, we have that: 

\begin{corollary}
There is a function $f:\zo^n\to\zo$ computable by depth-two polynomial-sized Sigmoidal circuits that requires $2^{\Omega(n^{1/4})}$-size to compute by $\THR\circ\MAJ$ circuits. 
\end{corollary}
\addtocontents{toc}{\protect\setcounter{tocdepth}{2}}

\section{Wire Complexity for Sigmoidal Circuits}

In this section, we explore interesting \emph{wire complexity} upper and lower bounds for sigmoidal circuits. Note that sparse sigmoidal circuits (i.e.~circuits with few wires) models \emph{sparse} neural networks, where some of the weights are fixed to be $0$. 

\subsection{Symmetric Functions using Linear Wires}

We prove a wire-complexity upper bound for \emph{every} symmetric Boolean function. Recall that $f:\zo^n\to\zo$ is \emph{symmetric} if $f(\vec x)$ depends only on the Hamming weight $|\vec x|$, i.e., $f(\vec x)=F(|\vec x|)$ for some $F:\{0,1,\dots,n\}\to\zo$.

\begin{theorem}\label{thm: symmetric upper bound}
    For every symmetric Boolean function $f:\zo^n\to\zo$, there is a depth-three sigmoidal circuit with $O(n)$ wires and polynomial weights that computes $f$ with separation $\Omega(1)$.
\end{theorem}

\begin{proof}
    Let $\gamma=1/6$. We can see that $\Sigmoid'(\xi)\ge\gamma$ for all $\xi\in[0,1]$, as $\Sigmoid'(1)=e/(1+e)^2>1/6$ and $\Sigmoid'$ is decreasing on $[0,\infty)$. By the mean value theorem, $\Sigmoid(v)-\Sigmoid(u)\ge \gamma(v-u)$ for all $0\le u\le v\le 1$. We construct the circuit layer by layer.

    \smallskip\noindent\textbf{Bottom layer.} A single gate $\Sigmoid_{\mathsf{sum}}(\vec x)\eqdef \Sigmoid\left(\tfrac1n\sum_{i=1}^n x_i\right)=\Sigmoid(|\vec x|/n)$ of fan-in $n$. Its output is a real-valued digest of the Hamming weight: it equals $\Sigmoid(m/n)$ when $|\vec x|=m$, and consecutive weights are separated by $\Sigmoid((m+1)/n)-\Sigmoid(m/n)\ge \gamma/n$.

    \smallskip\noindent\textbf{Middle layer.} Let $A\eqdef \lceil (2n/\gamma)\ln(10n)\rceil=O(n\log n)$, and for each $k\in[n]$ let $\tau_k\eqdef \Sigmoid((k-\tfrac12)/n)$ and
    \[
    V_k(\vec x)\eqdef \Sigmoid\bigl(A\cdot(\Sigmoid_{\mathsf{sum}}(\vec x)-\tau_k)\bigr),
    \]
    a gate of fan-in \emph{one} that reads only the output of the gate $\Sigmoid_{\mathsf{sum}}$. We first show that $|V_k(\vec x)-\Id[\,|\vec x|\ge k\,]|\le 1/(10n)$: 
    \begin{compactitem}
    \item If $|\vec x|\ge k$ then $\Sigmoid_{\mathsf{sum}}(\vec x)-\tau_k\ge \Sigmoid(k/n)-\Sigmoid((k-\tfrac12)/n)\ge \gamma/(2n)$. Then $V_k(\vec x)\ge\Sigmoid(\ln(10n))=1-\tfrac{1}{10n+1}\ge 1-\tfrac{1}{10n}$.
    \item Otherwise, symmetrically, $V_k(\vec x)\le \Sigmoid(-\ln(10n))\le 1/(10n)$.
    \end{compactitem} 

    \smallskip\noindent\textbf{Output gate.} Since $f$ is symmetric, $f(\vec x)=F(|\vec x|)$ for some function $F:\{0,1,\dots,n\}\to\zo$. Notice that $F(m)=F(0)+\sum_{k=1}^n c_k\,\Id[m\ge k]$ with $c_k\eqdef F(k)-F(k-1)\in\{-1,0,1\}$. Let
    \[
    C(\vec x)\eqdef \Sigmoid\Bigl(4\bigl(F(0)+\textstyle\sum_{k=1}^n c_k V_k(\vec x)-\tfrac12\bigr)\Bigr),
    \]
    a gate of fan-in $n$ reading the middle layer gates. Let $L(\vec x)\eqdef F(0)+\sum_k c_k V_k(\vec x)$. Since $\sum_k|c_k|\le n$, we have
    \[
    |L(\vec x)-F(|\vec x|)|=\Bigl|\sum_k c_k\bigl(V_k(\vec x)-\Id[\,|\vec x|\ge k\,]\bigr)\Bigr|\le n\cdot\tfrac{1}{10n}=\tfrac{1}{10}.
    \]
    
    We can verify that the circuit computes $f$ with separation at least $1/4$: if $f(\vec x)=1$ then $L(\vec x)\ge\tfrac{9}{10}$ and $C(\vec x)\ge\Sigmoid(8/5)>\tfrac34$; if $f(\vec x)=0$ then $L(\vec x)\le\tfrac{1}{10}$ and $C(\vec x)\le\Sigmoid(-8/5)<\tfrac12$. The circuit has depth three and $n+n+n=O(n)$ wires, and all weights and biases are at most $A=O(n\log n)$ in absolute value.
\end{proof}

Since the Parity function is symmetric, \Cref{thm: symmetric upper bound} applies to it.

\begin{corollary}
    There is a depth-three sigmoidal circuit with $O(n)$ wires and polynomial weights that computes $\oplus_n$ with separation $\Omega(1)$.
\end{corollary}

The construction illustrates a concrete advantage of sigmoidal gates: It is known that the parity function requires super-linear wires for threshold circuits \cite{IPS97} when the depth is an arbitrary constant and weights are arbitrary, and even for average-case computation \cite{CSS18}.

\subsection{Wire Complexity for Depth-2 Sigmoidal Circuits}
\label{sec: wire complexity depth two sigmoids}

To fully close the question of wire complexity for sigmoidal circuits, we prove tight upper and lower bounds for depth-two sigmoidal circuits (up to polylogarithmic factors). 

\begin{proposition}\label{prop: depth two upper parity}
    For every symmetric function $f:\zo^n\to\zo$, there is a depth-two polynomial-weight sigmoidal circuit with $O(n^2)$ wires that computes $f$ with separation $\Omega(1)$. 
\end{proposition}

\begin{proof}[Proof Sketch]
    For each $k\in\{0,1,\dots,n\}$ such that $f$ accepts strings with Hamming weight $k$, create a gate $V_k(\vec x)$ that outputs $\approx 1$ when $|\vec x|=k$, and $\approx 0$ otherwise. The output gate is a top sigmoidal gate that outputs $\approx 1$ when the sum of $V_k$ $\gg 0$, and $\approx 0$ otherwise. 
\end{proof}

Recall that Paturi and Saks \cite{PS94} proved an $\Omega(n^2/\log^2 n)$ wire complexity lower bound for depth-two threshold circuits with polynomially bounded integer weights computing Parity. They indeed show that depth-two threshold circuits with wire complexity $E$ and integer weights within $[-W,W]$ can be represented as the sign of a \emph{rational function} of degree $O(\sqrt E\log(nW))$. 

It can be observed that the result generalizes to sigmoidal circuits by combining \cite{PS94} with a rational approximation of $\Sigmoid$. We first state a generalized version of \cite[Theorem 3]{PS94}. Recall that the degree of a rational function $p/q$ is defined as $\max\{\deg p,\deg q\}$.

\begin{theorem}[Adapted from {\cite[Theorem 3]{PS94}}]\label{thm: PS94 restated}
    Let $c>0,c_0\ge 1$ be constants, and $\sigma:\R\to[0,1]$ be any non-decreasing $\kappa$-Lipschitz function satisfying the following conditions: For every $k\ge 1$, $a \ge 1$, there is a rational function $r_k(x)$ of degree at most $k$ with no pole in $[-a,a]$ such that for every $x\in [-a, a]$, $|r_k(x)-\sigma(x)|\le c_0a^{c_0}\cdot e^{-k^{1/c}/c_0}$.

    Then for any depth-two $\sigma$-circuit $C:\zo^n\to[0,1]$ with weights and biases within $[-W,W]$ and has at most $E$ wires, if $C$ computes a Boolean function $f:\zo^n\to\zo$ with separation $\delta$, there is a rational function $r:\zo^n\to\R$ of degree $O(\sqrt E\log^{c/2}(n\kappa EW/\delta))$ that sign-represent $f$, i.e., $f(x)=\sgn(r(x))$ for every $x\in\zo^n$.  
\end{theorem}

\begin{proof}[Proof Sketch]
    We follow the proof of \cite[Theorem 3]{PS94}. Fix the circuit $C$ and let $\sigma_1,\dots,\sigma_t$ be the gates in the bottom layer, $w_i$ be the weight for $\sigma_i$ in the top $\sigma$-gate, and $\ell(\vec x)$ be the affine function consisting of the bias of the top $\sigma$-gate and its wires from the input variables. As $\sigma$ is non-decreasing and $\kappa$-Lipschitz, there exists $\zeta_0,\zeta_1$ such that $\zeta_1-\zeta_0\ge \delta/\kappa$, and $f(\vec x)=1$ (resp.~$f(\vec x)=0$) iff $\sum_{i} w_i\sigma_i(\vec x)+\ell(\vec x)\ge \zeta_1$ (resp.~$\sum_{i} w_i\sigma_i(\vec x)+\ell(\vec x)\le \zeta_0$). 
    
    Let $k,d\ge 1$ be parameters to be determined. We consider gates with large and small fan-in: 
    \begin{compactitem}
    \item (\emph{Large Gates}). For each gate $\sigma_i(\vec x)$ with fan-in larger than $d$ (called \emph{large gates}), we approximate it using the provided degree-$k$ rational function $\hat\sigma_i$ with degree $k$. Note that $\sigma_i(\vec x)=\sigma(\langle\vec w,\vec x\rangle + \beta)$, where $|\langle\vec w,\vec x\rangle+\beta|\le (n+1)W$. It then suffices to instantiate the rational approximation with $a\eqdef (n+1)W$, so the error is $O((nW)^{c_0}\cdot e^{-k^{1/c}/c_0})$.  
    \item (\emph{Small Gates}). For each gate $\sigma_j(\vec x)$ with fan-in at most $d$ (called small gates), we use a degree-$d$ polynomial $\hat \sigma_j(\vec x)$ to compute it exactly over $\zo^n$. 
    \end{compactitem}
    Let $L\eqdef\{i\in[t]\mid \sigma_i\text{ is large}\}$. Since there are at most $E/d$ large gates, $\sum_{i\in L}w_i\hat\sigma_i(\vec x)$ is a rational function with degree at most $kE/d$. Meanwhile, we can observe that 
    \[ 
    \left|\sum_{i\in L} w_i\sigma_i(\vec x)-\sum_{i\in L} w_i\hat \sigma_i(\vec x)\right| =O\left(EW\cdot (nW)^{c_0}\cdot e^{-k^{1/c}/c_0}\right) = (nEW)^{O(1)}\cdot e^{-k^{1/c}/c_0}.
    \]  
    By setting $k\eqdef \Theta(\log^{c}(n\kappa EW/\delta))$, the error is at most $\delta/(3\kappa)$. This implies that $\sum_i w_i\hat\sigma_i(\vec x)$ approximates $\sum_i w_i\sigma_i(\vec x)$ within error $\delta/(3\kappa)$, and subsequently $R(x)\eqdef\sum_i w_i\hat \sigma_i(\vec x)+\ell(\vec x)-(\zeta_1+\zeta_0)/2$ sign-represent $f$, as desired. Note that the degree of $R(\vec x)$ is $kE/d + d$, which is minimized when we set $d\eqdef \sqrt{kE}$.  
\end{proof}

The following lemma gives the rational approximation of $\Sigmoid$ that we need. Its proof follows \cite{SRK94} and is based on Newman's theorem on the rational approximation of $|x|$ \cite{Newman64}; we defer it to \Cref{sec: sigmoid rational approx}.

\begin{restatable}[Rational approximation of $\Sigmoid$]{lemma}{SigmoidRationalApprox}\label{lmm: sigmoid rational approx}
    There is a constant $C_0$ such that for every $a\ge 1$ and $\eps\in(0,1/2]$, there is a rational function $r$ of degree at most $C_0\ln^2(a/\eps)$, with no pole in $[-a,a]$, such that $|r(x)-\Sigmoid(x)|\le\eps$ for every $x\in[-a,a]$.
\end{restatable}

\begin{corollary}\label{cor: sigmoid lower bound}
    Any function that can be computed by a depth-two sigmoidal circuit containing $E$ wires and using polynomial weights and biases with noticeable separation, can be sign-represented by a rational function of degree $O(\sqrt E \log n)$. Consequently, any depth-two sigmoidal circuit with polynomial weights and biases that compute Parity with noticeable separation requires $\Omega(n^2/\log^{2} n)$ wires. 
\end{corollary}

\begin{proof}
Let $\sigma\eqdef\Sigmoid$, which is non-decreasing and $\kappa$-Lipschitz for $\kappa=1/4$, and by \Cref{lmm: sigmoid rational approx} it satisfies the condition in \Cref{thm: PS94 restated} with $c=2$. Now let $C$ be a depth-two sigmoidal circuit with $E$ wires and weights within $[-W,W]$, where $1\le W\le n^{O(1)}$, that computes $f:\zo^n\to\zo$ with separation $\delta\ge n^{-O(1)}$. If $E\ge n^2$, then $f$ is sign-represented by a polynomial of degree at most $n\le\sqrt E$ (e.g., the multilinear extension of $f$ minus $1/2$). Otherwise, \Cref{thm: PS94 restated} yields a rational function of degree $O(\sqrt{E}\log(n\kappa EW/\delta))=O(\sqrt E\log n)$ that sign-represents $f$. 

For the second part, suppose that a rational function $P/Q$ sign-represents Parity, where $Q$ has no zero on $\zo^n$. Then the polynomial $P\cdot Q$ also sign-represents Parity, so $\deg P+\deg Q\ge n$ by \cite[Lemma 3]{PS94} (see also \cite[Corollary 1]{PS94}). Hence the degree of $P/Q$ is at least $n/2$, and subsequently $O(\sqrt E\log n)\ge n/2$, i.e., $E=\Omega(n^2/\log^2 n)$.
\end{proof}


\section{Lower Bounds via Random Restriction}
\label{sec: lb from random restriction}

In this section, we prove two lower bounds against sigmoidal circuits with noticeable separation via random restriction methods. 

\subsection{Random Restriction Lemmas for Sigmoid Functions}
\label{sec: random restriction}

We first develop lemmas that analyze the behavior of linear and sigmoidal functions under random restrictions, using random restriction lemmas for threshold functions \cite{KW16,KKL17} and the analytic property of the Sigmoid function. 

\subsubsection{Analytic Properties of Sigmoid Functions}
\label{sec: analytic sigmoid}

\newcommand{\di}{\mathrm{d}}

\begin{theorem}[Cauchy's coefficient formula, see, e.g., {\cite[Theorem IV.7]{FS09-book}}]\label{thm: expansion bound}
    Let $\lambda$ be a positively oriented simple loop around $0$, and $f(z)$ be analytic on and inside $\lambda$, then
    \[
    \frac{f^{(k)}(0)}{k!} = \frac{1}{2i\pi}\int_\lambda~\frac{f(z)}{z^{k+1}}~\di z. 
    \] 
\end{theorem}

\begin{lemma}[see, e.g., {\cite[Section 5.4]{Howie03complex}}]\label{lmm: estimation}
    Let $\lambda$ be a contour, and $f$ be a continuous complex function such that $|f(z)|\le M$ for all $z\in \lambda$, then 
    \[
    \left|\int_{\lambda}f(z)~\di z\right| \le M\cdot\ell(\lambda),
    \]
    where $\ell(\lambda)$ denotes the length of $\lambda$. 
\end{lemma}

\begin{corollary}\label{cor: sigmoid derivative}
    For any $k\ge 0$ and $\xi\in\R$, $|\Sigmoid^{(k)}(\xi)|\le k!/(\pi/2)^{k}$.  
\end{corollary}

\begin{proof}

    Let $g(z)=\Sigmoid(z+\xi)$ be the shifted function defined over $\C$; clearly, $g^{(k)}(0)=\Sigmoid^{(k)}(\xi)$. The function $g(z)$ is analytic at $0$. Moreover, the singular points of $g(z)$ are exactly $z = -(\xi+i\pi(2n+1))$ for $n\in\Z$, $|z|\ge \sqrt{\xi^2 + (2n+1)^2\pi^2}\ge \pi$, and thus $g(z)$ is analytic in $\{z\in\C\mid |z|<\pi\}$. 

    Let $\lambda$ be the positively oriented circle with radius $\pi/2$. Note that $|g(z)|\le 1$ within the circle. By Cauchy's coefficient formula, we can see that 
    \[
    \left|\frac{g^{(k)}(0)}{k!}\right| \le
    \frac{1}{2\pi}\left|\int_\lambda~\frac{g(z)}{z^{k+1}}~\di z\right| \le \frac{1}{2\pi}\left(\frac{\pi}{2}\right)^{-(k+1)} \pi^2 \le \left(\frac{\pi}{2}\right)^{-k}, 
    \] 
    where the second inequality follows from \Cref{lmm: estimation}. 
\end{proof}

\subsubsection{Random Restriction on Linear and Sigmoidal Functions}

We start by defining a few concepts. A \emph{restriction} is a pair $\rho=(I,\vec y)$, where $I\subseteq[n]$ is the set of free variables and $\vec y\in\zo^{[n]\setminus I}$. We use $f|_\rho$ to denote the function on the free variables obtained from $f$ by fixing $x_i$ to $y_i$ for every $i\notin I$. For any fixed partition $\sP$ of $[n]$ into $m$ blocks, we use $\cR_{\sP}$ to denote the distribution of restrictions in which $I$ contains exactly one uniformly random variable of each block, chosen independently for different blocks, and $\vec y$ is uniformly random.

\begin{lemma}[\cite{KW16}]\label{lmm: KW-LTF}
    Let $f:\zo^n\to\R$ be any linear function, and $g(\vec x)\eqdef \sgn(f(\vec x))$ be a linear threshold function. Then for any partition $\sP$ of $[n]$ into $m$ blocks of equal size,
    \[
    \Pr_{\rho\gets\cR_\sP}[g|_\rho\text{ is not a constant function}] \le O(m/\sqrt{n}). 
    \]
\end{lemma}

For a linear function $f(\vec x)=\langle \vec w,\vec x\rangle + \theta$ and $a \le b$, $\delta$, we say that $f(\vec x)$ is $\delta$-concentrated within $[a,b]$ if there exists $z\in[a,b]$ such that for every $\vec x\in\zo^n$, $|f(\vec x)-z|\le \delta$. The following lemma is a simple corollary of the restriction lemma above.

\begin{lemma}\label{lmm: KW-linear}
    For any linear function $f(\vec x)$ with $n$ variables and any partition $\sP$ of $[n]$ into $m$ blocks of equal size, $m\le n/16$, $\delta>0$, and $a,b$ such that $b-a\ge \delta$, over a block random restriction $\rho\gets\cR_{\sP}$, with probability $1-O(m(b-a)\cdot \delta^{-1}/\sqrt n)$, at least one of the following two conditions holds:
    \begin{compactenum}
    \item $f|_\rho(\vec z)\le a$ for every $\vec z\in\zo^m$, or $f|_\rho(\vec z)\ge b$ for every $\vec z\in\zo^m$;
    \item $f|_\rho$ is $\delta$-concentrated within $[a,b]$.
    \end{compactenum}
\end{lemma}

\begin{proof}
    Let $N\in\N$ such that $\delta/2\le (b-a)/N\le\delta$, let $\delta'\eqdef (b-a)/N$, and $\tau_i\eqdef a+i\delta'$ for $i=0,1,\dots,N$, so $\tau_0=a$ and $\tau_N=b$. For each $i$, let $g_i(\vec x)\eqdef\sgn(f(\vec x)-\tau_i)$, which is a linear threshold function. Applying \Cref{lmm: KW-LTF} to each $g_i$ and by union bound over $i$, with probability $1-O(Nm/\sqrt n)=1-O(m(b-a)\delta^{-1}/\sqrt n)$ every $g_i|_\rho$ is constant.

    Condition on this event. By monotonicity in $i$, the tuple $(g_0|_\rho,\dots,g_N|_\rho)\in\zo^{N+1}$ has the form $(1,\dots,1,0,\dots,0)$; let $i^*\in\{-1,\dots,N\}$ be the index of the last $1$ (with $i^*=-1$ if all entries are $0$). If $i^*=-1$ (resp.~$i^*=N$), then $f|_\rho\le\tau_0=a$ (resp.~$f|_\rho>\tau_N=b$) on $\zo^m$, yielding condition~1. Otherwise $\tau_{i^*}<f|_\rho\le\tau_{i^*+1}=\tau_{i^*}+\delta'$ on $\zo^m$, which satisfies condition~2 with $z\eqdef\tau_{i^*}+\delta'/2$.
\end{proof}

By a case analysis, we can also show that after a (block) random restriction, a sigmoid-like gate will likely be approximable by a low-degree polynomial. 

\begin{corollary}\label{cor: single sigmoid}
    Let $n,t\ge 20$, $m\le n/16$, $k\ge 0$, $\vec w\in\R^n$, $\theta\in\R$. Then for any partition $\sP$ of $[n]$ into $m$ blocks of equal size, over $\rho\gets\cR_{\sP}$, with probability at least $1-O(mt(k+1)\ln t/\sqrt{n})$, there is a degree-$k$ multilinear polynomial $p:\R^m\to\R$ such that for every $\vec x\in\zo^m$, 
    \[
    \big|\Sigmoid_{\vec w,\theta}|_\rho(\vec x)-p(\vec x)\big| \le t^{-(k+1)}, 
    \]
    where $O(\cdot)$ hides an absolute constant. Moreover, every coefficient of $p$ has absolute value at most $2^{k+1}$.
\end{corollary}

\begin{proof}
    Let $f(\vec x)\eqdef \langle\vec w,\vec x\rangle+\theta$, and $a> 0$ be determined later. By \Cref{lmm: KW-linear}, we know that over $\rho\gets\cR_\sP$, with probability at least $1-O(mta/\sqrt n)$, $f|_\rho$ is either $(1/t)$-concentrated within $[-a,a]$, or $f|_\rho\le -a$ on $\zo^m$, or $f|_\rho\ge a$ on $\zo^m$. We show that in either case, it can be approximated by a degree-$k$ polynomial.
    \begin{compactitem}
    \item (\emph{Case 1}). Suppose that $f|_\rho\le -a$ on $\zo^m$ (the case $f|_\rho\ge a$ is symmetric). Then $0<\Sigmoid_{\vec w,\theta}|_\rho(\vec z)\le e^{-a}$ for every $\vec z\in\zo^m$. We set $a\eqdef(k+1)\ln t$, so it suffices to choose $p\equiv 0$ (resp.~$p\equiv 1$ in the symmetric case).
    \item (\emph{Case 2}). Suppose that $f|_\rho$ is $(1/t)$-concentrated at $\mu\in[-a,a]$, i.e., $|f|_\rho(\vec z)-\mu|\le 1/t$ for every $\vec z\in\zo^m$. Let $q$ be the degree-$k$ Taylor approximation of $\Sigmoid$ at $\mu$. Taylor's theorem shows that for every $y$ with $|y-\mu|\le 1/t$, there exists a $\xi\in\R$ such that 
    \[
    \left|\Sigmoid(y)-q(y)\right|\le \frac{|\Sigmoid^{(k+1)}(\xi)|}{(k+1)!}\cdot |y-\mu|^{(k+1)}\le t^{-(k+1)},
    \]
    where the last inequality follows from \Cref{cor: sigmoid derivative}. It then suffices to take $p(\vec x)\eqdef q(f|_\rho(\vec x))$, which is a degree-$k$ polynomial in $\vec x$ as $f|_\rho$ is a linear function. 
    \end{compactitem}
    In both cases, we may assume $p$ is multilinear as the input is Boolean. For the moreover part, note that $|p(\vec z)|\le 1+t^{-(k+1)}\le 2$ for any $\vec z\in\zo^m$, as $\Sigmoid$ has range $(0,1)$. For every $S\subseteq[m]$ with $|S|\le k$, by the inclusion--exclusion principle, the coefficient of $\prod_{i\in S}z_i$ in $p$ is $\sum_{T\subseteq S}(-1)^{|S|-|T|}\,p(\vec 1_T)$, where $\vec 1_T\in\zo^m$ is the indicator vector of $T$. Therefore, its absolute value is at most $2^{|S|}\cdot 2\le 2^{k+1}$.
\end{proof}

\subsubsection{Random Restriction for Bounded Fan-in Gates}

We also need the following restriction lemma from \cite{KW16} that works for sigmoidal gates with bounded fan-in. A Boolean function $g:\zo^n\to\zo$ is said to be \emph{degenerated} if it depends on at most one of its input variables.

\begin{lemma}[{\cite[Lemma 3.1]{KW16}}]
    Let $f:\zo^n\to\R$ be any linear function that depends on at most $\ell$ of its input variables, and $g(\vec x)\eqdef \sgn(f(\vec x))$ be a linear threshold function. Then for any partition $\sP$ of $[n]$ into $m$ blocks of equal size,
    \[
    \Pr_{\rho\gets\cR_\sP}[g|_\rho\text{ is not degenerated}] \le O(m^{3/2}\ell/n^{3/2}).
    \]
\end{lemma}

\begin{lemma}\label{lmm: KW-linear wire}
    Let $f(\vec x):\zo^n\to\R$ be a linear function that depends on at most $\ell$ of its input variables, $\sP$ be a partition of $[n]$ into $m$ blocks of equal size, $m\le n/16$, $\delta>0$, and $a,b$ such that $b-a\ge\delta$. Then over a block random restriction $\rho\gets\cR_\sP$ with probability $1-O(m^{3/2}\cdot \ell \delta^{-1}(b-a)/n^{3/2})$, there is a free variable such that for each restriction $\sigma$ fixing it to $0$ or $1$, at least one of the following two conditions holds:
    \begin{compactenum}
    \item $f|_{\rho\circ\sigma}\le a$ on all inputs, or $f|_{\rho\circ\sigma}\ge b$ on all inputs;
    \item $f|_{\rho\circ\sigma}$ is $\delta$-concentrated within $[a,b]$. 
    \end{compactenum}
\end{lemma}

\begin{proof}[Proof Sketch]
    We follow the proof of \Cref{lmm: KW-linear}. The key observation is that if a non-constant $\sgn(f|_\rho-\tau)$ depends on only one variable, then this variable has the strictly largest absolute weight in $f|_\rho$. Subsequently if that variable is fixed arbitrarily, all threshold functions must be constant simultaneously.
\end{proof}

\begin{corollary}\label{cor: single sigmoid wire}
    Let $n,t\ge 20$, $m\le n/16$, $k\ge 0$, $\vec w\in\R^n$, $\theta\in\R$. Suppose that $\vec w$ has at most $\ell$ non-zero entries. Then for any partition $\sP$ of $[n]$ into $m$ blocks of equal size, over $\rho\gets\cR_{\sP}$, with probability at least $1-O(m^{3/2}t\ell(k+1)\ln t/n^{3/2})$, there is a degree-$(k+1)$ multilinear polynomial $p:\R^m\to\R$ such that for every $\vec x\in\zo^m$,
    \[
    \big|\Sigmoid_{\vec w,\theta}|_\rho(\vec x)-p(\vec x)\big| \le t^{-(k+1)}.
    \]
    Moreover, every coefficient of $p$ has absolute value at most $2^{k+2}$.
\end{corollary}

\begin{proof}[Proof Sketch]
    We follow the proof of \Cref{cor: single sigmoid} using \Cref{lmm: KW-linear wire}. The only difference is that after the restriction $\rho$, we find the free variable $x$ provided by \Cref{lmm: KW-linear wire} and construct two degree-$k$ polynomials $p_0,p_1$ that approximate $\Sigmoid_{\vec w,\theta}|_\rho$ when $x=0$ and $x=1$, respectively. It then suffices to define $p(\vec x)\eqdef (1-x)\cdot p_0(\vec x)+x\cdot p_1(\vec x)$. As in \Cref{cor: single sigmoid}, we may assume that $p$ is multilinear, and the moreover part follows in the same way since $p$ has degree at most $k+1$.
\end{proof}

\subsection{Lower Bounds for Generalized Andreev's Function}

We now prove a super-linear lower bound against depth-three sigmoidal circuits with a hard function, known as generalized Andreev's function, that is computable in polynomial time. Formally: 

\begin{restatable}{theorem}{KWSigmoid}\label{thm: depth-3 lower bound}
    There exists an absolute constant $c_0\ge 1$ such that the following holds. Let $a\ge 1$, let $n\in\N$ be sufficiently large (depending on $a$), and $d,W,\delta^{-1}\le n^a$. Suppose that $C$ is a $(2n)$-input depth-$3$ sigmoidal circuit, where
    \begin{compactitem}
    \item the output gate is a sigmoidal gate with fan-in $d$ and weights within $[-W,W]$;
    \item the middle layer consists of $d$ sigmoidal gates with weights and biases within $[-W,W]$;
    \item the bottom layer (i.e.~closest to input variables) has $s$ sigmoidal gates in total.
    \end{compactitem}
    Then if $C$ computes $\GA_{n,m}$ with separation $\delta\in(0,1)$, $s \ge n^{3/2} / (\log n)^{c_0\cdot a}$. 
\end{restatable}

\subsubsection{Counting Depth-2 Augmented Sigmoidal Circuits}

For the purpose of proving Andreev's function lower bound, we need to count the number of possible depth-2 sigmoidal circuits augmented with a low-degree polynomial. 

\begin{definition}[$\Sigmoid\circ(\Sigmoid\|\Poly)$] 
A depth-two augmented Sigmoidal circuit, denoted $\Sigmoid\circ(\Sigmoid\|\Poly)$, is of the following form: 
\begin{compactitem}
\item the bottom layer consists of $s$ sigmoidal gates $\sigma_1,\dots,\sigma_s$ and a real-valued multilinear polynomial $p:\R^n\to\R$, both fed by input variables;
\item the top layer is a sigmoidal gate $\sigma$ fed by $\sigma_1,\dots,\sigma_s$ and $p$, where the wire from $p$ has weight $1$ and the constant term of $p$ serves as the bias of $\sigma$, so it outputs $\Sigmoid(\sum_{i=1}^s v_i\sigma_i(\vec x)+p(\vec x))$ for some $v_1,\dots,v_s\in\R$.
\end{compactitem}
The \emph{size} of the circuit is defined as $s+1$, the \emph{wire complexity} is defined as the total number of wires of $\sigma_1,\dots,\sigma_s,\sigma$, and the \emph{degree} is defined as the degree of $p$. The \emph{weights} of the circuit consist of weights of $\sigma_1,\dots,\sigma_s,\sigma$ (including the biases of $\sigma_1,\dots,\sigma_s$) as well as the weights of monomials in $p$.
\end{definition}

\begin{lemma}\label{lmm: counting depth 2}
    Let $n,d,s,w\ge 1$, $W\ge 1$, $\delta\in(0,1/2]$. Then there is a list $\sL$ of $\AugSigmoid$ circuits such that for every $\AugSigmoid$ circuit $C$ with size at most $s+1$, wire complexity at most $w$, degree at most $d$, and weights within $[-W,W]$, there is a circuit $C'\in \sL$ such that $C$ and $C'$ are $\delta$-close to each other. Moreover,
    \[
    \log_2|\sL| = O\bigl(w\log(nsW/\delta)+p(n,d)\log(p(n,d)W/\delta)\bigr),
    \]
    where $p(n,d)=\sum_{i=0}^d\binom{n}{i}$.
\end{lemma}

\begin{proof}
    Let $C$ be such a circuit, where bottom gates are $\sigma_j(\vec x)=\Sigmoid(\langle\vec u_j,\vec x\rangle+b_j)$, $1\le j\le s$, and it computes $\Sigmoid(\sum_j v_j\sigma_j(\vec x)+p(\vec x))$. Let $C'$ be the circuit with the same topology obtained by rounding each weight and bias of the sigmoidal gates to the nearest integer multiple of $\eps\eqdef\delta/(4(n+1)sW)$, and each coefficient of $p$ to the nearest integer multiple of $\eps'\eqdef\delta/(4p(n,d))$. The list $\sL$ consists of all circuits obtained in this way. Such a circuit is determined by 
    \begin{compactitem}
    \item the number of bottom gates, the $\le w$ pairs $(j,i)$ such that $\sigma_j$ reads $x_i$, and $\le 2w$ rounded parameters of sigmoidal gates (one per wire and one bias per bottom gate; note that $s\le w$), each taking $O(nsW^2/\delta)$ values; 
    \item At most $p(n,d)$ rounded coefficients of $p$, each taking $O(p(n,d)W/\delta)$ values. 
    \end{compactitem}
    This gives the bound on $|\sL|$.

    We show that $C$ and $C'$ are $\delta$-close. Fix $\vec x\in\zo^n$. As $\Sigmoid$ is $(1/4)$-Lipschitz and each bottom gate has at most $n+1$ parameters, $|\sigma_j(\vec x)-\sigma'_j(\vec x)|\le (n+1)\eps/4\le \delta/(16sW)$. Since $\sigma_j(\vec x)\in[0,1]$, $|v'_j|\le 2W$, and every multilinear monomial takes values in $\zo$, the preactivation of the output gate changes by at most
    \[
    \sum_{j}\bigl(|v_j-v'_j|\cdot\sigma_j(\vec x)+|v'_j|\cdot|\sigma_j(\vec x)-\sigma'_j(\vec x)|\bigr)+p(n,d)\cdot\eps' \;\le\; \frac{\delta}{4}+\frac{\delta}{8}+\frac{\delta}{4} \;<\; \delta,
    \]
    so the output changes by less than $\delta/4$.
\end{proof}

\Cref{lmm: counting depth 2} only counts circuits with bounded weights, while the bottom layer of the circuits in our lower bounds may have unbounded weights. After a random restriction, however, these gates depend on few variables, and the next lemma bounds their parameters up to a small error.

\begin{restatable}[Parameter compression]{lemma}{SigmoidCompression}\label{lmm: sigmoid compression}
    Let $m\ge 1$, $\eps\in(0,1)$, $\vec w\in\R^m$, and $b\in\R$. Then there are $\vec w'\in\R^m$ and $b'\in\R$, whose entries have absolute values at most $(m+1)!\cdot\ln(3/\eps)$, such that for every $\vec z\in\zo^m$, $\bigl|\Sigmoid(\langle\vec w,\vec z\rangle+b)-\Sigmoid(\langle\vec w',\vec z\rangle+b')\bigr|\le\eps$. 
\end{restatable}

\Cref{lmm: sigmoid compression} is an approximate analogue of the classical fact that every threshold function on $m$ variables can be realized with integer weights of absolute value $2^{O(m\log m)}$, and we prove it by the classical argument (see, e.g., \cite[Section 3]{Hastad94}). The proof is deferred to \Cref{sec: compression}.

\subsubsection{A Discriminator Lemma for Sigmoidal Functions}

We generalize the standard discriminator lemma (see, e.g., \cite{HMPSMT93}) to sigmoidal gates that compute a Boolean function with noticeable separation. 

\begin{lemma}\label{lmm: disc}
    Let $\sigma:\R\to\R$ be a non-decreasing and $\kappa$-Lipschitz function. Let $g_1,\dots,g_m:\zo^n\to[0,1]$, $\vec w\in[-W,W]^m,\theta,\delta_0,\delta_1\in\R$, $f:\zo^n\to\{0,1,*\}$ be the partial function defined as 
    \[
    f(\vec x)\eqdef \begin{cases}
        1 & \sigma_{\vec w,\theta}(g_1(\vec x),\dots,g_m(\vec x)) \ge \delta_1 \\ 
        0 & \sigma_{\vec w,\theta}(g_1(\vec x),\dots,g_m(\vec x)) \le \delta_0 \\ 
        * & \text{otherwise}
    \end{cases}
    \]
    Assume that $\delta_1>\delta_0$ and $f^{-1}(0),f^{-1}(1)\ne\varnothing$. Then for every integer $B\ge 1$, there exist $t\in\{1,\dots,B\}$ and $i\in[m]$ such that for $g'_{i,t}(\vec x)\eqdef \sgn(g_i(\vec x)-t/B)\in\zo$,
    \[
    \left|\Pr_{\vec x\gets\zo^n}[g'_{i,t}(\vec x)=1\mid f(\vec x)=1]-\Pr_{\vec x\gets\zo^n}[g'_{i,t}(\vec x)=1\mid f(\vec x)=0]\right| \ge \frac{\delta_1-\delta_0}{\kappa m W}-\frac{2}{B}.
    \]
    In particular, for $B=\lceil 4\kappa m W/(\delta_1-\delta_0)\rceil$, the lower bound becomes at least $(\delta_1-\delta_0)/(2\kappa m W)$.
\end{lemma}

\begin{proof}
    Throughout the proof, all expectations and probabilities are taken over a uniform $\vec x\gets\zo^n$; we omit the subscript for brevity. By definition, $g'_{i,t}(\vec x)=1$ iff $g_i(\vec x)> t/B$.

    Recall that $g_i(\vec x)\in[0,1]$ for every $i\in[m]$ and $\vec x\in\zo^n$. Subsequently, for each $i\in[m]$, we have
    \[
    \left|g_i(\vec x) - \frac{1}{B}\sum_{t=1}^B g_{i,t}'(\vec x)\right| \le \frac{1}{B}.
    \]
    Let $\tilde g_i(\vec x)\eqdef \tfrac{1}{B}\sum_{t=1}^B g'_{i,t}(\vec x)$, $L(\vec x)\eqdef \sum_{i=1}^m w_i g_i(\vec x)+\theta$, and $\tilde L(\vec x)\eqdef \sum_{i=1}^m w_i \tilde g_i(\vec x)+\theta$. Since $|w_i|\le W$,
    \begin{equation}\label{eq:disc-linear}
    |L(\vec x)-\tilde L(\vec x)|\le \sum_{i=1}^m |w_i|\cdot |g_i(\vec x)-\tilde g_i(\vec x)|\le \frac{mW}{B}.
    \end{equation}

    Let $a_1\eqdef\min\{L(\vec x)\in\R\mid \vec x\in\zo^n,~\sigma(L(\vec x)) \ge \delta_1\}$ and $a_0\eqdef\max\{L(\vec x)\in\R\mid \vec x\in\zo^n,~\sigma(L(\vec x)) \le \delta_0\}$; the existence of $a_0,a_1$ follows from $f^{-1}(0),f^{-1}(1)\ne\varnothing$. Note that by the definition of $f$, $L$, and $\tilde{L}$,   
    \begin{align*}
    f(\vec x) = 1\; \Rightarrow \; L(\vec x) \ge a_1 \;\Rightarrow\; \tilde{L}(\vec x)\ge a_1 - \frac{mW}{B}; \\ 
    f(\vec x) = 0\; \Rightarrow \; L(\vec x) \le a_0 \;\Rightarrow\; \tilde{L}(\vec x)\le a_0 + \frac{mW}{B}. 
    \end{align*}
    Moreover, as $\sigma$ is non-decreasing and $\kappa$-Lipschitz, we have $a_1-a_0\ge (\delta_1-\delta_0)/\kappa$. Subsequently, we have 
    \[
    \Ex[\tilde L(\vec x)\mid f(\vec x)=1]-\Ex[\tilde L(\vec x)\mid f(\vec x)=0] \ge \frac{\delta_1-\delta_0}{\kappa} - \frac{2mW}{B}. 
    \]
    
    Finally, by linearity of expectation and the fact that $g'_{i,t}(\vec x)\in\zo$,
    \[
    \Ex[\tilde L(\vec x)\mid f(\vec x)=1] - \Ex[\tilde L(\vec x)\mid f(\vec x)=0] = \frac{1}{B}\sum_{i=1}^m\sum_{t=1}^B w_i\cdot \Delta_{i,t},
    \]
    where $\Delta_{i,t}\eqdef \Pr[g'_{i,t}(\vec x)=1\mid f(\vec x)=1]-\Pr[g'_{i,t}(\vec x)=1\mid f(\vec x)=0]$. Since $|w_i|\le W$, the absolute value of the RHS is at most $mW\cdot \max_{i\in[m],t\in[B]}|\Delta_{i,t}|$, hence
    \[
    \max_{i\in[m],\,t\in[B]}|\Delta_{i,t}| \ge \frac{1}{mW}\left(\frac{\delta_1-\delta_0}{\kappa} - \frac{2mW}{B}\right) = \frac{\delta_1-\delta_0}{\kappa m W} - \frac{2}{B},
    \]
    as required. The ``in particular'' statement follows from $B\ge 4\kappa mW/(\delta_1-\delta_0)$, which gives $2/B\le (\delta_1-\delta_0)/(2\kappa mW)$.
\end{proof}

\subsubsection{Proof of the Size Lower Bound}

We need the hard function $F:\zo^n\times\zo^m\to\zo$ constructed in \cite{KW16}, with a simple modification so it runs in $\poly(n,m)$ time (rather than $\poly(n,2^m)$ as stated in \cite{KW16}). We first recall a lemma from \cite{KW16}. 

\begin{lemma}[{\cite[Lemma 6.1]{KW16}}]\label{lmm: KW list}
    There is an absolute constant $C\ge 1$ such that the following holds. Let $\eps>0$, and let $\mathcal{S}$ be a set of Boolean functions on $m$ variables such that $\Cor(g,g')\le\eps$ for any two distinct $g,g'\in\mathcal{S}$. Then for every Boolean function $f$ on $m$ variables, at most $C/\eps$ functions $g\in\mathcal{S}$ satisfy $\Cor(f,g)>C\sqrt{\eps}$.
\end{lemma}

\begin{lemma}[implicit in {\cite[Corollary 6.2]{KW16}}] \label{lmm: KW code}
Let $n$ be sufficiently large and $m$ be an even integer such that $n^2 \le 2^m \le 2^n$. There is a function $F:\zo^n\times \zo^{m}\to\zo$ computable in $\poly(n)$ time such that given any class of Boolean functions $\sL$ on $m$ variables with $|\sL|\le 2^{n/2}$, there is an $\vec x\in \zo^n$ such that for any $f\in\sL$, we have $\Cor( f,F(\vec x,\cdot))\le O(\sqrt{n})/2^{m/4}$.
\end{lemma}

\begin{proof}[Proof Sketch]
    Let $\ell\eqdef m/2$, and identify $\zo^\ell$ with the field $\mathbb{F}_{2^\ell}$ using an irreducible polynomial of degree $\ell$ over $\mathbb{F}_2$, which can be found deterministically in $\poly(\ell)$ time \cite{Shoup90}. For $\vec x\in\zo^n$ and $\vec z=(u,v)\in\mathbb{F}_{2^\ell}^2$, let
    \[
    F(\vec x,\vec z)\eqdef \Bigl\langle \sum_{i=1}^n x_i u^{i-1},\;v\Bigr\rangle \bmod 2,
    \]
    where $\langle\cdot,\cdot\rangle$ is the inner product of the binary representations; it can be computed with $O(n)$ field operations. This is indeed the powering construction of small-biased sets from \cite{AGHP92}. 
    
    For $\vec x\ne\vec x'$, the polynomial $q(u)\eqdef\sum_{i=1}^n(x_i\oplus x'_i)u^{i-1}$ is nonzero and has fewer than $n$ roots, and $F(\vec x,(u,v))\oplus F(\vec x',(u,v))=\langle q(u),v\rangle\bmod 2$ is a uniformly random bit over $v$ when $q(u)\ne 0$. Hence $F(\vec x,\cdot)$ and $F(\vec x',\cdot)$ agree on a $1/2+O(n/2^{m/2})$ fraction of inputs. By \Cref{lmm: KW list} with $\eps\eqdef n/2^{m/2}$, each $f\in\sL$ has correlation more than $O(\sqrt n)/2^{m/4}$ with $F(\vec x,\cdot)$ for at most $O(2^{m/2}/n)$ strings $\vec x$. Since $|\sL|\cdot O(2^{m/2}/n)\le O(2^n/n)<2^n$, some $\vec x\in\zo^n$ satisfies the lemma.
\end{proof}

Fix $m\eqdef 16\lceil\log_2 n\cdot\log_2\log_2 n\rceil$. For simplicity of presentation, we assume $n/m$ is an integer; the general case follows by a simple padding argument. Let $F(\vec x,\vec z)$ be the polynomial time computable function above. We will use the following function $\GA_{n,m}(\vec x,\vec y)$ with $n + (n/m)\cdot m = 2n$ input variables, where $\vec x\in\zo^{n}$ and $\vec y\in\zo^{n}$. The string $\vec y$ is parsed as $\vec y_1,\dots,\vec y_{m}\in\zo^{n/m}$, and we define 
\[
\GA_{n,m}(\vec x,\vec y)\eqdef F(\vec x,\vec z),
\]
where $z_i\eqdef \oplus(\vec y_i)$ is the parity of bits in $\vec y_i$. By \Cref{lmm: KW code}, $\GA_{n,m}$ is computable in polynomial time. The function is introduced by Kane and Williams \cite{KW16} and is a generalization of Andreev's function \cite{Andreev87}; the variants of Andreev's function have found many applications in proving circuit lower bounds, see, e.g., \cite{IN93,PZ93,Hastad98,IMZ12,KR13,KW16,KKL17}.

\KWSigmoid*

\begin{proof}
    Suppose that $C$ computes $\GA_{n,m}$ with separation $\delta$. Let $\vec x\in\zo^{n}$ be a string and $t,k\in\N$ be parameters to be determined later~\defermark{$(\diamond)$}{size-diamond}, $k\le n/2$, and $C_{\vec x}(\vec y)\eqdef C(\vec x,\vec y)$ be the circuit after fixing the first $n$ input bits to be $\vec x$. As mentioned above, we assume for simplicity of presentation that $n/m$ is an integer.
    
    \paragraph{Behavior under random restriction.} Parse $\vec y=(\vec y_1,\dots,\vec y_{m})$, where $\vec y_i\in\zo^{n/m}$ for every $i\in[m]$, and let $\sP$ be the partition of $[n]$ into $m$ blocks so the $i$-th block corresponds to $\vec y_i$. Then by \Cref{cor: single sigmoid} and Markov inequality, there is a restriction $\rho$ over $\cR_\sP$ such that all but
    \begin{equation}\label{eq: s1 def}
    s_1\eqdef O\left(\frac{s|\sP|\cdot t(k+1)\ln t}{\sqrt n}\right) = O\left(\frac{s m t(k+1)\ln t}{\sqrt{n}}\right)
    \end{equation}
    sigmoidal gates in the first layer are $t^{-(k+1)}$-close to a degree-$k$ real coefficient polynomial. Fix any restriction that satisfies this condition, and let $n_1\eqdef m$ be the number of remaining input variables of $C_{\vec x}|_\rho$. Let $C_1\eqdef C_{\vec x}|_\rho$. Note that $C_1$ computes the function $\vec z\mapsto F(\vec x,\vec z\oplus\vec\eta)$ with separation $\delta$, where $\eta_i\in\zo$ is the parity of the bits of $\vec y_i$ fixed by $\rho$.

    \begin{proposition}\label{prop: simulation under restriction}
    Let
    \begin{equation}\label{eq: W1 def}
    W_1\eqdef \max\bigl\{(2^{k+1}s+2n+1)W,\;(n_1+1)!\cdot\ln(3t^{k+1})\bigr\}.
    \end{equation}
    For each subcircuit $G$ in the middle layer of $C_1$, there exists a $\AugSigmoid$ circuit $G'$ of size at most $s_1+1$, degree $k$, and weights within $[-W_1,W_1]$ such that $|G(\vec z)-G'(\vec z)|\le sW\cdot t^{-(k+1)}$ on any $\vec z\in\zo^{n_1}$.
    \end{proposition}

    The proof is straightforward, and is deferred to the end of the proof. 

    \paragraph{Applying the discriminator lemma.} Let $C_1'(\vec z)$ be the circuit obtained from $C_1$ by replacing each subcircuit $G$ in the middle layer with $G'$ in \Cref{prop: simulation under restriction}. Then we know that $|C_1'(\vec z)-C_1(\vec z)|\le (dW)\cdot sW\cdot t^{-(k+1)}\le sdW^2/t^{k+1}$. Assume that $sdW^2/t^{k+1}\le \delta /10$ (which we will verify later~\defermark{$(\star)$}{size-star}), $C_1'(\vec z)$ computes $F(\vec x,\vec z\oplus\vec\eta)$ with separation at least $\delta - 2sdW^2/t^{k+1} \ge 4\delta/5$ (an additive error of $e$ can increase the outputs on $0$-inputs and decrease the outputs on $1$-inputs by $e$ each, so it decreases the separation by at most $2e$).

    Note that $C_1'(\vec z)$ is a sigmoidal gate taking at most $d$ inputs (each of which is a $\AugSigmoid$ circuit of size $s_1+1$, degree $k$, and weights within $[-W_1,W_1]$). We now define another circuit $C''_1(\vec z)$ obtained from $C'_1(\vec z)$. 
    
    Let $\sL$ be the list of circuits in \Cref{lmm: counting depth 2} for approximating $\AugSigmoid$ circuit of size $s_1+1$, degree $k$, and weights within $[-W_1,W_1]$ up to precision $\delta/(10 d W)$. We replace each $\AugSigmoid$ subcircuit that feeds the output sigmoidal gate of $C_1'(\vec z)$ with its best approximation in $\sL$, so that
    \[
    |C''_1(\vec z)-C'_1(\vec z)| \;\le\; (dW)\cdot \frac{\delta}{10dW} \;\le\; \frac{\delta}{10}.
    \]
    Therefore, $C''_1(\vec z)$ still computes $F(\vec x,\vec z\oplus\vec\eta)$ with separation at least $4\delta/5 - 2\cdot\delta/10 = 3\delta/5$. Moreover, we know that each subcircuit in $\sL$ can be completely specified with
    \begin{equation}\label{eq: logL formula}
    \log_2|\sL| = O\left(n_1s_1\log(dn_1s_1\delta^{-1} W_1)+k^2 \left(\frac{n_1e}{k}\right)^k \log(kdn_1\delta^{-1}W_1)\right)
    \end{equation}
    bits, where we use the approximation $\binom{n}{k}\le(ne/k)^k$.
    
    Then by \Cref{lmm: disc} with $\sigma=\Sigmoid$ (which is $1/4$-Lipschitz) and $B=\lceil 4\kappa dW/(3\delta/5)\rceil = O(dW/\delta)$, which applies since $F(\vec x,\cdot)$ is not constant for the string $\vec x$ we fix below, there exist a $\AugSigmoid$ circuit $G'\in\sL$ and an index $\tau\in[B]$ such that, defining $G'_{\tau}(\vec z)\eqdef \sgn(G'(\vec z)-\tau/B)$,
    \begin{align}
    & \left|\Pr_{\vec z\gets\zo^{n_1}}[G'_{\tau}(\vec z)=1\mid F(\vec x,\vec z\oplus\vec\eta)=1]-\Pr_{\vec z\gets\zo^{n_1}}[G'_{\tau}(\vec z)=1\mid F(\vec x,\vec z\oplus\vec\eta)=0]\right| \nonumber \\
    & \quad \ge\; \frac{3\delta/5}{2\kappa d W} \;=\; \frac{6\delta}{5 d W}\;\ge\; \frac{\delta}{dW}. \label{equ: discriminator promise in sigmoid}
    \end{align}
    Note that $G'_\tau(\vec z)$ can be completely specified with $O(\log_2|\sL|+\log_2 B)$ bits.

    \paragraph{Counting argument.} We now set parameters \resolvemark{$(\diamond)$}{size-diamond} and prove that $\log_2(|\sL|\cdot B+1)=o(n)$, where:
    \[
    t\eqdef \lceil(\log_2 n)^{6a}\rceil \qquad\text{and}\qquad k\eqdef \lceil\log_2 n/\log_2\log_2 n\rceil.
    \]
    By assumption we have $s\le n^{3/2}$. Combined with $d,W,1/\delta\le n^a$ we have $sdW^2/\delta\le n^{4a+3/2}$. Note that
    \[
    t^{k+1}\;\ge\; t^k \;\ge\; 2^{k\cdot 6a\log_2\log_2 n}\;\ge\; 2^{6a\log_2 n} \;=\; n^{6a},
    \]
    which is at least $10\cdot sdW^2/\delta$ for every $a\ge 1$ and $n$ sufficiently large, as required by \resolvemark{$(\star)$}{size-star}.

    We now instantiate \Cref{eq: logL formula} with the chosen parameters. Recall that $n_1=m=\Theta(\log n\cdot\log\log n)$, and let $\delta_{\sL}\eqdef\delta/(10dW)$, so $\log_2(1/\delta_{\sL})=O(a\log n)$. Subsequently,
    \begin{compactitem}
    \item By \Cref{eq: s1 def}, $s_1 = O\!\big(s m t(k+1)\ln t/\sqrt n\big) = O\big(a\, s\,(\log n)^{6a+2}\log\log n/\sqrt n\big)$.
    \item By \Cref{eq: W1 def}, $\log_2 W_1 = O(a\log n+n_1\log n_1) = O\big(a(\log n)^2/\log\log n\big)$.
    \item The binomial term $(n_1 e/k)^k$ in \Cref{eq: logL formula} satisfies
    \[
    \big(n_1 e/k\big)^k \;\le\; \bigl(17e(\log_2\log_2 n)^2\bigr)^{\log_2 n/\log_2\log_2 n+1} \;=\; n^{o(1)}.
    \]
    \end{compactitem}

    The bounds above give $\log_2(d n_1 s_1 W_1/\delta_{\sL}) = O(a(\log n)^2/\log\log n)$. Plugging them into \Cref{eq: logL formula},
    \[
    \log_2|\sL|
    \;=\; O\!\left(\frac{a^2\,s\,(\log n)^{6a+5}\,\log\log n}{\sqrt n}\right) + n^{o(1)},
    \qquad \log_2 B \;=\; O(\log(dW/\delta)) \;=\; O(a\log n).
    \]
    Set $c_0\eqdef 12$, and suppose toward contradiction that $s\le n^{3/2}/(\log n)^{c_0\cdot a}$. Since $c_0 a-(6a+5)=6a-5\ge 1$ for every $a\ge 1$,
    \[
    \frac{a^2\,s\,(\log n)^{6a+5}\,\log\log n}{\sqrt n}
    \;\le\; \frac{a^2\,n\,\log\log n}{(\log n)^{c_0 a-(6a+5)}}
    \;=\; \frac{a^2\,n\,\log\log n}{(\log n)^{6a-5}}
    \;\le\; \frac{a^2\,n\,\log\log n}{\log n}\;=\;o(n),
    \]
    and so $\log_2(|\sL|\cdot B+1)=o(n)$, as desired.

    \paragraph{Applying \Cref{lmm: KW code}.} Define the augmented list $\sL'\eqdef \big\{\,\vec z\mapsto G'_\tau(\vec z\oplus\vec\eta')\,:\,G'\in\sL,\,\tau\in[B],\,\vec\eta'\in\zo^m\,\big\}\;\cup\;\{\mathbf{0}\}$, where $\mathbf{0}:\zo^{n_1}\to\zo$ is the constant zero function. Then $|\sL'|\le 2^m\cdot|\sL|\cdot B+1\le 2^{o(n)}\le 2^{n/2}$, as $m=o(n)$. Note also that $n^2\le 2^{m}\le 2^n$ since $16\log_2 n\le m\le n$. Subsequently, by \Cref{lmm: KW code} applied to $\sL'$, there exists $\vec x^*\in\zo^n$ such that for every $f\in\sL'$,
    \begin{equation}
    \Cor\!\big(f,\,F(\vec x^*,\cdot)\big) \;\le\; \frac{O(\sqrt n)}{2^{m/4}} \;\le\; O(\sqrt n)\cdot n^{-4\log_2\log_2 n}.\label{equ: cor small in list}
    \end{equation}
    Let $\eps$ be the largest correlation between $f\in\sL'$ and $F(\vec x^*,\cdot)$ and set $\vec x\eqdef \vec x^*$. 
    
    Recall that the discriminator lemma yields (for some $G'\in \sL$ and $\tau\in[B]$) the inequality \eqref{equ: discriminator promise in sigmoid}. Replacing $\vec z$ by $\vec z\oplus\vec\eta$ there, the function $g(\vec z)\eqdef G'_\tau(\vec z\oplus\vec\eta)$ in $\sL'$ satisfies
    \begin{equation}\label{eq: disc lower bound}
    \big|p_1-p_0\big| \;\ge\; \frac{\delta}{dW},\qquad\text{where }p_b\eqdef \Pr_{\vec z\gets\zo^{n_1}}[g(\vec z)=1\mid F(\vec x^*,\vec z)=b].
    \end{equation}
    Since $\mathbf{0}\in\sL'$, \eqref{equ: cor small in list} also holds for $\mathbf{0}$, which implies that $F(\vec x^*,\cdot)$ is $\eps$-balanced; as $\eps<1/2$, $F(\vec x^*,\cdot)$ is not constant, so the discriminator lemma indeed applies. Combining \eqref{eq: disc lower bound} with \Cref{lmm: discriminator to correlation}, we have
    \[
    \Cor\!\big(g,F(\vec x^*,\cdot)\big)\;\ge\; \frac{\delta}{2dW}-\eps.
    \]
    On the other hand, $g\in\sL'$ gives $\Cor(g,F(\vec x^*,\cdot))\le\eps$. Subsequently, $\delta/(dW)\le 4\eps = O(\sqrt n)\cdot n^{-4\log_2\log_2 n}$. However, $\delta/(dW)\ge n^{-a}\cdot n^{-2a} = n^{-3a}$, which is much larger than $O(\sqrt n)\cdot n^{-4\log_2\log_2 n}$ for $n$ sufficiently large. This leads to a contradiction.
\end{proof}

\begin{proof}[Proof of \Cref{prop: simulation under restriction}]
Let $G$ be a subcircuit in the middle layer of $C_1$, and $F_1,\dots,F_s$ be the gates in the first layer, so that
\[
    G(\vec z)= \Sigmoid\left(\sum_{i=1}^s w_i\cdot F_i(\vec z) + \ell(\vec z)\right)
\]
for some $w_1,\dots,w_s\in[-W,W]$ and an affine function $\ell$ that collects the bias of $G$ and its wires from the (restricted) input variables. As $G$ has at most $2n$ input wires, the absolute values of the coefficients of $\ell$ sum to at most $(2n+1)W$. 

Let $\sA$ be the set of indices of the (at most $s_1$) first-layer gates that are not approximated by polynomials. For $i\notin\sA$, let $p_i$ be the multilinear polynomial given by \Cref{cor: single sigmoid}, whose coefficients have absolute values at most $2^{k+1}$. For $i\in\sA$, let $F'_i$ be the sigmoid gate given by \Cref{lmm: sigmoid compression} for $F_i$ (as a function of the $n_1$ free variables) and $\eps\eqdef t^{-(k+1)}$, whose weights and bias have absolute values bounded by $(n_1+1)!\cdot \ln(3 t^{k+1})\le W_1$. Let
\[
G'(\vec z)\eqdef \Sigmoid\left(\sum_{i\in\sA} w_i\cdot F'_i(\vec z)+P(\vec z)\right),\qquad\text{where }P\eqdef\sum_{i\notin\sA}w_ip_i+\ell.
\]
Then $G'$ is a $\AugSigmoid$ circuit of size at most $s_1+1$ and degree $k$, and each coefficient of $P$ has absolute value at most $2^{k+1}sW+(2n+1)W\le W_1$. Moreover, every first-layer gate $F_i$ is replaced by a function that differs at most $t^{-(k+1)}$ from $F_i$ on $\zo^{n_1}$; as $\Sigmoid$ is $(1/4)$-Lipschitz, $|G(\vec z)-G'(\vec z)|\le \frac14\sum_i|w_i|\cdot t^{-(k+1)}\le sW\cdot t^{-(k+1)}$.
\end{proof}

\subsubsection{Proof of the Wire Lower Bound}

Now we prove a similar lower bound on the total number of wires in the first layer. It follows closely the proof of \Cref{thm: depth-3 lower bound}; we describe only the changes (marked in \diff{blue}).

\begin{theorem}\label{thm: depth-3 wire lower bound}
    There exists an absolute constant $c_0\ge 1$ such that the following holds. Let $a\ge 1$, let $n\in\N$ be sufficiently large (depending on $a$), and $d,W,1/\delta\le n^a$. Suppose that $C$ is a $(2n)$-input depth-$3$ sigmoidal circuit, where
    \begin{compactitem}
    \item the output gate is a sigmoidal gate with fan-in $d$ and weights within $[-W,W]$;
    \item the middle layer consists of $d$ sigmoidal gates with weights and biases within $[-W,W]$;
    \item the bottom layer consists of sigmoidal gates, each with fan-in at least one, \diff{whose total fan-in is $w$}.
    \end{compactitem}
    If $C$ computes $\GA_{n,m}$ with separation $\delta\in(0,1)$, then
    $
    \diff{w} \;\ge\; \diff{n^{5/2}}/(\log n)^{c_0\cdot a}.
    $
\end{theorem}

\begin{proof}[Proof Sketch]
    The crucial step is to replace \Cref{cor: single sigmoid} with the wire-complexity variant \diff{\Cref{cor: single sigmoid wire}}: the failure probability of the $j$-th first-layer gate with fan-in $w_j$ under random restriction $\rho\gets\cR_\sP$ becomes \diff{$O(m^{3/2}\, t\, w_j(k+1)\ln t/n^{3/2})$}. Since $w = \sum_j w_j$, by linearity of expectation, the expected number of bad gates is at most \diff{$O(m^{3/2}\, t\, w(k+1)\ln t/n^{3/2})$}, and Markov's inequality then yields a restriction $\rho\in\cR_\sP$ such that all but
    \begin{equation}\label{eq: s1 wire def}
    s_1\;\eqdef\; O\!\left(\diff{\frac{m^{3/2}\, t\, w\,(k+1)\ln t}{n^{3/2}}}\right)
    \end{equation}
    first-layer gates are $t^{-(k+1)}$-close to a degree-\diff{$(k+1)$} polynomial (rather than degree $k$, by \Cref{cor: single sigmoid wire}). Analogously to \Cref{prop: simulation under restriction}, for each middle-layer subcircuit $G$, there is a $\AugSigmoid$ circuit $G'$ of size $s_1+1$, degree \diff{$k+1$}, and weights within $[-W_1, W_1]$ for $W_1\eqdef \max\{(2^{\diff{k+2}}s+2n+1)W,\,(n_1+1)!\cdot\ln(3t^{k+1})\}$, with the same per-gate error $|G-G'|\le sW\cdot t^{-(k+1)}$. We can then apply the discriminator lemma to obtain a $\AugSigmoid$ discriminator for $F(\vec x,\cdot\oplus\vec\eta)$, provided that $t^{k+1}\ge 10\cdot sdW^2/\delta$ \defermark{($\star$)}{wire-star}.

    Assume for contradiction that we have $w\le n^{5/2}/(\log n)^{c_0\cdot a}$; in particular, let $s$ be the number of bottom-layer gates, then $s\le w\le n^{5/2}$. Combined with $d,W,1/\delta\le n^a$, this gives \diff{$sdW^2/\delta\le n^{4a+5/2}$}. We can set
    \[
    t\eqdef \lceil(\log_2 n)^{\diff{7a}}\rceil \qquad\text{and}\qquad k\eqdef \lceil\log_2 n/\log_2\log_2 n\rceil
    \]
    so that $t^{k+1}\ge 10\cdot sdW^2/\delta$ for every $a\ge 1$ and $n$ sufficiently large (the condition \resolvemark{$(\star)$}{wire-star}).

    Following the counting argument of \Cref{thm: depth-3 lower bound}, the dominant term $n_1 s_1\log_2 W_1$ becomes $O(\diff{w\,a^{2}(\log n)^{7a+11/2}\,(\log\log n)^{3/2}/n^{3/2}})$. By setting \diff{$c_0\eqdef 14$}, it follows that $\log_2(|\sL|\cdot B+1)=o(n)$. We can then apply \Cref{lmm: KW code} with the list $\sL'$ defined as in the proof of \Cref{thm: depth-3 lower bound} to conclude a contradiction.
\end{proof}

\subsection{Correlation Bounds with Parity via Average Sensitivity}

In this section, we prove a correlation bound for polynomial-weight small sigmoidal circuits and the parity function. Our main result is the following lower bound that matches the correlation bound for linear threshold circuits (see, e.g., \cite{CSS18}). 

\begin{restatable}{theorem}{CorParityGate}\label{thm: correlation parity gate}
    For any $a,d\ge 1$ and sufficiently large $n$, let $C:\zo^n\to[0,1]$ be any depth-$d$ sigmoidal circuit with $s$ gates and weights within $[-n^a,n^a]$. Then for any $0\le \delta_0<\delta_1\le 1$ with $\delta_1-\delta_0\ge n^{-a}$, we have
    \[
    \Pr_{\vec x\gets\{0,1\}^n}\left[C(\vec x)>\delta_1\land\oplus_n(\vec x)=1\right] + \Pr_{\vec x\gets\{0,1\}^n}\left[C(\vec x)<\delta_0\land\oplus_n(\vec x)=0\right]\le \frac{1}{2} + s^{1-1/d} \cdot n^{-1/(2d)+o(1)}.
    \]
    In particular, for $d\ge 2$ and $s \le n^{1/(2(d-1))-\Omega(1)}$, if $C$ computes a Boolean function $f$ with separation at least $n^{-a}$, then $\Cor(f,\oplus_n)\le n^{-\Omega(1)}$.
\end{restatable}

Indeed, we also prove an average sensitivity upper bound for any Boolean function computed by polynomial-weight small sigmoidal circuits with noticeable separation. Formally: 

\begin{restatable}{theorem}{CorAS}\label{thm: AS sigmoid}
For any $a,d\ge 1$ and sufficiently large $n$, let $C:\zo^n\to[0,1]$ be any depth-$d$ sigmoidal circuit with $s$ gates and weights within $[-n^a,n^a]$. Suppose that it computes a Boolean function $f:\zo^n\to\zo$ with separation $\delta\ge n^{-a}$, then $\AS(f)\le s^{1-1/d}\cdot n^{1-1/(2d)+o(1)}$. 
\end{restatable}

\subsubsection{Random Restriction on Polynomial Sigmoid Functions}

To prove the correlation bound for parity, we need to analyze the behavior of \emph{polynomial sigmoid functions} under random restriction.

\begin{definition}[Polynomial Sigmoid Functions]
    A \emph{polynomial sigmoid function} is a function of the form 
    \[
    \PSigmoid_{p}(\vec x) = \Sigmoid(p(\vec x)),
    \] 
    where $p$ is a real-weighted multi-linear polynomial. The \emph{degree} of the function is defined as the degree of the polynomial $p$. The function is said to have \emph{weight} at most $W$ if each coefficient of monomials in $p$ has absolute value at most $W$.
\end{definition}

Recall the definition of restrictions $\rho=(I,\vec y)$ in \Cref{sec: random restriction}. We use $\cR_{n,m}$ to denote the distribution of restrictions in which $I$ is a uniformly random subset of $[n]$ of size $m$ and $\vec y$ is uniformly random. When the input length $n$ is clear from the context, we may drop it and simply use $\cR_m$ to denote the distribution.

\begin{restatable}{lemma}{PSigRestriction}\label{lmm: psigmoid restriction}
    For any degree-$d$ multilinear polynomial $p(\vec x)$ with $n$ variables, $t\ge 2$, $k\ge 1$, $m\le n/16$, $\delta\in(0,1/16)$ with $\delta\le\sqrt{m/n}$, over a random restriction $\rho\gets\cR_m$, with probability at least
    \[
    1 - O\left(\sqrt{m/n}\cdot kt\cdot \bigl(\log(n/m)\cdot \log(tk/\delta)\bigr)^{O(d^2)}\right)
    \]
    there is a degree-$(dk)$ polynomial $q:\R^m\to\R$ such that $\PSigmoid_p|_\rho,q:\zo^m\to\R$ are $\delta$-almost $t^{-(k+1)}$-close to each other.
\end{restatable}

The proof of the lemma is close to its linear version (see \Cref{cor: single sigmoid}), with a crucial difference that we need to apply the random restriction lemma for polynomials in \cite{KKL17}. We defer the proof to the appendix (see \Cref{sec: KKL}). 

\subsubsection{Average Gap Sensitivity}

We will prove the correlation bound by considering the \emph{average sensitivity} of functions computed by Sigmoidal circuits. We start with the definition of average gap sensitivity for real-valued Boolean functions $f:\zo^n\to[0,1]$. 

\begin{definition}[average gap sensitivity]
    Let $f:\zo^n\to[0,1]$ be a function, its \emph{$(\delta_0,\delta_1)$-gap sensitivity} at $x\in\zo^n$, denoted $\sfS_{\delta_0,\delta_1}(f,x)$, is defined as 
    \[
    \sfS_{\delta_0,\delta_1}(f,x)\eqdef \sum_{i\in[n]}\Id[\max\{f(\vec x),f(\vec x^{(i)})\} > \delta_1 \land \min\{f(\vec x),f(\vec x^{(i)})\} < \delta_0],
    \]
    where $\vec x^{(i)}$ denotes the input with the $i$-th bit flipped. The \emph{average $(\delta_0,\delta_1)$-gap sensitivity} of $f$, denoted $\AS_{\delta_0,\delta_1}(f)$, is 
    \[
    \AS_{\delta_0,\delta_1}(f)\eqdef \Ex_{x\gets\zo^n}[\sfS_{\delta_0,\delta_1}(f,x)]. 
    \]
\end{definition}

The following lemmas show that a function $f:\zo^n\to[0,1]$ with small gap sensitivity, when we view it as a binary classifier with gap $\delta$, has small correlation with the parity function over $n$ bits. 

\begin{lemma}\label{lmm: sensitivity to correlation}
    Let $\delta_0,\delta_1\in\R$, $\delta_0<\delta_1$, $s\in\R$, and $f:\zo^n\to[0,1]$ be any function with $\AS_{\delta_0,\delta_1}(f)\le s$. Then
    \[
    \Pr_{x\gets\zo^n}[f(\vec x)>\delta_1\land \oplus_n(\vec x)=1] + \Pr_{x\gets\zo^n}[f(\vec x)<\delta_0\land \oplus_n(\vec x)=0] \le \frac{1}{2} + \frac{s}{2n}
    \]
\end{lemma}

\newcommand{\cE}{\mathcal{E}}
\newcommand{\cG}{\mathcal{G}}

\begin{proof}
    By the definition of average $(\delta_0,\delta_1)$-gap sensitivity, we know that 
    \begin{equation}\label{equ: prob interp of AS}
    \frac{\AS_{\delta_0,\delta_1}(f)}{n} \eqdef \Pr_{\vec x,i}[\max\{f(\vec x),f(\vec x^{(i)})\} > \delta_1 \land \min\{f(\vec x),f(\vec x^{(i)})\} < \delta_0],
    \end{equation}
    where $x\gets\zo^n$ and $i\gets[n]$. Let $\cG$ be the event in $\Pr[\cdot]$ above, $\cE$ be the event that either $f(\vec x)>\delta_1\land \oplus_n(\vec x)=1$ or $f(\vec x)<\delta_0\land \oplus_n(\vec x)=0$, and $\cE'$ be the event that either $f(\vec x^{(i)})>\delta_1\land \oplus_n(\vec x^{(i)})=1$ or $f(\vec x^{(i)})<\delta_0\land \oplus_n(\vec x^{(i)})=0$. By the definition of the events, $\cE\land\cE'$ implies $\cG$, since $\oplus_n(\vec x)\ne\oplus_n(\vec x^{(i)})$. Moreover, as the marginal distribution of both $\vec x$ and $\vec x^{(i)}$ are uniformly random, we know that $\Pr[\cE]=\Pr[\cE']$, and subsequently
    \[
    \frac{s}{n}\ge \frac{\AS_{\delta_0,\delta_1}(f)}{n} =\Pr_{\vec x,i}[\cG] \ge \Pr_{\vec x,i}[\cE\land \cE'] \ge 1 - 2\Pr_{\vec x,i}[\neg\cE].
    \]
    This implies that $\Pr_{\vec x,i}[\cE] = 1 - \Pr_{\vec x,i}[\neg\cE] \le 1/2 + s/(2n)$.
\end{proof}

Next, we prove two useful lemmas on average sensitivity. The first lemma is an adaptation of \cite[Lemma 3]{Boppana97}.
\begin{lemma}\label{lmm: average-sensitivity restriction}
    Let $f:\zo^n\to[0,1]$ be a function and $1\le m\le n$. Then
    \[ 
    \frac{\AS_{\delta_0,\delta_1}(f)}{n} = \frac{\Ex_{\rho\gets\cR_m}[\AS_{\delta_0,\delta_1}(f|_\rho)]}{m}.
    \]  
\end{lemma}
\begin{proof}
    For $\vec x\in\zo^n$ and $i\in[n]$, we say $(\vec x,i)$ is good if $\max\{f(\vec x),f(\vec x^{(i)})\} > \delta_1$ and $\min\{f(\vec x),f(\vec x^{(i)})\} < \delta_0$.  For a random restriction $\rho\gets\cR_m$, we can consider it as a random subset $I\subseteq [n]$ of size $m$ (the free variables) and a random assignment $\vec y\in\zo^{n-m}$ to variables outside of $I$. It can be verified that for each $I\subseteq[n]$,
    \begin{align*}
    \Ex_{\vec y\gets\zo^{n-m}}[\AS_{\delta_0,\delta_1}(f|_\rho);\rho\eqdef (I,\vec y)] &= \sum_{i\in I}\Ex_{\vec x\gets\zo^n}[\Id[(\vec x,i)\text{ is a good pair}]] \\ 
    & = \frac{1}{2^n} \sum_{i\in I}\sum_{\vec x\in\zo^n} \Id[(\vec x,i)\text{ is a good pair}].
    \end{align*}
    Subsequently, 
    \begin{align*}
    \Ex_{\rho\gets\cR_m}[\AS_{\delta_0,\delta_1}(f|_\rho)] &= \Ex_{I\subseteq[n],|I|=m}\left[\frac{1}{2^n} \sum_{i\in I}\sum_{\vec x\in\zo^n} \Id[(\vec x,i)\text{ is a good pair}]\right] \\ 
    &= \Ex_{\vec x\gets\zo^n} \left[\sum_{i\in[n],(\vec x,i)\text{ good}} \Ex_{I\subseteq[n],|I|=m}[i\in I]\right] \\
    &= \Ex_{\vec x\gets\zo^n} \left[\sum_{i\in[n],(\vec x,i)\text{ good}} \frac{m}{n}\right] = \frac{m}{n}\cdot \AS_{\delta_0,\delta_1}(f). \qedhere
    \end{align*}
\end{proof}

\begin{lemma}\label{lmm: as for close}
    Let $f,g:\zo^n\to[0,1]$ be $\gamma$-almost $\delta$-close to each other. If $\AS_{\delta_0+\delta,\delta_1-\delta}(f)\le s$, where $\delta_1-\delta_0>\delta$, then $\AS_{\delta_0,\delta_1}(g)\le s + 2\gamma n$.
\end{lemma}

\begin{proof}
    Let $\delta_0'\eqdef\delta_0+\delta$ and $\delta_1'\eqdef\delta_1-\delta$. Let $B\subseteq\zo^n$ be the set of inputs on which $f$ and $g$ differ by more than $\delta$, and thus $|B|\le \gamma\cdot 2^n$. Call a pair $(\vec x,i)\in\zo^n\times[n]$ \emph{good} if both $\vec x,\vec x^{(i)}\notin B$, and \emph{bad} otherwise; the number of bad pairs is at most $2|B|\cdot n\le 2\gamma\cdot n\cdot 2^n$.

    For a good pair $(\vec x,i)$ contributing to $\sfS_{\delta_0,\delta_1}(g,\vec x)$, i.e.,
    \[ 
    \max\{g(\vec x),g(\vec x^{(i)})\}>\delta_1 \;\text{and}\; \min\{g(\vec x),g(\vec x^{(i)})\}<\delta_0, 
    \] 
    we can see that $(\vec x,i)$ also contributes to $\sfS_{\delta_0',\delta_1'}(f,\vec x)$. Subsequently, $\AS_{\delta_0,\delta_1}(g) \le \AS_{\delta_0',\delta_1'}(f) + 2\gamma n \le s + 2\gamma n$.
\end{proof}

\subsubsection{An Upper Bound on Average Sensitivity}
\label{sec: as upper}

By \Cref{lmm: sensitivity to correlation}, it suffices to analyze the average gap sensitivity of a sigmoidal circuit to prove a correlation lower bound with the parity function. We consider a strengthened model called \emph{polynomial-adjusted sigmoidal circuits}.

\begin{definition}[polynomial-adjusted sigmoidal circuit]\label{def: poly sigmoid circuit}
A \emph{polynomial-adjusted sigmoidal circuit} of degree $r$ on $n$ input variables is a circuit with a single output gate, where each gate $v$ computes
\[
G_v(\vec x)=\Sigmoid\Bigl(p_v(\vec x)+\sum_{u\to v}w_{uv}\cdot G_u(\vec x)\Bigr)
\]
for a multilinear polynomial $p_v$ of degree at most $r$ in the input variables (with arbitrary real coefficients), and the sum is over the gates $u$ that feed $v$. 

The \emph{size} of the circuit is the number of its gates, and its \emph{weights} are the coefficients $w_{uv}$ of the wires between gates. The \emph{depth} of a gate is the maximum number of gates on a path from an input variable to it (so the gates of depth $1$ are fed only by input variables), and the depth of the circuit is the depth of its output gate.
\end{definition}

In particular, a depth-$d$ sigmoidal circuit whose weights are within $[-W,W]$ is a depth-$d$ polynomial-adjusted sigmoidal circuit of degree $1$ whose weights are within $[-W,W]$. 

\begin{lemma}\label{lmm: as upper bound}
    Fix any $d\ge 1$, there exists $c=c(d)\ge 1$ such that the following is true for any $a\ge 1$ and any sufficiently large $n$ and any $1\le r\le n$. 
    
    Suppose that $C$ is an $n$-input polynomial-adjusted sigmoidal circuit of depth $d$, degree $r$, and size at most $s$, whose weights are within $[-W,W]$ for $W\le n^a$. Then for every $\delta\ge n^{-a}$ and every $\delta_0,\delta_1\in[0,1]$ such that $\delta_1-\delta_0\ge \delta$,
    \[
    \AS_{\delta_0,\delta_1}(C)\le s^{1-1/d}\cdot n^{1-1/(2d)}\cdot n^{2/\log\log n}\cdot 2^{r^3\cdot (\log \log n)^c}.
    \]
\end{lemma}

\begin{proof}
    Let $\Phi_d(s,n,r)$ denote the right-hand side of the bound. If $\Phi_d(s,n,r)\ge n$, the bound holds trivially as $\AS_{\delta_0,\delta_1}(C)\le n$, so we may assume that $\Phi_d(s,n,r)<n$. Let $c=c(d)\ge 1$ be determined later. We prove the lemma by induction on the depth $d\ge 1$.

    \paragraph{Base Case.} In the base case, i.e., $d=1$, the circuit is simply a $\PSigmoid_p(\vec x)$ gate, where $p(\vec x)$ is a multilinear polynomial of degree at most $r$ (so $s=1$).

    Fix $\delta_0,\delta_1\in\R$. By the monotonicity of $\Sigmoid$, there is $\theta\in\R$ such that $\PSigmoid_p(\vec x)>\delta_1\Rightarrow p(\vec x)>\theta$, and $\PSigmoid_p(\vec x)<\delta_0\Rightarrow p(\vec x) < \theta$. Let $f(\vec x)=\sgn(p(\vec x)-\theta)$, then by \Cref{thm: PTF AS},\footnote{It is worth noting that there is a claimed improvement to this upper bound by OpenAI, which may slightly improve our results in certain parameter regimes; see \href{https://github.com/openai/math/blob/main/preprints/Average-Sensitivity-of-Polynomial-Threshold-Functions-September-25-2026/main.pdf}{https://github.com/openai/math/blob/main/preprints/Average-Sensitivity-of-Polynomial-Threshold-Functions-September-25-2026/main.pdf}.} when $c$ is larger than an absolute constant and $n$ is sufficiently large, we have
    \begin{equation}\label{equ: base case bound}
    \AS_{\delta_0,\delta_1}(C) \le \AS(f) \le \sqrt n(\log n)^{O(r\log r)}2^{O(r^2\log r)}\le n^{1/2}\cdot 2^{r^3 (\log \log n)^c},
    \end{equation}
    where the first inequality follows from the definition of average (gap) sensitivity.

    \newcommand{\cV}{\mathcal{V}}

    \paragraph{Applying Random Restriction.} Now we consider the induction case: assume that the lemma holds for $d$, we will prove that it also holds for $d+1$. Let $C$ be a circuit of depth $d+1$, and $\cV$ be the set of gates of depth $1$ in $C$; as they are not fed by any gate, each $u\in\cV$ computes $\PSigmoid_{p_u}$ for a polynomial $p_u$ of degree at most $r$. Let $m \le n/16$ be determined later \defermark{($\heartsuit$)}{smaller-length-cor}, and we apply a random restriction $\rho\gets\cR_m$ to the circuit $C$. Let $t\ge 2,k\ge 1,\gamma\in(0,1/16)$ with $\gamma\le\sqrt{m/n}$ be parameters to be determined \defermark{($\diamond$)}{approx-para}. By \Cref{lmm: psigmoid restriction} and union bound, we know that with probability at least $1-\beta$, where
    \begin{equation}
    \beta \eqdef s \sqrt{m/n}\cdot kt \left(\log(n/m)\cdot \log(tk/\gamma)\right)^{O(r^2)},\label{equ: fail prob cor}
    \end{equation}
    every $G_u|_\rho$ with $u\in\cV$ is $\gamma$-almost $t^{-(k+1)}$-close to a polynomial $q_u$ of degree at most $rk$; where $O(\cdot)$ hides an absolute constant. 

    Let $\cE$ be the above event, and fix any $\rho$ such that $\cE$ happens. Let $C'$ be the circuit on $\zo^m$ obtained from $C|_\rho$ by removing the gates in $\cV$ and, for each remaining gate $v$, replacing $p_v$ by $p_v|_\rho+\sum_{u\in\cV,\,u\to v}w_{uv}\cdot q_u$, which has degree at most $r'\eqdef\min\{rk,m\}$. We may remove duplicated variables in each monomial so it is multilinear. The depth of every remaining gate decreases by exactly one, so $C'$ is a polynomial-adjusted sigmoidal circuit on $m$ inputs of depth $d$, degree $r'$, and size at most $s$, whose weights are within $[-W,W]$.

    Let $S\subseteq\zo^m$ be the set of $\vec z$ such that $|G_u|_\rho(\vec z)-q_u(\vec z)|\le t^{-(k+1)}$ for every $u\in\cV$; by union bound, $|S|\ge(1-s\gamma)\cdot 2^m$. Fix $\vec z\in S$. At $\vec z$, the preactivations of a remaining gate $G_v$ in $C|_\rho$ and in $C'$ share the term $p_v|_\rho(\vec z)$ and differ only through the gates feeding $v$, where each $u\in\cV$ is replaced by $q_u$. As $\Sigmoid$ is $(1/4)$-Lipschitz, the error of $v$ is at most $sW/4$ times the maximum error of the gates feeding $v$, and the error of each $u\in\cV$ is at most $t^{-(k+1)}$. By induction on the depth, $|C|_\rho(\vec z)-C'(\vec z)|\le\delta_\rho\eqdef(sW+1)^d\cdot t^{-(k+1)}$. As a result, $C|_\rho$ and $C'$ are $s\gamma$-almost $\delta_\rho$-close.

    \paragraph{Applying Induction Hypothesis.} We instantiate the deferred parameters \resolvemark{($\heartsuit$)}{smaller-length-cor}, \resolvemark{($\diamond$)}{approx-para} as
    \[
    \gamma\eqdef n^{-(a+2)},\quad t\eqdef \lceil n^{1/\log_2\log_2 n}\rceil,\quad k\eqdef \lceil 10ad\log_2\log_2 n\rceil,\quad m\eqdef \lceil n^{d/(d+1)}/s^{2/(d+1)}\rceil,
    \]
    taking $n$ sufficiently large compared to $a,d$ throughout. We can verify that: 
    \begin{compactitem}
        \item $m\le n/16$ and $m\ge n^{(d-1)/d}$: As $\Phi_{d+1}(s,n,r)\ge s^{d/(d+1)}\cdot n^{1-1/(2(d+1))}$ and $\Phi_{d+1}(s,n,r)<n$, we have $s<n^{1/(2d)}$, and thus $m\ge n^{d/(d+1)}\cdot n^{-1/(d(d+1))}=n^{(d-1)/d}$.
        \item \emph{$\delta_\rho\le \delta/4$.} Since $t^{k+1}\ge n^{10ad}$, $s<n$, $W\le n^a$, and $\delta\ge n^{-a}$, we have 
        \[\delta_\rho\le (2n^{a+1})^d/n^{10ad}\le n^{-a-2}\le \delta/4.\] 
        \item $\beta\le s^{1-1/(d+1)}\cdot n^{-1/(2(d+1))}\cdot n^{2/\log\log n}\cdot (\log n)^{O(r^2)}$ for sufficiently large $n$; see \Cref{equ: fail prob cor} for the definition of $\beta$. To see this, notice that $\sqrt{m/n}\le 2n^{-1/(2(d+1))}/s^{1/(d+1)}$, $kt\le n^{2/\log\log n}$, and $\log(n/m)\cdot\log(tk/\gamma) = (\log n)^{O(1)}$.
        \item \emph{$\gamma\le\sqrt{m/n}$ and $\gamma<1/16$}. Since $m\ge 1$, $a\ge 1$, we have $\gamma=n^{-(a+2)}\le n^{-1/2}\le\sqrt{m/n}$.
    \end{compactitem}
    First assume that $d\ge2$, so that $m\ge\sqrt n$. Set $\delta_0'\eqdef \delta_0+\delta_\rho$, $\delta_1'\eqdef \delta_1-\delta_\rho$, $a'\eqdef 2(a+1)$, $\delta'\eqdef \delta/2$. It follows that $nW\le n^{a+1}\le m^{a'}$, $\delta'=\delta/2\ge n^{-a}/2\ge m^{-a'}$, and $\delta_1'-\delta_0'\ge \delta'$. Hence by the induction hypothesis,
    \begin{equation}\label{equ: AS Cprime IH}
    \AS_{\delta_0',\delta_1'}(C')\le s^{1-1/d}\cdot m^{1-1/(2d)}\cdot m^{2/\log\log m}\cdot 2^{(rk)^3(\log \log m)^{c'}},
    \end{equation}
    where $c'\eqdef c'(d)$ is the constant from IH. As $k=\Theta(ad\log\log n)$, provided that $c\ge c'+5$ and $n$ is sufficiently large, we can obtain that
    \begin{equation}\label{equ: AS Cprime small}
    \AS_{\delta_0',\delta_1'}(C')\le s^{1-1/d}\cdot m^{1-1/(2d)}\cdot n^{2/\log\log n}\cdot 2^{r^3(\log \log n)^{c}}.
    \end{equation}

    Note that conditioned on $\cE$, $C'$ and $C|_\rho$ are $(s\gamma)$-almost $\delta_\rho$-close, and thus \Cref{lmm: as for close} gives $\AS_{\delta_0,\delta_1}(C|_\rho)\le \AS_{\delta_0',\delta_1'}(C')+2s\gamma m$. Subsequently we have 
    \begin{align}
    \AS_{\delta_0,\delta_1}(C) & =\frac{n}{m}\cdot\Ex_{\rho\gets\cR_m}[\AS_{\delta_0,\delta_1}(C|_\rho)] \tag{\Cref{lmm: average-sensitivity restriction}}\\
    & \le \frac{n}{m}\left(\Ex_{\rho\gets\cR_m}[\AS_{\delta_0,\delta_1}(C|_\rho)\mid \cE] + m\cdot \Pr[\lnot\cE]\right)\notag\\
    & \le \frac{n}{m}\cdot\AS_{\delta_0',\delta_1'}(C')+2s\gamma n+\beta  n. \tag{\Cref{lmm: as for close}}
    \end{align}
    Plugging \Cref{equ: AS Cprime small} into the first term and using $m\ge n^{d/(d+1)}/s^{2/(d+1)}$,
    \begin{align*}
    \frac{n}{m}\cdot\AS_{\delta_0',\delta_1'}(C') & \le s^{1-1/d}\cdot n\cdot m^{-1/(2d)}\cdot n^{2/\log\log n}\cdot 2^{r^3(\log\log n)^c}\\
    & \le s^{1-1/(d+1)}\cdot n^{1-1/(2(d+1))}\cdot n^{2/\log\log n}\cdot 2^{r^3(\log\log n)^c},
    \end{align*}
    where the last step uses the fact that $1-1/d+1/(d(d+1))=1-1/(d+1)$. The remaining terms satisfy $2s\gamma n\le 2n^{-a}\le 2$ and $\beta n\le s^{1-1/(d+1)}\cdot n^{1-1/(2(d+1))}\cdot n^{2/\log\log n}\cdot (\log n)^{O(r^2)}$. Subsequently, we can conclude that for some $c=c(d)$ and sufficiently large $n$, $\AS_{\delta_0,\delta_1}(C)\le s^{1-1/(d+1)}\cdot n^{1-1/(2(d+1))}\cdot n^{2/\log\log n}\cdot 2^{r^3(\log \log n)^{c}}$.

    For $d=1$, the circuit $C'$ is a single
    $\PSigmoid_Q$ gate with $\deg Q\le r'$.
    We therefore apply the base-case argument directly,
    which requires no bounds on the weights or separation.
    Using $r'\le rk=O_a(r\log\log n)$ in the PTF estimate
    of \eqref{equ: base case bound}, we obtain
    \[
    \AS_{\delta_0',\delta_1'}(C')
    \le \sqrt m\,2^{O_a(r^3(\log\log n)^3)}.
    \]
    Here $\delta_0'$ and $\delta_1'$ are defined as above.
    For $c>3$ and sufficiently large $n$, this also implies
    \eqref{equ: AS Cprime small}, since its leading factor
    $s^{1-1/d}m^{1-1/(2d)}$ equals $\sqrt m$ when $d=1$.
\end{proof}

By instantiating $r=1$ to \Cref{lmm: as upper bound}, we can obtain the average-sensitivity bounds for Boolean functions computed by sigmoidal circuits. %

\CorAS*

By combining \Cref{lmm: sensitivity to correlation} and \Cref{lmm: as upper bound} (for $r=1$), we can immediately derive the correlation bound with parity for sigmoidal circuits. %

\CorParityGate*


\section{On Sigmoidal Circuits with Large Weight}
\label{sec: unbounded weight}

In this section, we study sigmoidal circuits with large or even unbounded weight. As the simplest case to attack, we first consider depth-two circuits: $\Sigmoid\circ\Sigmoid$. We establish an architecture-preserving weight bound and a gate lower bound for depth-two circuits. We then allow the depth to grow and show that unrestricted weight sigmoidal circuits can compute inner product mod \(2\) with sublinear size. 

\subsection{Weight Upper Bounds on Depth 2}
\label{sec: weights depth two}

Recall that for linear threshold gates (i.e.~$\THR(\vec x)\eqdef \Id[\langle \vec w,\vec x\rangle + \beta\ge 0]$), it is known that integer weights $\vec w$ and bias $\beta$ of $O(n\log n)$-bits suffices (see, e.g., \cite{Hastad94}). This upper bound naturally generalizes to polynomial threshold gates; see, e.g., \cite{Saks93}. However, it does not apply to sigmoidal gates (or other standard activation functions) as the output is non-Boolean, so arbitrary changes on the current gate could alter the output of downstream gates.

In this section, we prove a weight bound for depth-$2$ Sigmoid circuits to compute a Boolean function with no separation. Specifically, every function computed by depth-two sigmoidal circuits with no separation can be computed by a depth-two sigmoidal circuit of the same architecture and bounded weights.\footnote{Note that every Boolean function on $n$ inputs can be computed
by a depth-two sigmoidal circuit with $O(2^n)$ gates and
polynomially bounded weights and biases, by simulating its
disjunctive normal form. This construction requires
exponentially many gates, whereas our theorem bounds the
parameters for much smaller circuit while maintaining the gate count.}

\begin{theorem}\label{thm: Sigmoid weight bound depth 2}
Suppose that $f:\{0,1\}^n\to\{0,1\}$ is computed by a depth-$2$ sigmoidal circuit with $h\geq 1$ gates in the bottom layer and no separation. Then $f$ is computed by a circuit with the same architecture whose weights and biases have absolute values at
most \[(h(n+1)+1)^{O(h(n+2)+1)}
    =2^{O(hn\log(hn))}\] 
with no separation. %
\end{theorem}

On Boolean inputs, exponentiating the bottom-layer parameters turns sigmoid outputs into rational functions of the new exponentiated parameters. Clearing positive denominators expresses correct classification as polynomial inequalities. Recall that a subset of $\mathbb R^k$ is semi-algebraic if
it is defined by a finite Boolean combination of polynomial
equalities and inequalities; see \cite[Section~2.3]{BSP06}.

To prove \Cref{thm: Sigmoid weight bound depth 2}, we will use a standard result due to Basu, Pollack, and Roy \cite{BPR96} that shows the existence of rational points contained in semi-algebraic sets. 

\begin{theorem}[{\cite[Theorem 4.1.2]{BPR96}}, see also {\cite[Theorem 13.16]{BSP06}}] 
\label{thm: small rational point} 
    Let $\mathcal S\subseteq \R^k$ be a nonempty semi-algebraic set defined by only strict inequalities in integer polynomials $\{P_i\}$. If each $P_i$ has degree at most $d\geq 2$ and coefficient bit size at most $\tau$, then $\mathcal S$ contains a point in $\mathbb Q^k$ whose coordinates have bit-size at most $\tau d^{O(k)}$, where $O(\cdot)$ hides a constant independent of the number of equalities and inequalities.
\end{theorem}

\begin{proof}[Proof of \Cref{thm: Sigmoid weight bound depth 2}]
    Let $\sigma(x)\eqdef\Sigmoid(x)$ for simplicity. Suppose our depth $2$ circuit denoted as $N: \{0, 1\}^n\to \mathbb R$ has $h$ gates in the first layer. Write $c(x)=c_0+\sum_{j=1}^n u_jx_j$, where $u_j$ is the weight of the direct wire from $x_j$ to the output, or zero if that wire is absent. By definition, any Boolean function computed by the circuit (with no separation) may be parameterized by
    \begin{align*}
        f(x) = \Id \left[c(x) + \sum_{i=1}^{h}v_{i}\sigma \left(b_i + \sum_{j=1}^{n}w_{ij}x_j\right) > 0\right].
    \end{align*}
    (The top sigmoidal gate is removed by the monotonicity of $\Sigmoid(x)$.) By shifting $c_0$ slightly, we may assume that the output of the inner function in $f(x)$ is always non-zero.

    Rewriting variables $B_i = \exp(-b_i)$ and $W_{ij} = \exp(-w_{ij})$, we can express
    \begin{align*}
        \sigma \left(b_i + \sum_{j=1}^{n}w_{ij}x_j\right) &= \left( 1 + \exp\left(-b_i - \sum_{j=1}^{n} w_{ij}x_j\right) \right)^{-1}\\
        &= \left( 1 + B_i\prod_{j\in[n]}W_{ij}^{x_j}  \right)^{-1}.
    \end{align*}
    Note that $M_i(x) = B_i\prod_{j\in[n]}W_{ij}^{x_j}$ is a monomial term in the variables $B_i$ and $W_{ij}$. Then
    \begin{align*}
        f(x) &= \Id \left[c(x) + \sum_{i=1}^{h}v_{i}\sigma \left(b_i + \sum_{j=1}^{n}w_{ij}x_j\right) > 0\right]\\
        &= \Id \left[c(x) + \sum_{i=1}^{h}v_{i}(1+ M_i(x))^{-1} > 0\right]\\
        &= \Id \left[c(x)\prod_{i\in [h]}(1+M_i(x)) + \sum_{i=1}^{h}v_{i} \prod_{j\neq i}(1+M_j(x)) > 0\right].
    \end{align*}
    Let $P_x \eqdef c(x)\prod_{i\in [h]}(1+M_i(x)) + \sum_{i=1}^{h}v_{i} \prod_{j\neq i}(1+M_j(x))$ be a polynomial in variables $B_i, W_{ij}, c_0,u_j, v_i$ with $\deg(P_x)\leq h(n+1) + 1$. To preserve the architecture, substitute $W_{ij}=1$ for absent bottom wires and $u_j=0$ or $v_i=0$ for absent output wires, we don't treat these entries as free variables.

    Let $\epsilon_x=2f(x)-1$, and let $\mathcal S_f$ be the
    set of parameters where the model classifies the input correctly. Thus
    \[
    \mathcal S_f=
    \left\{(c_0,u_j,v_i,B_i,W_{ij}):
    \epsilon_xP_x>0\text{ for every }x\in\{0,1\}^n,\;
    B_i,W_{ij}>0
    \right\}.
    \]
    We further restrict $\mathcal S_f$ by fixing the signs of $W_{ij}-1$, $u_j$, and $v_i$ on the nonzero wires to their original signs.
    The parameters obtained after shifting $c_0$ satisfy these
    inequalities, so $\mathcal S_f\ne\varnothing$. Subsequently, this is a nonempty open semi-algebraic set
    in $k\le h(n+2)+n+1$ variables, with each defining
    polynomial of degree at most $d=h(n+1)+1$.
    After expanding $P_x$, every coefficient is either
    $0$ or $1$. Hence the signed constraints have coefficients
    in $\{-1,0,1\}$ and coefficient bit size $O(1)$.

    By \Cref{thm: small rational point}, $\mathcal S_f$ contains a rational point $(\hat c_0,\hat u_j, \hat v_i, \hat B_i, \hat W_{ij})$ whose coordinates have numerator and denominator bit lengths at most $L \leq d^{O(k)}$. Since $B_i, W_{ij}$ are positive, it implies that $
    2^{-L}\leq \hat B_i,\hat W_{ij}\leq 2^L.$ Then we know that the circuit $\hat N$ with the architecture of $N$ and parameters $\hat c_0,\hat u_j,\hat v_i$, $
    \hat b_i=-\ln \hat B_i,
    \hat w_{ij}=-\ln \hat W_{ij}$
    computes $f$ with output threshold $1/2$, where
    \[
    |\hat b_i|,|\hat w_{ij}|
    \leq L\ln 2=d^{O(k)}.
    \]
    That is, the bottom layer weights are bounded by $d^{O(k)}$.

    We may normalize the top layer weights $\hat c_0,\hat u_j,\hat v_i$ by dividing them by their maximum magnitude $M\eqdef\max\bigl(|\hat c_0|,\max_{j\in[n]}|\hat u_j|,\max_{i\in[h]}|\hat v_i|\bigr)$. As $P_x$ is linear in $(\hat c_0,\hat u_j,\hat v_i)$, dividing by $M>0$ preserves every sign, and the top layer weights become at most $1$ in magnitude. Hence every parameter is bounded by
    \[
    d^{O(k)}
    =\bigl(h(n+1)+1\bigr)^{O(h(n+2)+1)}
    =2^{O(hn\log(hn))}.
    \]
    This completes the proof.
\end{proof}

\subsection{Lower Bounds for Depth-2 Circuits}
\label{sec: lb depth two}

We next prove an $\Omega(n/\log n)$ gate lower bound for
depth-two sigmoidal circuits computing the selection
function. For depth-2 circuits with sigmoid gates and discrete inputs, better VC dimension upper bounds are known. We use these to obtain a lower bound on the gate complexity of neural networks with sigmoids computing an explicit Boolean function using ideas from \cite{Koiran96}. Throughout this section, we follow the notation and terminology of \cite{Koiran96}, take logarithms in base $2$, and assume for simplicity that $n$ is a power of $2$.

We define the \emph{selection Boolean function} $f_n(\vec{x}, \vec{y})$ on inputs $\vec{x} \in \{0,1\}^{\log n}$ and $\vec{y} \in \{0,1\}^n$, where $f_n(\vec{x}, \vec{y}) = y_{\vec{x}}$ is simply the $\vec{x}$-th entry of $\vec{y}$. Here, we identify $\vec{x}$ with the integer in $\{0, \dots, n-1\}$ whose binary representation is $\vec{x}$, and index the coordinates of $\vec{y}$ from $0$ to $n-1$. For brevity, we write $f$ instead of $f_n$ from now on.

The selection function defines a family $\mathcal{F}$ of $2^n$ Boolean functions on $\{0,1\}^{\log n}$ parameterized by $\vec{y}^* \in \{0,1\}^n$, where each $f_{\vec{y}^*}$ is given by $f_{\vec{y}^*}(\vec{x}) = f(\vec{x}, \vec{y}^*)$. As observed in \cite{Koiran96}, every Boolean function $g$ on $\log n$ inputs belongs to $\mathcal{F}$, since $g$ is identical to $f_{\vec{y}^*}$ for $\vec{y}^* = (g(0), g(1), \dots, g(n-1))$. Thus $\mathcal{F}$ shatters $\{0,1\}^{\log n}$ and $\operatorname{VCdim}(\mathcal{F}) = n$. 

For a fixed network architecture, we can view some of its parameters (weights and biases) as indeterminate \emph{programmable parameters}, which gives rise to a family of Boolean functions as those parameters vary over $\mathbb{R}$ \cite{Koiran96,KS97}. To bound the VC dimension of our restricted architecture, we use a special case of \cite[Theorem~8.3]{AB02} (originally due to Goldberg and Jerrum \cite{goldberg1993bounding}). Recall that $\operatorname{sgn}(\alpha) = 1$ if $\alpha \ge 0$ and $\operatorname{sgn}(\alpha) = 0$ otherwise.

\begin{theorem}[Special case of {\cite[Theorem~8.3]{AB02}}]\label{thm:ab-poly}
Let $X$ be a set and let $p: \mathbb{R}^d \times X \to \mathbb{R}$ be such that, for every fixed $\vec{x} \in X$, the map $\vec{a} \mapsto p(\vec{a}, \vec{x})$ is a polynomial of degree at most $l$. Then the class $H = \{\vec{x} \mapsto \Id[p(\vec{a}, \vec{x}) > 0] : \vec{a} \in \mathbb{R}^d\}$ satisfies $\operatorname{VCdim}(H) \le 2d \log(12l)$.
\end{theorem}

\cite[Theorem 8.3]{AB02} is stated for $\operatorname{sgn}(p(\vec{a}, \vec{x}))$ (and more generally for fixed Boolean combinations of $k$ such classes, with bound $2d\log(12kl)$). The strict inequality version above follows immediately by applying their theorem to $-p$, since $\Id [p > 0] = 1 - \operatorname{sgn}(-p)$ and replacing every function $h \in H$ with $1 - h$ does not change the VC dimension of the class.

We now upper bound the VC dimension of depth-2 sigmoid architectures whose programmable parameters are restricted to the biases.

\begin{lemma}\label{lem:vc-biases}
Let $C$ be a depth-2 sigmoid circuit with $s$ gates in which all weights are fixed real numbers and every gate other than the output gate takes only input coordinates as inputs. For each bias vector $\vec{t} = (t_1, \dots, t_s) \in \mathbb{R}^s$, let $C_{\vec{t}}$ denote the circuit obtained by assigning the bias $t_i$ to the $i$-th gate. For any fixed threshold $\theta \in (0,1)$, the class $H = \{h_{\vec{t}} : \vec{t} \in \mathbb{R}^s\}$ of Boolean functions $h_{\vec{t}}(\vec{x}) = \mathbbm{1}[C_{\vec{t}}(\vec{x}) > \theta]$ satisfies
\[
  \operatorname{VCdim}(H) \le 2s \log(12s).
\]
The same bound holds if the output gate is a linear threshold gate instead of a sigmoid gate.
\end{lemma}

\begin{proof}
Number the gates so that gates $1, \dots, s-1$ are in the first layer and gate $s$ is the output gate. Let $\vec{w}_i$ be the weight vector of gate $i < s$, let $v_1, \dots, v_{s-1} \in \mathbb{R}$ be the weights on the wires from gates $1, \dots, s-1$ to the output gate, and let $\vec{u}$ be the vector of weights on the direct wires from the input coordinates to the output gate (wires not present in $C$ are assigned weight zero).

Fix an input $\vec{x}$. For each $i < s$, the output of gate $i$ is
\[
  \sigma(\vec{w}_i \cdot \vec{x} + t_i) = \frac{1}{1 + e^{-\vec{w}_i \cdot \vec{x}} e^{-t_i}} = \frac{1}{1 + c_i a_i}, \quad \text{where } c_i := e^{-\vec{w}_i \cdot \vec{x}} > 0 \text{ and } a_i := e^{-t_i} > 0,
\]
so that $c_i$ depends on $\vec{x}$ but not on $\vec{t}$. The preactivation of the output gate is therefore
\begin{equation}\label{eq:preact}
  z(\vec{x}, \vec{t}) := \sum_{i < s} \frac{v_i}{1 + c_i a_i} + \vec{u} \cdot \vec{x} + t_s.
\end{equation}
Since $\sigma$ is strictly increasing, $C_{\vec{t}}(\vec{x}) = \sigma(z(\vec{x}, \vec{t})) > \theta$ holds if and only if $z(\vec{x}, \vec{t}) > \tau$, where $\tau := \sigma^{-1}(\theta) = \ln(\theta/(1-\theta))$. (For a linear threshold output gate, simply take $\tau$ to be its threshold.) Multiplying \eqref{eq:preact} by the strictly positive quantity $\prod_{j < s} (1 + c_j a_j) > 0$, the condition $z(\vec{x}, \vec{t}) > \tau$ is equivalent to $p(a_1, \dots, a_{s-1}, t_s; \vec{x}) > 0$, where
\[
  p(a_1, \dots, a_{s-1}, t_s; \vec{x}) := \sum_{i < s} v_i \prod_{\substack{j < s \\ j \neq i}} (1 + c_j a_j) + (\vec{u} \cdot \vec{x} + t_s - \tau) \prod_{j < s} (1 + c_j a_j).
\]
For each fixed $\vec{x}$, $p(\vec{b}; \vec{x})$ is a polynomial of degree at most $s$ in the $s$ real variables $\vec{b} = (a_1, \dots, a_{s-1}, t_s)$, with coefficients depending only on $\vec{x}$ and the fixed weights of $C$. Setting $\vec{b} = (e^{-t_1}, \dots, e^{-t_{s-1}}, t_s) \in \mathbb{R}^s$ shows that every $h_{\vec{t}} \in H$ is of the form $\vec{x} \mapsto \Id[p(\vec{b}; \vec{x}] > 0]$. Thus $H$ is a subclass of the class in Theorem~\ref{thm:ab-poly} with $d = l = s$, which yields $\operatorname{VCdim}(H) \le 2s \log(12s)$.
\end{proof}

We now combine Lemma~\ref{lem:vc-biases} with Koiran's method \cite{Koiran96} to derive the gate complexity lower bound for computing $f_n$.

\begin{theorem}\label{thm:depth2-lb}
Let $C$ be an arbitrary depth-2 sigmoid circuit (with no restriction on the weights) that computes the selection Boolean function $f_n$ with zero separation. Then the number $m$ of gates in $C$ satisfies $n \le 2m \log(12m)$; in particular, $m = \Omega(n / \log n)$.
\end{theorem}

\begin{proof}
Gates that lie on no directed path to the output gate can be deleted without changing the function computed, so we may assume $C$ has none; since $C$ has depth $2$, every gate other than the output gate takes only input coordinates as inputs. Zero separation means that there is a threshold $\theta \in [0,1]$ such that $C(\vec{x}, \vec{y}) > \theta$ whenever $f_n(\vec{x}, \vec{y}) = 1$ and $C(\vec{x}, \vec{y}) \le \theta$ whenever $f_n(\vec{x}, \vec{y}) = 0$. Since $\sigma$ takes values in $(0,1)$ and $f_n$ is not constant, we necessarily have $\theta \in (0,1)$.

Let gates $1, \dots, m-1$ be the first-layer gates and gate $m$ be the output gate. In $C(\vec{x}, \vec{y})$, each first-layer gate $i < m$ computes $\sigma(\vec{w}_i \cdot \vec{x} + \vec{\beta}_i \cdot \vec{y} + \gamma_i)$, and the output gate $m$ computes $\sigma\bigl(\sum_{i < m} v_i g_i + \vec{u} \cdot \vec{x} + \vec{\beta}_m \cdot \vec{y} + \gamma_m\bigr)$, where $g_i$ is the output of gate $i$. Fixing $\vec{y} = \vec{y}^* \in \{0,1\}^n$ simply replaces each bias $\gamma_i$ with $t_i = \gamma_i + \vec{\beta}_i \cdot \vec{y}^*$ while keeping all weights $\vec{w}_i, v_i, \vec{u}$ fixed. Thus, the resulting circuit $C^{\vec{y}^*}$ on input $\vec{x} \in \{0,1\}^{\log n}$ is precisely the circuit $C_{\vec{t}}$ of Lemma~\ref{lem:vc-biases} (with $s = m$).

Consequently, every function $f_{\vec{y}^*} \in \mathcal{F}$ belongs to the class $H$ of Lemma~\ref{lem:vc-biases} (with $s = m$ and threshold $\theta$), meaning $\mathcal{F} \subseteq H$. Since $\operatorname{VCdim}(\mathcal{F}) = n$, Lemma~\ref{lem:vc-biases} gives
\[
  n = \operatorname{VCdim}(\mathcal{F}) \le \operatorname{VCdim}(H) \le 2m \log(12m),
\]
which implies $m = \Omega(n / \log n)$.
\end{proof}

\subsection{A Sublinear Upper Bound for IP Mod 2}
\label{sec: IP mod 2 lower bound}

Inner product mod $2$ (we refer to as $\IPtwo$) is a classical circuit lower bound target. Given inputs $x, y\in \{0,1\}^n$, $IP_2$ computes $\sum_{i\in [n]}x_iy_i$ mod $2$. For both Threshold gate circuits and ReLU gate circuits, $\IPtwo$ requires $\Omega(n)$ gates to compute, under unrestricted weights (Due to \cite{ROS94} and \cite{VRPS21} respectively).

However, we prove a surprising \emph{upper bound} for Sigmoid circuits computing the Inner Product Mod 2 function: $\IPtwo(\vec x,\vec y)\eqdef \langle\vec x,\vec y\rangle\bmod 2$. Let $n$ be the length of $\vec x$ (and $\vec y$), i.e., $2n$ be the input length. 

\IPtwoUpper*

The construction built on a bit-decoding technique due to Koiran \cite{Koiran96}, and an observation from \cite{DS96} that multiplication can be approximately computed from a linear combination of $\Sigmoid$. We first explain these two tricks.

Let $\sigma(x)\eqdef\Sigmoid(x)$. The following lemma (implicit in \cite{DS96}) shows how to compute multiplication of two real numbers approximately using a linear combination of four $\sigma$-units. 

\begin{lemma}\label{lem: sigmoid 4 gate multiplication}
Fix $t\in \mathbb R$. Suppose $|x|, |y| < \varepsilon$, then the $4$ gate construction \[
p' = \sigma (t+x+y) - \sigma(t+x) - \sigma(t+y) + \sigma(t)
\] 
satisfies $p' - \sigma''(t)xy=O(\varepsilon^3)$.
\end{lemma}

\begin{proof}
    Writing $\sigma(x+t) = \sigma(t) +x\sigma'(t) + \frac{x^2}{2}\sigma''(t) + O(x^3)$. Calculating directly:
    \begin{align*}
    p'
    &= \sigma (t+x+y) - \sigma(t+x) - \sigma(t+y) + \sigma(t)\\
    &= \bigl((x+y)-x-y\bigr)\sigma'(t)
       + \frac{\sigma''(t)}{2}
         \bigl((x+y)^2-x^2-y^2\bigr)
       + O(\varepsilon^3)\\
    &= \frac{\sigma''(t)}{2}\cdot 2xy
       + O(\varepsilon^3)\\
    &= \sigma''(t)xy+O(\varepsilon^3).
    \end{align*}
    Thus taking the difference we obtain $p' -  \sigma''(t)xy = O(\varepsilon^3).$
\end{proof}

Applying the lemma after scaling arbitrarily large $x, y$ results in the following corollary showing that for any pair of reals $x, y$, we can approximate $xy$ up to arbitrary accuracy using Sigmoid gates.

\begin{corollary}
    Suppose $x, y\in \mathbb R$, $\max(|x|, |y|)\leq M$, $h > 0$, the $4$ gate construction \[
    Q_{t, h}(x, y) = \frac{\sigma(t + hx + hy) - \sigma(t + hx) - \sigma(t + hy) + \sigma(t)}{\sigma''(t)h^2}
    \]
    satisfies $|Q_{t, h}(x, y) - xy|\leq O(hM^3)$, where the constant scales with $1/\sigma''(t)$.
\end{corollary}
\begin{proof}
    Writing $p = \sigma(t)$, we have
    \[
    \sigma'''(u)=p(1-p)(1-6p+6p^2),
    \] 
    we can see that $\sigma''' (t)\leq \frac{1}{4}$ for all $t$.

    Expanding the difference as an integral we get \[
    \sigma(t+a+b)-\sigma(t+a)-\sigma(t+b)+\sigma(t)
    =
    \int_0^a\int_0^b \sigma''(t+u+v)\,dv\,du.
    \]
    So,
    \[
    Q_{t,h}(x,y)-xy
    =
    \frac{1}{\sigma''(t)h^2}
    \int_0^{hx}\int_0^{hy}
    \bigl(\sigma''(t+u+v)-\sigma''(t)\bigr)\,dv\,du.
    \]
    Thus, for $h>0$,
    \begin{align*}
    |Q_{t,h}(x,y)-xy|
    &=
    \frac{1}{|\sigma''(t)|h^2}
    \left|
    \int_0^{hx}\int_0^{hy}\int_0^1
    \sigma'''\bigl(t+s(u+v)\bigr)(u+v)
    \,ds\,dv\,du
    \right|\\
    &\leq
    \frac{1}{4|\sigma''(t)|h^2}
    \int_0^{h|x|}\int_0^{h|y|}(u+v)\,dv\,du.\\
    &\leq \frac{h|x||y||x+y|}{4|\sigma''(t)|} = O\left(\frac{hM^3}{\sigma''(t)}\right)
    \end{align*}
    since $|u+v|\leq h|x+y|$, thus showing our corollary.
\end{proof}

The above derivation only works because of the accuracy of the degree-$2$ Taylor approximation to $\Sigmoid(x)$; indeed, this technique generalizes to any function $\gamma(x)$ that is 3 times continuously differentiable in a neighborhood of $a\in\R$ satisfying $\gamma''(a)\ne 0$ (see, e.g., \cite[Section 2]{DS96}). This approximation is impossible for $\ReLU$ and $\THR$, which is why \Cref{thm: IP2 sublinear Sigmoid} does not generalize to $\ReLU$ and $\THR$ circuits (in fact, near-linear lower bounds hold for both classes \cite{ROS94,VRPS21}). 

For a threshold circuit or sigmoidal circuit $C(\vec x)$ with $n$-input and $m$-output, we can naturally the function $f:\R^n\to\R^m$ by allowing the input to take values from $\R$ (rather than $\zo$). The following lemma explain a bit decoding trick in \cite{KS97}.

\begin{lemma}\label{lem: Threshold bit decoding}
    For every $n\ge 1$, there is a threshold circuit
    $C:\R\to\{0,1\}^n$ with $n$ gates such that, for every
    $x\in[0,1)$ with binary expansion $x=0.x_1x_2\dots$,
    its $i$-th gate outputs $x_i$ for every $i\in[n]$. Trailing digits are treated as $0$s.
\end{lemma}
\begin{proof}
    We construct the gates in order. Suppose $G_i(x)=x_i$ for all $i<d$,
    and let
    \[
        z_d=2^d x-\sum_{i=1}^{d-1}2^{d-i}G_i(x)
           =x_d+\sum_{j>d}2^{d-j}x_j.
    \]
    The last sum lies in $[0,1)$, so the threshold gate
    $G_d(x)=\Id[z_d\ge 1]$ outputs $x_d$. For $d=1$ the first
    sum is empty, giving the base case. We used $1$ gate for each digit, totaling $n$ gates.
\end{proof}

By choosing large enough weights, we may simulate threshold gates by sigmoidal gates within an arbitrarily small error (see \Cref{lmm: threshold by sigmoidal}). Thus we expect a similar statement to hold for sigmoidal circuits. 

Note that binary digits cannot be uniformly approximated on $[0,1)$ within error
less than $1/2$ by continuous sigmoid circuits. We therefore restrict to
a fixed finite input set, as arises from a circuit on Boolean inputs, and formulate the corollary below. 

\begin{corollary}\label{cor: Sigmoid bit extraction}
    Fix a finite set $T\subseteq[0,1)$, $\eps>0$, and $n\ge1$. There is an $n$-gate sigmoidal circuit such that, for every $x\in T$, its $i$-th gate approximates the $i$-th binary digit of $x$ to within $\eps$, for every $i\in[n]$.
\end{corollary}

This follows by applying the finite-input simulation argument of \Cref{lmm: threshold by sigmoidal} to the threshold bit decoder on $T$.

Now we explain the upper bound. At a high level, we group the input into $k$ groups of equal size. By \emph{encoding} the input appropriately, we can compute $\IPtwo$ in each group using the multiplication gadget in \Cref{lem: sigmoid 4 gate multiplication} and then decode the $t$-bit from the result $p_i$ of the multiplication for some $t=\Theta(n\log n/k)$. The key observation is that, instead of decoding for each group and then compute the XOR, we can simply add $\sum_ip_i$ and decode the $t$-th bit of that number. This leads to a square-root saving in total circuit size by choosing $k=\sqrt{n\log n}$.

\begin{proof}[Proof of \Cref{thm: IP2 sublinear Sigmoid}]
    Suppose our input bits are
    $(x_1, x_2\dots x_n, y_1, y_2\dots y_n).$ Define \[
    l = \lceil\log_{2}(n+1)\rceil,\qquad k = \lceil\sqrt{n\log n}\rceil
    \]
    We group the bits into $k$ buckets, each containing at most
    $\lceil\frac{n}{k}\rceil$ bits. Write $r=\lceil n/k\rceil$ and pad each bucket with zeroes to length $r$.

    Consider any such bucket with input bits
    $(x_{m_0}, x_{m_1}\dots x_{m_{r-1}})$ and
    $(y_{m_0}, y_{m_1}\dots y_{m_{r-1}})$. Let 
    \begin{equation}
    X\eqdef \sum_{i=0}^{r-1}2^{il}x_{m_i},\; Y\eqdef \sum_{i=0}^{r-1}2^{(r-1-i)l}y_{m_i}\label{equ: X and Y encoding}
    \end{equation}
    We define $p=XY$ be the product of $X$ and $Y$, so 
    \begin{align*}
    p = XY
    =
    \sum_{j=0}^{2r-2}2^{jl}
    \left(
      \sum_{\substack{0\leq a,c<r\\a+r-1-c=j}}
      x_{m_a}y_{m_c}
    \right).
    \end{align*}
    The coefficient for the term $2^{(r-1)l}$ is directly
    $\sum_{i=0}^{r-1}x_{m_i}y_{m_i}$.
    Moreover, if $2^l>r$, this would imply in the binary representation of $p$, the bits in the region $[(r-1)l, rl - 1]$ is directly the binary representation of $\sum_{i=0}^{r-1} x_{m_i}y_{m_i}$, and in particular, the parity of it is the bit of $p$ of weight $2^{(r-1)l}$.

    We construct the first layer using four gates for each of the $k$ buckets, padding each bucket with zeros to length $r$. Let $X_b,Y_b,p_b$ be \eqref{equ: X and Y encoding} for the basket $b$. Fix arbitrary $t$ with $\sigma''(t) > 0$ and put $M = 2^{rl}$, we write $\tilde p_b$ following \Cref{lem: sigmoid 4 gate multiplication}: 
    \[
    \tilde p_b
    =\frac{
    \sigma(t+hX_b+hY_b)-\sigma(t+hX_b)
    -\sigma(t+hY_b)+\sigma(t)
    }{\sigma''(t)h^2}
    =p_b+O_t(hM^3).
    \]
    Choose $h > 0$ such that $|\tilde p_b - p_b|\leq \frac{1}{8k}$ for every bucket and input.
    Summing up the product from each group, we see that
    \[
    P=\sum_{b=1}^k p_b
     =\sum_{j=0}^{2r-2}C_j2^{jl},
    \qquad
    C_{r-1}=\sum_{i=1}^n x_i y_i,
    \qquad
    0\leq C_j\leq n.
    \]
    Choose $l = \lceil\log_{2}(n+1)\rceil$. Since $0\le C_j\le n<2^l$, no carries occur between the base-$2^l$ digits of $P$. Thus the digits from $2^{(r-1)l}$ to $2^{rl-1}$ are exactly the digits of the binary representation of $C_{r-1}$. And note by our choice that $|\tilde p_d - p_d|\leq \frac{1}{8k}$ we have that \[|\tilde P - P| = |\sum_{d}\tilde p_d - p_d|\leq \sum_d |\tilde p_d - p_d|\leq \frac{1}{8}.\]

    Let $L = (2r-1)l$. Since $P$ has at most $2r-1$ base-$2^l$ digits, $0\le P<2^L$. Set $z=(\tilde P+\tfrac12)/2^L$. Also by the error bound on $\tilde P$, we have that $\frac{P+3/8}{2^L}\leq z\leq\frac{P+5/8}{2^L}.$ Thus the first $L$ fractional bits of $z$ are the $L$-bit binary representation of $P$. The bit of $P$ in position $(r-1)l$ is therefore the $L-(r-1)l=rl$-th fractional bit of $z$. By \Cref{cor: Sigmoid bit extraction}, applied to $z$ with error $1/8$, thresholding the last of these $rl$ gates at $1/2$ gives $\IPtwo(x,y)$.

    The circuit requires a total of \[rl + 4k = \left\lceil\frac{n}{k}\right\rceil\lceil\log_2 (n+1)\rceil + 4k\]
    gates for a depth of $\left\lceil\frac{n}{k}\right\rceil\lceil\log_2 (n+1)\rceil + 1$. Choosing $k = \lceil\sqrt{n\log n}\rceil$ obtains an asymptotic gate count of $O(\sqrt {n\log n})$, concluding our construction.
\end{proof}

It is worth noting that ReLU circuits computing $\IPtwo$ require a minimum of $\Omega(n/\log n)$ gates \cite{VRPS21}, and threshold circuits require $\Omega(n)$ gates \cite{ROS94}; thus our result shows a separation between Sigmoid circuits and ReLU circuits. 

\paragraph{Upper bound for slightly super-constant depth.} It is worth noting that if we adjust our block sizes according to our allocated depth and decode in parallel, we obtain the following theorem.

\begin{theorem}[Sublinear gate for growing depth]\label{thm: sublinear gates IP2 growing depth}
    For any unbounded increasing $d:\mathbb N \to \mathbb N$, there exists a set of Sigmoid circuits $S$ computing $\IPtwo$ on $n$ coordinates with depth at most $d(n)$ such that the asymptotic gates of $S_i$ is $o(n)$.
\end{theorem}

\begin{proof}
    Write $D = d(n) \geq 4$ and choose \[
    m = \left\lfloor\sqrt{\min(n, D-3)}\right\rfloor,\qquad B = \left\lceil\frac{n}{m}\right\rceil
    \]
    Divide the input into $B$ blocks of size $m$, padding the final block, we run \cref{thm: IP2 sublinear Sigmoid} algorithm on each block $b$ in parallel, denoted as the \emph{decoder algorithm} of block $b$. A block $b$ requires $l\geq \lceil\log_2 (m+1)\rceil$, where $l$ denotes the exponent of $2$ in the multiplication expansion. Divide the bits in this block into $r = \lceil\sqrt{\frac{m}{l}}\rceil$ groups for multiplication. This means that this block requires a total gate count of \[
    G_b = 3 \lceil m/r\rceil + rl = O(m/r + rl)
    \]
    gates, and a depth of $1 + rl \leq 1 + \sqrt{ml} + l$. 
    
    Let $p_b$ be the parity of $\IPtwo$ on the inputs to that block. As the algorithm in $\cref{thm: IP2 sublinear Sigmoid}$ can compute $p_b$ up to arbitrary accuracy, we assume the output $F_b$ of the decoder on block $b$ has \[
    |F_b - p_b|\leq \frac{1}{16B}
    \]

    Now say $P = \sum_{b\in[B]}p_b$ and $F = \sum_{b\in [B]}F_b$, we have that \[
    |P - F|\leq \frac{1}{16}
    \]
    Now we simply compute the parity of $P=round(F)$ using the depth $3$ circuit from \cref{thm: symmetric upper bound} thresholding at half integers instead of integers. Since $F$ is between $0$ and $B$, requires a total of $B+2$ gates to output $PF$ whose distance to the parity of $P$, denoted $P_F$ has \[
    |PF - P_F|\leq \frac{1}{16}.
    \]
    thus we obtain a constant separation for calculating $IP_2$. 

    The total gate count is \[O(B(m/r+rl)) + 3 = O(n\log m/\sqrt m) = o(n),\] since $m\to\infty$ as $D\to\infty$. And the depth is \[4 + \sqrt{ml} + l = O(\sqrt{\min(n, D)}\log(\min(n, D))\] showing the theorem.
\end{proof}


\section{Further Simulations between Threshold and Sigmoidal Circuits} 
\label{sec: further simulation}

Given that unrestricted sigmoidal circuits are very hard to analyze, we may consider an intermediate model between polynomial weights sigmoidal circuits with noticeable separation and unrestricted sigmoidal circuits. We now study the connection between this model and threshold circuits.  

\subsection{Connection to Threshold-of-Majority}

One natural candidate is to assume that all weights are small, but it is allowed to compute a function with no separation. We show that this model is strong enough to simulate $\THR\circ\MAJ$; indeed, a further restricted model is equivalent to $\THR\circ\MAJ$. Formally:

\SigmoidEquiv*

The key observation underlying (2) $\Rightarrow (1)$ is that for any function $f:T\to\R$, where $T\subseteq\R$ is a finite set, we can represent a small multiple of $f$ as a sum of differences of pairs of sigmoids with nearby biases. Formally:

\begin{lemma}\label{lmm: exact difference}
Let $T\subseteq\R$ be a finite set. There are $\beta_1,\dots,\beta_{|T|}\in(-1/2,1/2)$ such that for every $\psi:T\to\R$ and every sufficiently small $\delta>0$ (depending on $T$ and $\psi$), there are $\gamma_1,\dots,\gamma_{|T|}\in(-1,1)$ satisfying
\[
\sum_{j=1}^{|T|}\Bigl(\Sigmoid(u+\gamma_j)-\Sigmoid(u+\beta_j)\Bigr)=\delta\cdot\psi(u)\quad\text{for every }u\in T.
\]
\end{lemma}

\begin{proof}
Recall that $\Sigmoid(u+t)=e^{u}/(e^{u}+e^{-t})$. The matrix $(\Sigmoid(u+t))_{u,t\in T}\in\R^{|T|\times |T|}$ is invertible as it is the product of two matrices: 
\begin{compactitem}
    \item The diagonal matrix with entries $e^u$, which is invertible.
    \item The matrix $(1/(e^u+e^{-t}))_{u,t\in T}$. This is a Cauchy matrix over $\R$ (see, e.g., \cite[Section 0.9]{HJ13}) with lists $(e^{u})_{u\in T}$ and $(e^{-t})_{t\in T}$ as parameters, where both lists contain distinct elements, and $e^u+e^{-t}\ne 0$. It is known that Cauchy matrices with distinct elements are invertible.  
\end{compactitem} 

Let $\vec v_\beta\eqdef(\Sigmoid'(u+\beta))_{u\in T}\in\R^T$, we show that the vectors $\{\vec v_\beta\}_{\beta\in(-1/2,1/2)}$ span $\R^T$. Suppose for contradiction that $\vec w\in\R^T$ is orthogonal to every $\vec v_\beta$ for $\beta\in(-1/2,1/2)$. Let $g(\beta)\eqdef\sum_{u\in T}w_u\Sigmoid(u+\beta)$. Then $g'(\beta)=\langle\vec w,\vec v_\beta\rangle$ vanishes for $\beta\in(-1/2,1/2)$. As $g'(\beta)$ is a rational function of $e^\beta$, it vanishes for all $\beta\in\R$. Therefore, $g$ is a constant, which must be $0$ as $g(\beta)\to0$ when $\beta\to-\infty$. In particular, $g(t)=0$ for every $t\in T$, and thus $\vec w=\vec 0$ by the invertibility of the matrix $(\Sigmoid(u+t))_{u,t\in T}\in\R^{|T|\times |T|}$. 
    
Fix $\beta_1,\dots,\beta_{|T|}$ such that the vectors for $\beta=\beta_j$ form a basis. Consider the continuously differentiable vector-valued function $F:\R^{|T|}\to\R^T$ defined by 
\[ F(\vec\eta)_u\eqdef\sum_{j=1}^{|T|}(\Sigmoid(u+\beta_j+\eta_j)-\Sigmoid(u+\beta_j)). \] 
Then $F(\vec 0)=\vec 0$, and the Jacobian matrix of $F$ at $\vec 0$ is the matrix $(\Sigmoid'(u+\beta_j))_{u\in T,j\in[|T|]}$, which we have proved to be invertible.

Subsequently, by the inverse function theorem (see, e.g., \cite[Theorem 9.24]{Rudin76}), there are neighborhoods $V,W$ of $\vec 0$ such that $F:V\to W$ has a continuous inverse $F^{-1}:W\to V$. Then for every sufficiently small $\delta>0$, there is $\vec\eta$ satisfying that $F(\vec\eta)=\delta\psi$ and $\|\vec\eta\|_\infty<1/2$. The lemma then holds by setting $\gamma_j\eqdef\beta_j+\eta_j$. 
\end{proof}

\begin{proof}[Proof of \Cref{thm: equiv sigmoid threshold}]
    (1) $\Rightarrow$ (2): Assume that $f$ can be computed by a depth-two sigmoidal circuit of size $h\eqdef s\cdot\poly(n,W)$ with no separation, where weights are integers from $[-W,W]$. Each gate in the bottom layer is of form $\Sigmoid_{\vec w,\beta}(\vec x)\eqdef \Sigmoid(\langle\vec w,\vec x\rangle+\beta)$. As $\vec x\in\zo^n$, $\langle\vec w,\vec x\rangle$ takes values in $T\eqdef\{-nW,\dots,nW\}$, and subsequently $\Sigmoid_{\vec w,\beta}$ can be exactly simulated by a linear combination of the $2nW+1$ functions $\Id[\langle\vec w,\vec x\rangle = t]$, $t\in T$, each of which is the difference of the two $\MAJ$ gates $\Id[\langle\vec w,\vec x\rangle \ge t]$ and $\Id[\langle\vec w,\vec x\rangle \ge t+1]$.

    We can also replace the top $\Sigmoid$ gate by a $\THR$ gate, since $\Sigmoid$ is monotone, and the original circuit computes $f$ with no separation. The resulting $\THR\circ\MAJ$ circuit also computes $f$, and is of size at most $2(2nW+1)\cdot h = s\cdot \poly(n,W)$.

    (2) $\Rightarrow$ (1): Let $K\eqdef nW$ and $T\eqdef\{-K,\dots,K\}$. 
    Now assume that $f(\vec x)=\Id[\sum_{i=1}^m\alpha_iG_i(\vec x)\ge\theta]$, $m\le s\cdot\poly(n,W)$, where each $G_i(\vec x)=\Id[\langle\vec a_i,\vec x\rangle\ge b_i]$ is a $\MAJ$ gate with $\vec a_i\in\{-W,\dots,W\}^n$. By slightly decreasing $\theta\in\R$, we may assume that $\sum_i\alpha_iG_i(\vec x)\neq\theta$ for every $\vec x$. As $\langle\vec a_i,\vec x\rangle\in T$, we can apply \Cref{lmm: exact difference} to the function $\psi_i(u)\eqdef\alpha_i\cdot\Id[u\ge b_i]$ to obtain biases $\beta_{i,1},\dots,\beta_{i,|T|}\in(-1,1)$ such that for every sufficiently small $\delta>0$, there are $\gamma_{i,1}^{(\delta)},\dots,\gamma_{i,|T|}^{(\delta)}\in(-1,1)$ such that 
    \[ 
    \delta\alpha_i\cdot G_i(\vec x)=\sum_{j=1}^{|T|}\big(\Sigmoid(\langle\vec a_i,\vec x\rangle+\gamma_{i,j}^{(\delta)})-\Sigmoid(\langle\vec a_i,\vec x\rangle+\beta_{i,j})\big)
    \] 
    for every $\vec x\in\zo^n$. For $\delta>0$, let
    \[
    S_\delta(\vec x)\eqdef\sum_{i=1}^m\sum_{j=1}^{|T|}\big(\Sigmoid(\langle\vec a_i,\vec x\rangle+\gamma_{i,j}^{(\delta)})-\Sigmoid(\langle\vec a_i,\vec x\rangle+\beta_{i,j})\big) = \sum_{i=1}^m \delta \alpha_i \cdot G_i(\vec x).
    \]
    As $\zo^n$ is finite and $\sum_i\alpha_iG_i(\vec x)\neq\theta$ for every $\vec x$, we can fix a small $\delta>0$ such that $|\delta\theta|<1$ and for every $\vec x\in\zo^n$, $S_\delta(\vec x)-\delta\theta>0$ if and only if $f(\vec x)=1$. Hence the depth-two sigmoidal circuit $C(\vec x)\eqdef\Sigmoid(S_\delta(\vec x)-\delta\theta)$ satisfies $C(\vec x)>1/2$ if $f(\vec x)=1$ and $C(\vec x)<1/2$ otherwise.

    Finally, $C$ has $2m(2nW+1)+1=s\cdot\poly(n,W)$ gates. Its bottom gates have weights $\vec a_i\in\{-W,\dots,W\}^n$ and biases in $(-1,1)$; its output gate has weights $\pm1$ and bias in $(-1,1)$.     
\end{proof}

By applying the same argument, we may simulate $\THR\circ\MAJ^{d}$ by depth-$(d+1)$ sigmoidal circuits with small integer weights and small biases for any $d\ge 1$. 

\begin{theorem}\label{thm: sigmoid simulates THR MAJ d}
Let $d\ge 1$ and $W\in\N$. Suppose that $f:\zo^n\to\zo$ is computed by a $\THR\circ\MAJ^{d}$ circuit $C$ of size $s$ in which the $\MAJ$ gates have integer weights in $[-W,W]$. Then $f$ is computed with no separation by a depth-$(d+1)$ sigmoidal circuit of size $\poly(n,s,W)$ in which all gates have integer weights in $[-W,W]$ and biases in $(-1,1)$.
\end{theorem}

\begin{proof}
Without loss of generality, we assume $C$ is layered, i.e., every gate reads from the previous layer. For a gate $H$ in layer $\ell\in[d]$ of $C$, write $H(\vec x)=\Id[\langle\vec a,\vec x\rangle\ge b]$ if $\ell=1$ and $H(\vec x)=\Id[\sum_i c_iH_i(\vec x)\ge b]$ if $\ell\ge 2$, where $H_1,\dots,H_r$ are the inputs of $H$. Let $v_H(\vec x)$ be the integer
$\langle\vec a,\vec x\rangle$ for $\ell=1$, or
$\sum_i c_iH_i(\vec x)$ for $\ell\ge2$. Then $H=\phi_H\circ v_H$, where $\phi_H(v)\eqdef\Id[v\ge b]$ is a Boolean function on $T_\ell\eqdef\{-K_\ell,\dots,K_\ell\}$, with
$K_\ell=(n+s)W$.

Let $\delta_1\eqdef 1$ and $\delta_2,\dots,\delta_{d+1}>0$ be sufficiently small parameters. For every gate $H$ in layer $\ell$ and every $\beta\in(-1,1)$, we will construct a sigmoid gate $S_{H,\beta}$ in layer $\ell$ of the sigmoidal circuit that outputs \emph{exactly} $\Sigmoid(\beta+\delta_\ell v_H(\vec x))$. For $\ell=1$, let $S_{H,\beta}\eqdef\Sigmoid(\langle\vec a,\vec x\rangle+\beta)$. For $\ell\ge 2$, assume that such gates have been constructed for depth $\ell-1$, we apply \Cref{lmm: exact difference} to the set $\delta_{\ell-1}T_{\ell-1}$ and the functions $\psi_i(\delta_{\ell-1}v)\eqdef c_i\phi_{H_i}(v), i\in[r]$. Then for any sufficiently small $\delta_\ell$, there are $\beta_j,\gamma_{i,j}\in(-1,1)$ such that
\[
\sum_{i=1}^r\sum_{j=1}^{|T_{\ell-1}|}\Bigl(S_{H_i,\gamma_{i,j}}(\vec x)-S_{H_i,\beta_j}(\vec x)\Bigr)=\delta_\ell\sum_{i=1}^r\psi_i\bigl(\delta_{\ell-1}v_{H_i}(\vec x)\bigr)=\delta_\ell v_H(\vec x). 
\]
We define $S_{H,\beta}(\vec x)$ as the sigmoid of the LHS plus $\beta$. Note that we may assume $\delta_\ell$ is small enough so such $\beta_j,\gamma_{i,j}\in(-1,1)$ exist for all gates in layer $\ell$. Let $G$ be the output gate of $C$ that outputs $G(\vec x)=\Id[\sum_k\alpha_kH_k(\vec x)\ge\theta]$ with real $\alpha_k,\theta$. The final sigmoidal circuit outputs $S_{G,0}(\vec x)=\Sigmoid(\delta_{d+1}(\sum_k\alpha_kH_k(\vec x)-\theta))$. Assuming without loss of generality that $\sum_k\alpha_kH_k(\vec x)\ne\theta$ for every $\vec x$, the circuit computes $f$ with no separation. 

The weights and biases are as claimed by construction. Let $w=\poly(n,s)$ be the total number of wires in $C$. For each wire that connects the gate $H$ in layer $\ell\le d$ to the upper layer, we use $2|T_\ell|$ sigmoidal gates of form $S_{H,\beta_j}$ and $S_{H,\gamma_{i,j}}$. So the final circuit is of size at most $O(\max_\ell|T_\ell|\cdot(s+w))=\poly(n,s,W)$.
\end{proof}

\subsection{Connection to Threshold Circuits}

Using a slightly different approach, we can simulate $\THR\circ\THR$ using $\Sigmoid\circ\Sigmoid$ with the restriction that the top $\Sigmoid$ gate has small weights.

\SigSimTHR*

\begin{proof}
Assume without loss of generality that the output gate is not fed directly by input variables. Write $f(\vec x)=\Id[\sum_{i=1}^m\alpha_iG_i(\vec x)\ge\theta]$, where $G_1,\dots,G_m$ ($m\le n+s$) are the $\THR$ gates. By slightly decreasing $\theta$, we may assume that $\sum_i\alpha_iG_i(\vec x)\ne\theta$ for every $\vec x$. Each gate $G_i$ can be written as a disjoint OR, and therefore a sum, of at most $p=\poly(n)$ $\ETHR$ gates \cite[Theorem 7]{HP10}(by padding with the constant $\ETHR$ gate $\Id[1=0]$, we may assume that there are exactly $p$ of them). Denote $G_i(\vec x)=\sum_{j=1}^{p}\Id[L_{i,j}(\vec x)=0]$, where each $L_{i,j}$ is an affine function with integer coefficients. Hence $f(\vec x)=\Id[P(\vec x)>0]$, where
\[
P(\vec x)\eqdef\sum_{i=1}^m\alpha_i\sum_{j=1}^{p}\Id[L_{i,j}(\vec x)=0]-\theta
\]
satisfies that $P(\vec x)\ne 0$ for every $\vec x\in\zo^n$. 

Recall that $\Sigmoid'(u)=\Sigmoid(u)\Sigmoid(-u)\le e^{-|u|}$. Then the function $\Sigmoid'(\lambda L_{i,j}(\vec x))$ equals $\Sigmoid'(0)=1/4$ if $L_{i,j}(\vec x)=0$, and is at most $e^{-\lambda}$ otherwise (since $L_{i,j}$ has integer coefficients). In other words, $4\cdot \Sigmoid'(\lambda L_{i,j}(\vec x))$ converges to the $\ETHR$ gate $\Id[L_{i,j}(\vec x)=0]$ as $\lambda\to\infty$. For $\lambda,\delta>0$, let
\begin{align}
& D_{\lambda,\delta}(\vec x)\eqdef\sum_{i=1}^m\sum_{j=1}^{p}\Bigl(\Sigmoid(\lambda L_{i,j}(\vec x)+4\delta\alpha_i)-\Sigmoid(\lambda L_{i,j}(\vec x))\Bigr)-\delta\theta \label{equ: difference sim}\\ 
& P_\lambda(\vec x)\eqdef 4\sum_{i=1}^m\alpha_i\sum_{j=1}^{p}\Sigmoid'(\lambda L_{i,j}(\vec x))-\theta.\label{equ: derivative sim}
\end{align}
By the definition of the derivative, $\lim_{\delta\to 0^+}D_{\lambda,\delta}(\vec x)/\delta=P_\lambda(\vec x)$ for every $\lambda$ and $\vec x$. By the discussion above, $\lim_{\lambda\to\infty}P_\lambda(\vec x)=P(\vec x)$ for every $\vec x$. 

As $\zo^n$ is finite and $P(\vec x)\ne 0$, we can fix a large enough $\lambda$ such that $P_\lambda(\vec x)$ has the sign of $P(\vec x)$ for every $\vec x$,  then a small enough $\delta$ such that $D_{\lambda,\delta}(\vec x)$ has the sign of $P_\lambda(\vec x)$ for every $\vec x$. Hence the depth-two sigmoidal circuit $C(\vec x)\eqdef\Sigmoid(D_{\lambda,\delta}(\vec x))$ satisfies $C(\vec x)>1/2$ if $f(\vec x)=1$ and $C(\vec x)<1/2$ otherwise. The circuit $C$ has $2pm+1\le 2(n+s)p+1=s\cdot\poly(n)$ gates. Its output gate has weights $\pm1$ and bias $-\delta\theta$.
\end{proof}

Combined with the simulation of threshold circuits with sigmoidal circuits in \Cref{lmm: threshold by sigmoidal}, we can generalize this simulation to larger depth. 

\begin{theorem}\label{thm: sigmoid simulates THR d}
Let $d\ge 1$. Suppose that $f:\zo^n\to\zo$ is computed by a depth-$(d+1)$ circuit $C$ of size $s$ consisting of $\THR$ gates. Then $f$ is computed with no separation by a depth-$(d+1)$ sigmoidal circuit of size $\poly(n,s)$, where the output gate has weights in $\{\pm 1\}$.
\end{theorem}

\begin{proof}[Proof Sketch]
Without loss of generality assume that the circuit is layered. The proof is almost identical to that of \Cref{thm: sigmoid simulates THR THR} with two changes. First, we write each gate $G_i$ at depth $d$ as a sum of $\ETHR$ gates, and then simulate each exact threshold gate using the difference of two sigmoids; see \eqref{equ: difference sim}. Using this method, we can replace the top two layers by two layers of sigmoid gates, where the output gate has weights in $\{\pm 1\}$. Second, the first $d-1$ layers of $C$ are simulated approximately. By perturbing the biases, we may assume that the linear form of every gate in $C$ is nonzero on every input. For each threshold gate $H:\vec z\mapsto\Id[\langle\vec w,\vec z\rangle+\beta\ge 0]$, we replace it with a sigmoidal gate 
\[
S_{H,\lambda_0}(\vec z)\eqdef \Sigmoid(\lambda_0\langle\vec w,\vec z\rangle+\lambda_0\beta). 
\]
Note that $\lim_{\lambda_0\to\infty} S_{H,\lambda_0}(\vec z)=H(\vec z)$ (this is similar to the standard simulation in \Cref{lmm: threshold by sigmoidal}). We replace each gate in the first $d-1$ layer by $S_{H,\lambda_0}$, where $\lambda_0$ will be determined later. For each gate $H$ in the $d-1$ layer, we use $\tilde H(\vec x)$ to denote the subcircuit at $H$ in the original circuit, and $\tilde S_{H,\lambda_0}(\vec x)$ to denote the subcircuit at $S_{H,\lambda_0}$ after the transformation. We can see by induction on the layers that for every input $\vec x\in\zo^n$, $\lim_{\lambda_0\to\infty}\tilde S_{H,\lambda_0}(\vec x)=\tilde H(\vec x)$. 

Let $D_{\lambda,\delta}(\vec z)$ and $P_{\lambda}(\vec z)$ be the functions as in \eqref{equ: difference sim} and \eqref{equ: derivative sim} to approximate the top two layers, where $\vec z\in\zo^{\ell}$ denotes the input of the top two layers, i.e., the outputs of the gates in layer $d-1$. Let $H_1,\dots,H_\ell$ be the gates in layer $d-1$. The final circuit is of form 
\[
C(\vec x)\eqdef \Sigmoid\Big(D_{\lambda,\delta}(\tilde S_{H_1,\lambda_0}(\vec x),\dots,\tilde S_{H_\ell,\lambda_0}(\vec x))\Big). 
\]
Let $\vec z = \vec z(\vec x) \eqdef (\tilde H_1(\vec x),\dots,\tilde H_\ell(\vec x))$. Fix $\lambda$ and $\delta$ as before, so that $\sgn(D_{\lambda,\delta}(\vec z))=\sgn(P(\vec z))$ for every $\vec z\in\zo^\ell$, where $\sgn(P(\vec z(\vec x)))=f(\vec x)$ for every $\vec x\in\zo^n$. As $D_{\lambda,\delta}$ is continuous and $\zo^n$ is finite, we can then fix a large enough $\lambda_0\in\R$ such that the approximation error between $\tilde S_{H_i,\lambda_0}(\vec x)$ and $\tilde H_i(\vec x)$ in the first $d-1$ layers does not change the sign of $D_{\lambda,\delta}(\vec z)$ for every $\vec x\in\zo^n$. Subsequently, $C(\vec x)$ computes $f(\vec x)$ with no separation, and $C$ has $\poly(n,s)$ gates.
\end{proof}


\newcommand{\etalchar}[1]{$^{#1}$}

\appendix
\clearpage
\section{Further Related Work}

\paragraph{ReLUs, RBMs, and piece-wise polynomial activations.}
Martens, Chattopadhyay, Pitassi, and Zemel \cite{martens2013representational} studied the representational efficiency of Restricted Boltzmann Machines (RBMs) \cite{hinton1986learning}, characterizing the RBM free energy as a depth-$2$ network of ``softplus'' ($\ln(1+e^y)$) units with unit output weights and relating it to ``hardplus'' (ReLU) networks and threshold circuits. %
While they showed that size-$O(n^2)$ RBMs can represent any symmetric distribution, they proved $2^{\Omega(n)}$ size lower bounds for computing $\mathrm{IP2}_n$ assuming weights are polynomially bounded. %
Similarly, Maass \cite{Maass97} showed that for networks with piece-wise polynomial activations, polynomial rational weights suffice up to a polynomial size overhead. %

In the unbounded-weight regime, Vardi, Reichman, Pitassi, and Shamir \cite{VRPS21} (see also \cite{vardi2020neural}) proved an $\Omega(n/\log n)$ gate lower bound for constant-depth ReLU networks computing $\IPtwo_n$, contrasting with our $O(\sqrt{n\log n})$ sigmoidal upper bound (Theorem~\ref{thm: IP2 sublinear Sigmoid}).

For functions with \emph{real-valued} inputs, Maass \cite{maass1997networks} established a stronger $\Omega(n)$ gate lower bound against unrestricted sigmoidal networks of arbitrary depth (with zero separation, and even allowing a mix of sigmoidal and threshold gates) computing the Element Distinctness function $\mathrm{ED}_n : (\mathbb{R}^+)^n \to \{0,1\}$, which outputs $1$ if $x_i = x_j$ for some $i \neq j$ and $0$ if $|x_i - x_j| \ge 1$ for all $i \neq j$. %
Crucially, Maass's proof exploits the continuous input domain $(\mathbb{R}^+)^n$ to scale the first input coordinate by a single programmable multiplier $\lambda \in \mathbb{R}^+$, allowing a network with $k$ gates to shatter \emph{every} subset $S \subset \mathbb{R}^+$ of size $n-1$ using $k+1$ programmable parameters; invoking Sontag's $(2w+1)$ upper bound \cite{Sontag97} on the ``Sontag dimension'' (the maximum $d$ such that \emph{all} sets of $d$ points in general position are shattered by $w$ parameters) immediately yields $k \ge (n-4)/2$. 

Because this scaling and general-position argument requires \emph{continuous} inputs, the best known gate lower bound for \emph{Boolean-input} functions against unrestricted sigmoidal circuits of general depth remains $\Omega(n^{1/4})$ \cite{Koiran96}. 

Size lower bounds for neural networks with continuous inputs and outputs are known. These apply for a variety of activation functions including signoids. 
Eldan and Shamir~\cite{eldan2016power} provide a construction of a function that can be approximated in the $\ell^2$-norm by a depth $3$ network but requires exponential size to be approximated by a depth-2 network. Safran, Reichman and Valiant~\cite{safran2024many} prove a linear size lower bounds for neural networks approximating the maximum function: their lower bound holds regardless of the depth of the network and includes sigmoidal networks. These lower bounds rely on the continuity of the domain in essential ways~\cite{SRV25} and do not imply size lower bounds for sigmooidal networks computing Boolean functions. 

\paragraph{Communication complexity.}
Communication complexity \cite{KN97} is a standard tool for depth-$2$ threshold circuit lower bounds, yielding $\Omega(n)$ gate lower bounds for $\mathrm{IP2}_n$ against $\mathsf{THR}\circ\mathsf{THR}$ \cite{Nisan93, ROS94} and $2^{\Omega(n)}$ lower bounds against $\mathsf{MAJ}\circ\mathsf{THR}$ \cite{Razborov92-THR, HMPSMT93} and $\mathsf{THR}\circ\mathsf{MAJ}$ \cite{FKLMSS01, Forster02, LSS08, Sherstov13, ACHS24}. %
While real communication complexity has been studied in connection with analog networks \cite{Koiran96}, it does not directly yield gate lower bounds for Boolean-input sigmoidal circuits with unrestricted weights, since a single real-valued gate output can encode $\Omega(n)$ bits of information (as each boolean input could have a different output, communicating one output out of $2^n$ would require $n$ bits).%
\section{Missing Proofs}
\label{sec: missing proofs}

\subsection{Proof of the Parameter Compression Lemma}
\label{sec: compression}

\SigmoidCompression*

\begin{proof}
    Let $g(\vec z)\eqdef\Sigmoid(\langle\vec w,\vec z\rangle+b)$ and $B\eqdef\lceil 1/\eps\rceil+1$, so that $1/B<\eps$ and $B\le 3/\eps$. For every $\vec z\in\zo^m$, pick $j_{\vec z}\in\{0,1,\dots,B-1\}$ such that $g(\vec z)\in[j_{\vec z}/B,(j_{\vec z}+1)/B]$. As $\Sigmoid$ is increasing, $\Sigmoid(\langle\vec w',\vec z\rangle+b')$ lies in the same interval if and only if
    \[
    \Sigmoid^{-1}(j_{\vec z}/B)\le\langle\vec w',\vec z\rangle+b'\le\Sigmoid^{-1}((j_{\vec z}+1)/B),
    \]
    where for the boundary cases, the left (resp.~right) inequality is dropped if $j_{\vec z}=0$ (resp.~$j_{\vec z}=B-1$). As $\Sigmoid^{-1}(y)=\ln(y/(1-y))$, these bounds have absolute values at most $\ln(B-1)$. 
    
    Let $P\subseteq\R^{m+1}$ be the set of $(\vec w',b')$ satisfying all these inequalities, and in addition having every coordinate of the same sign as the corresponding coordinate of $(\vec w,b)$ (e.g., $w'_i\ge 0$ if $w_i\ge 0$, and $w'_i\le 0$ otherwise). Then $P$ is a polyhedron that contains $(\vec w,b)$. Because of the sign constraints, $P$ contains no line, so it has a vertex. Each defining inequality of $P$ has the form $\langle\vec a,(\vec w',b')\rangle\le c$ with $\vec a\in\{0,\pm 1\}^{m+1}$ and $|c|\le\ln(B-1)$, and each vertex is the unique solution of $m+1$ linearly independent such inequalities taken with equality. Fix $(\vec w',b')$ to be any vertex, and write the corresponding $m+1$ tight inequalities as $A\vec u=\vec c$, where $\vec u\eqdef(\vec w',b')$, $A\in\{0,\pm 1\}^{(m+1)\times(m+1)}$ is invertible, and every entry of $\vec c$ has absolute value at most $\ln(B-1)$. By Cramer's rule, $u_i=\det(A_i)/\det(A)$, where $A_i$ is obtained from $A$ by replacing its $i$-th column with $\vec c$. Here $\det(A)$ is a nonzero integer. Moreover, by expanding $\det(A_i)$ along its $i$-th column, we know that
    \[
    |\det(A_i)|\;\le\;\sum_{r=1}^{m+1}|c_r|\cdot m!\;\le\;(m+1)!\cdot\ln(B-1)\;\le\;(m+1)!\cdot\ln(3/\eps),
    \]
    where the second inequality follows from the fact that every $m\times m$ minor of $A$ has entries in $\{0,\pm 1\}$ and hence absolute value at most $m!$. Therefore, for every $\vec z$, both $g(\vec z)$ and $\Sigmoid(\langle\vec w',\vec z\rangle+b')$ lie in an interval of length $1/B<\eps$.
\end{proof}

\subsection{Proof of the Restriction Lemma}
\label{sec: KKL}

\begin{lemma}[{implied by \cite[Lemma 1.6]{KKL17}}]\label{lmm: KKL-PTF}
    Let $f(\vec x)=\sgn(p(\vec x))$, where $p(\vec x)$ is a degree-$d$ multilinear polynomial. Then for every $0<\delta\le 1/16$ and $m\le n/16$,
    \[
    \Pr_{\rho\gets\cR_{m}}\left[f|_\rho\text{ is not }\delta\text{-almost a constant}\right] \le (\sqrt{m/n}+\delta)\cdot (\log(n/m)\cdot \log \delta^{-1})^{O(d^2)}.
    \]
\end{lemma}

Recall that a Boolean function is said to be $\delta$-almost a constant if for some $b\in\zo$, all but a $\delta$ fraction of input evaluates to $b$. We say that a polynomial $p(\vec x)$ is $\delta$-almost, $\eps$-concentrated within $[a,b]$ if there exists $z\in[a,b]$ such that all but a $\delta$ fraction of input $\vec x$, $|p(\vec x)-z|\le \eps$. The following lemma is a simple corollary of the restriction lemma. 

\begin{lemma}\label{lmm: KKL-corollary}
    For any degree-$d$ multilinear polynomial $p(\vec x)$ with $n$ variables, $m\le n/16$, $\delta\in(0,1)$ with $\delta\le\sqrt{m/n}$, $\eps>0$, $a,b\in\R$ such that $b-a\ge\eps$, over a random restriction $\rho\gets \cR_m$, with probability
    \[
    1-O\left(\sqrt{m/n}\cdot (b-a)\eps^{-1}\bigl(\log(n/m)\cdot \log((b-a)\eps^{-1}\delta^{-1})\bigr)^{O(d^2)}\right),
    \]
    at least one of the following three conditions holds:
    \begin{compactitem}
    \item $p|_\rho(\vec z)\le a$ for all but $\delta$-fraction of $\vec z\in \zo^m$,
    \item $p|_\rho(\vec z)\ge b$ for all but $\delta$-fraction of $\vec z\in \zo^m$,
    \item $p|_\rho$ is $\delta$-almost $\eps$-concentrated within $[a,b]$.
    \end{compactitem}
\end{lemma}

\begin{proof}
    Set $N\eqdef \lceil (b-a)/\eps\rceil$, $\eps'\eqdef (b-a)/N\in[\eps/2,\eps]$, and $\tau_i\eqdef a+i\eps'$ for $i=0,1,\dots,N$, so $\tau_0=a$ and $\tau_N=b$. We adopt the convention $\sgn(x)=1$ if $x>0$ and $\sgn(x)=0$ otherwise. For each $i$, let $g_i(\vec x)\eqdef\sgn(p(\vec x)-\tau_i)$, the sign of a degree-$d$ multilinear polynomial. Let $\delta'$ be a parameter to be determined later.

    Applying \Cref{lmm: KKL-PTF} to each $g_i$ and by union bound, with probability at least
    \begin{equation}
    1-(N+1)\cdot (\sqrt{m/n}+\delta')\cdot \bigl(\log(n/m)\cdot \log(\delta'^{-1})\bigr)^{O(d^2)},\label{equ: failure prob A1}
    \end{equation}
    every $g_i|_\rho$ is $\delta'$-almost a constant; that is, for each $i$ there exists $c_i\in\zo$ such that the set $B_i\eqdef\{\vec z\in\zo^m:g_i|_\rho(\vec z)\ne c_i\}$ has $|B_i|/2^m\le \delta'$.

    Condition on this event. Let $B\eqdef\bigcup_{i=0}^{N} B_i$, where $|B|/2^m\le (N+1)\delta'$. We set $\delta'\eqdef\min\{\delta/(N+1),1/16\}$ so $|B|/2^m\le\delta$, and \Cref{lmm: KKL-PTF} applies as $\delta'\le 1/16$. In particular, there exists a $\vec z_0\notin B$. Subsequently, the tuple $(c_0,\dots,c_N)$ is non-increasing: assume not, i.e., $c_{i}=0$ but $c_{i+1}=1$, we will have
    \[
    g_i|_\rho(\vec z_0) = c_i = 0, \quad g_{i+1}|_\rho(\vec z_0) = c_{i+1} = 1,  \tag{by $\vec z_0\notin B$}
    \]
    which is impossible by the definition of $g_i$'s. Let $i^*\in\{-1,0,\dots,N\}$ be the index of the last $1$ in $(c_0,\dots,c_N)$ (with $i^*=-1$ if all entries are $0$). If $i^*=-1$, then $p|_\rho(\vec z)\le\tau_0=a$ for every $\vec z\notin B$, satisfying the first bullet. If $i^*=N$, then $p|_\rho(\vec z)>\tau_N=b$ for every $\vec z\notin B$, satisfying the second bullet. Otherwise $i^*\in\{0,\dots,N-1\}$, $\tau_{i^*}<p|_\rho(\vec z)\le \tau_{i^*+1}$ for any $\vec z\notin B$, satisfying the third bullet. It then follows by invoking the parameters to \Cref{equ: failure prob A1}, using $\delta'\le\delta\le\sqrt{m/n}$ so that $\sqrt{m/n}+\delta'\le 2\sqrt{m/n}$.
\end{proof}

\PSigRestriction*

\begin{proof}
    Apply \Cref{lmm: KKL-corollary} to $p$ with interval $[-(k+1)\ln t, (k+1)\ln t]$, $\eps\eqdef 1/t$, and $\delta\eqdef \delta$. Over $\rho\gets\cR_m$, with probability at least
    \begin{align*}
    & 1-O\left(\sqrt{m/n}\cdot tk\ln t\cdot \bigl(\log(n/m)\cdot \log(tk\ln t\cdot \delta^{-1})\bigr)^{O(d^2)}\right)\\ 
    & \; \ge 1 - O\left(\sqrt{m/n}\cdot kt\cdot \bigl(\log(n/m)\cdot \log(tk/\delta)\bigr)^{O(d^2)}\right),
    \end{align*}
    at least one of the three conditions of \Cref{lmm: KKL-corollary} holds. We show that in each case, $\PSigmoid_p|_\rho$ admits a degree-$(dk)$ polynomial $q:\R^m\to\R$ that is $\delta$-almost $t^{-(k+1)}$-close.
    \begin{compactitem}
    \item (\emph{Case 1}.) Suppose $p|_\rho(\vec z)\le -(k+1)\ln t$ for all but $\delta$-fraction of $\vec z\in\zo^m$. Then $\Sigmoid(p|_\rho(\vec z))\le e^{-(k+1)\ln t}=t^{-(k+1)}$ for these $\vec z$. It suffices to choose the constant polynomial $q(\vec z)\equiv 0$.
    \item (\emph{Case 2}.) Similarly, it suffices to choose the constant polynomial $q(\vec z)\eqdef 1$. 
    \item (\emph{Case 3}.) Suppose $p|_\rho$ is $\delta$-almost $(1/t)$-concentrated within $[-(k+1)\ln t, (k+1)\ln t]$, i.e., there exists $\mu\in[-(k+1)\ln t, (k+1)\ln t]$ such that $|p|_\rho(\vec z)-\mu|\le 1/t$ for all but $\delta$-fraction of $\vec z\in\zo^m$. Let $q_T$ be the degree-$k$ Taylor approximation of $\Sigmoid$ at $\mu$; by Taylor's theorem and \Cref{cor: sigmoid derivative}, $|\Sigmoid(y)-q_T(y)|\le(2/\pi)^{k+1}|y-\mu|^{k+1}\le t^{-(k+1)}$ whenever $|y-\mu|\le 1/t$. Therefore, the degree-$dk$ polynomial $q(\vec z)\eqdef q_T(p|_\rho(\vec z))$ satisfies the condition.
    \end{compactitem}
    This completes the proof. 
\end{proof}

\subsection{Rational Approximation of \texorpdfstring{$\Sigmoid$}{Sigmoid}}
\label{sec: sigmoid rational approx}

In this subsection, we prove \Cref{lmm: sigmoid rational approx}. We will need Newman's theorem on the rational approximation of the absolute value function $|y|$.

\begin{theorem}[Newman's theorem {\cite{Newman64}}; see also {\cite[Section 4.1]{PP87}}]\label{thm: Newman}
    There is an absolute constant $C_N$ such that for every integer $n\ge 1$, there is a rational function $N_n$ of degree at most $n$ with $\bigl|N_n(y)-|y|\bigr|\le C_N e^{-\sqrt{n}}$ for all $y\in[-1,1]$.
\end{theorem}

Similar to \cite[Corollary 3]{PS94}, we need an approximation of the sign function via rational functions when the input is bounded away from $0$. Formally: 

\begin{proposition}[Approximate sign function]\label{prop: approximate indicator}
    For every $B\ge 2$ and $\eta\in(0,1)$, there is a rational function $S$ of degree $O(\ln^2(B/\eta))$ with no pole in $[-B,B]$ such that for every $y\in[-B,B]$, $S(y)\in[-\eta,1+\eta]$; moreover, $S(y)\in[1-\eta,1+\eta]$ if $y\ge 1$, and $S(y)\in[-\eta,\eta]$ if $y\le -1$.
\end{proposition}

\begin{proof}
    Let $n\eqdef\lceil\ln^2(C_N(B+1)/(2\eta))\rceil=O(\ln^2(B/\eta))$ and $N(y)\eqdef(B+1)\cdot N_n(y/(B+1))$, where $N_n$ is the rational function of \Cref{thm: Newman}. Then $\bigl|N(y)-|y|\bigr|\le C_N(B+1)e^{-\sqrt{n}}\le 2\eta$ for $|y|\le B+1$; in particular, $N$ has no pole on $[-B-1,B+1]$. Define
    \[
    S(y)\eqdef\frac12+\frac{N(y+1)-N(y-1)}{4},
    \]
    a rational function of degree at most $2n$. For $y\in[-B,B]$, we have $|y\pm1|\le B+1$, and hence $S(y)$ approximates within an error of $\eta$ the function $\frac12+\frac{|y+1|-|y-1|}{4}$, which equals $1$ for $y\ge 1$, equals $0$ for $y\le -1$, and lies in $[0,1]$ for all $y$.
\end{proof}

We are now ready to prove \Cref{lmm: sigmoid rational approx}, restated below. It follows essentially the proof of \cite[Theorem 10]{SRK94}; for completeness, we provide a self-contained proof. 

\SigmoidRationalApprox*

\begin{proof}
    Let $\eta=\Omega(\eps)$ be a parameter to be determined, and $L\eqdef O(\ln(1/\eps))$ so that $e^{-L+1}\le\eta$. At a high level, we approximate $\Sigmoid(x)$ on $[-a,L]$ using the Taylor expansion of $e^{-x}$, on $(L,a]$ by the constant $1$; the two approximations are combined using the rational approximation of the sign function in \Cref{prop: approximate indicator}. Note that when $L\ge a$, it suffices to use only Taylor approximation. 
    
    Let $T_k(y)=\sum_{j=0}^{k-1} (-y)^j/j!$ be the degree-$(k-1)$ Taylor approximation of $e^{-y}$. It follows that for any $y\in [-(L+1),L+1]$, there exists $\xi$ between $0$ and $y$ such that
    \[
    |T_k(y) - e^{-y}| \le e^{-\xi}\cdot\frac{|y|^{k}}{k!} \le e^{L+1}\cdot\frac{(L+1)^{k}}{(k/e)^k}.
    \]
    Let $k$ be the least odd integer with $k\ge 10(L+1)$, so that $k=O(L)$. Then the error term is at most $e^{L+1}\cdot(e/10)^{10(L+1)}\le e^{-L+1}\le\eta$. Note that $\Sigmoid(x)=\frac{1}{1+e^{-x}}$, and the function $x\mapsto \frac{1}{1+x}$ is $1$-Lipschitz over $\R_+$. Also, since $k-1$ is even, $T_k(y)>0$ for all $y\in\R$. Therefore, the function $r_0(x)\eqdef 1/(1+T_k(x))$ is a degree-$(k-1)$ rational function without poles in $\R$ that approximates $\Sigmoid(x)$ with error $\eta$ for every $x\in[-(L+1),L+1]$.  

    Moreover, for $x\le -(L+1)$, all terms of $T_k(x)$ are non-negative, so $T_k(x)\ge T_k(-(L+1))\ge e^{L+1}-\eta$. In that case, $r_0(x)\le e^{-(L+1)}$ and $\Sigmoid(x)\le e^{-(L+1)}$, and thus $|r_0(x)-\Sigmoid(x)|\le \eta$. 
    
    Now we combine the approximations above. Let $S$ be the rational function of \Cref{prop: approximate indicator} with $B\eqdef a+L$ and error $\eta$, and let 
    \[ 
    r(x)\eqdef r_0(x)+S(x-L)\cdot\bigl(1-r_0(x)\bigr).
    \] 
    For $x\in[-a,a]$ we have $|x-L|\le B$, so $r$ is a rational function of degree at most $k+O(\ln^2(B/\eta))=O(\ln^2(a/\eps))$ with no pole in $[-a,a]$. Write $S_0 = S(x-L)$. 
    \begin{compactitem}
    \item $S_0\in[-\eta,1+\eta]$;
    \item $S_0\in[1-\eta,1+\eta]$ if $x\ge L+1$ and $S_0\in[-\eta,\eta]$ if $x\le L-1$. 
    \end{compactitem} 
    For every $x\in[-a,a]$, we will show that $|r(x)-\Sigmoid(x)|\le \eps$ by a case analysis. Note that 
    \[
    r(x) - \Sigmoid(x) = (r_0(x)-\Sigmoid(x))(1-S_0) + S_0(1-\Sigmoid(x)).
    \]
    For the first term: if $x\le L+1$, then $|r_0(x)-\Sigmoid(x)|\le\eta$ and $|1-S_0|\le 1+\eta$; if $x>L+1$, then $|r_0(x)-\Sigmoid(x)|\le 1$ (as $r_0(x),\Sigmoid(x)\in(0,1)$) and $|1-S_0|\le\eta$. In both cases the first term is at most $\eta(1+\eta)\le 2\eta$. For the second term: if $x\le L-1$, then $|S_0|\le\eta$ and $0\le 1-\Sigmoid(x)\le 1$; if $x>L-1$, then $|S_0|\le 1+\eta$ and $0\le 1-\Sigmoid(x)\le e^{-x}\le e^{-(L-1)}\le\eta$. Therefore, for any $x\in[-a,a]$, 
    \[ |r(x)-\Sigmoid(x)| \le 4\eta.  \] 
    It then suffices to set $\eta \eqdef \eps /4$. 
\end{proof}

\newpage

\section*{AI Methodology}

Following the recommendation of ``AI and TCS: The Next Six Months'' white paper, we describe the usage of AI tools in this research project in detail.  

\begin{flushleft}
\renewcommand{\arraystretch}{1.3}
\begin{tabular}{@{}l@{\qquad}c@{}}
\textbf{Activity} & \textbf{AI used} \\
Asking the research question & No \\
Coming up with the approach & No \\
Proof development & Yes \\
Writing and exposition & No \\
Checking for bugs & Yes \\
Other supporting tasks & Yes \\
\end{tabular}
\end{flushleft}

\paragraph{Details.} We acknowledge the use of various Generative AI tools (including Anthropic Fable 5, Fable 5.1, Opus 5.5, GPT 5.6, and GPT 6) for brainstorming, literature research, solving sub-problems, and composing parts of proofs; all generated contents are reviewed and edited by the authors for clarity. Intellectual contributions of Generative AI tools are summarized below. 
\begin{compactitem}
\item The authors had a (wrong) proof sketch of a super-linear wire lower bound for parity; the idea leads us to the results in \Cref{sec: lb from random restriction}. After that, the upper bound \Cref{thm: symmetric upper bound} (disproving our super-linear wire lower bound) was observed by Fable 5 when authors asked it to formalize the (wrong) proof sketch. The proof was edited extensively by the authors for clarity. 
\item The lower bound in \Cref{sec: wire complexity depth two sigmoids} is first observed by GPT 5.6 Sol in interaction with authors. The current proof (with an explicit comparison to \cite{PS94,SRK94}) is written by the authors.  
\item \Cref{thm: equiv sigmoid threshold} was observed in interaction with GPT 6 Astra. The authors constructed a function separating $\Sigmoid\circ\Sigmoid$ with small weights and $\MAJ\circ\MAJ$, and asked GPT 6 Astra to find more functions separating these classes. The authors observed that all functions are from $\THR\circ\MAJ$, and after asked by the authors, GPT 6 Astra proved the equivalence result. \Cref{thm: sigmoid simulates THR THR} was further proved by GPT 6 Astra as requested to prove similar results for $\THR\circ\THR$ using the same approach. The current proof is simplified and rewritten by the authors with the help of Claude Opus 5.5 and Fable 5.1.
\end{compactitem}
Most of the queries to LLMs are given in the context of an earlier draft of the paper, which we do not attempt to describe in detail. 

Other parts of the paper (e.g.~\Cref{sec: lb from random restriction} and \Cref{sec: weights depth two}) are not significantly affected by Generative AI tools.  
Generative AI tools (ranging from Claude Opus 4.7 to Fable 5.1, as well as GPT 5.5 to GPT 6) are also used in routine tasks (calculating the parameters with a given proof skeleton, checking bugs, typesetting, etc) that do not constitute substantial conceptual contributions.

All AI-generated content was independently verified and edited by the authors, who take full responsibility for the correctness of all claims and proofs in this paper.

\end{document}